\documentclass[10pt]{IEEEtran}

\usepackage{times,amsmath,epsfig}
\usepackage[table]{xcolor}%
\usepackage{textcomp}
\usepackage{mathtools}
\usepackage[super]{nth}
\usepackage{color}
\usepackage{colortbl}
\usepackage{multirow}
\usepackage{graphicx}
\usepackage{textcomp}
\usepackage{cases}
\usepackage{multirow}
\usepackage{multicol}
\usepackage{todonotes}
\usepackage{tablefootnote}
\usepackage{booktabs}
\usepackage{textgreek}
\usepackage{latexsym}
\usepackage{amsfonts,amssymb,amsmath}
\usepackage{hyperref}
\usepackage[flushleft]{threeparttable}

\usepackage{xcolor}

\usepackage{amssymb}
\usepackage{pifont}
\usepackage{tablefootnote}
\usepackage{booktabs, tabularx}
    \usepackage[flushleft]{threeparttable}

\usepackage{graphicx}
\usepackage{cite}
\usepackage{amsmath,amssymb,amsfonts}
\usepackage{algorithmic}
\usepackage{gensymb}
\usepackage{subfigure}
\usepackage{graphicx}
\usepackage{textcomp}
\usepackage{cases}
\usepackage{multirow}
\usepackage{multicol}
\usepackage{todonotes}
\usepackage{tablefootnote}
\usepackage{booktabs}
\usepackage{textgreek}
\usepackage{threeparttable}
\usepackage{latexsym}
\usepackage{amsfonts,amssymb,amsmath}
\usepackage{hyperref}
\usepackage{xcolor}
\newcommand{\xmark}{\ding{55}}%
\newcolumntype{P}[1]{>{\centering\arraybackslash}p{#1}}

\renewcommand{\baselinestretch}{0.95}

\begin{document}

\title{Comparative Analysis of  Radar Cross Section Based UAV Classification Techniques\\
}

\author{%
Martins Ezuma\IEEEauthorrefmark{1}, Chethan Kumar Anjinappa\IEEEauthorrefmark{1}, Vasilii Semkin\IEEEauthorrefmark{2}, and Ismail Guvenc\IEEEauthorrefmark{1}\\
\IEEEauthorblockA{\IEEEauthorrefmark{2}Dept. of Electrical and Computer Engineering, North Carolina State University, Raleigh, NC 27606\\
\{mcezuma, canjina, iguvenc\}@ncsu.edu.}\\
\IEEEauthorblockA{\IEEEauthorrefmark{2}
VTT Technical Research Centre of Finland, 02150 Espoo, Finland ({vasilii.semkin@vtt.fi})}\vspace{-2mm}%
 \thanks{This work has been supported in part by the National Aeronautics and Space Administration (NASA) under the Federal Award ID number NNX17AJ94A. The work of V. Semkin was supported in part by the Academy of Finland.}
 \thanks{Also, the authors are grateful to
Mr. Kenneth Ayotte and the management of the Ohio State
University Electroscience Laboratory for helping out with
the experiments.}}


\maketitle

\begin{abstract}
This work investigates the problem of unmanned aerial vehicles (UAVs) identification using their radar cross-section~(RCS) signature. The RCS of six commercial UAVs are measured at 15 GHz and 25 GHz in an anechoic chamber, for both vertical-vertical and  horizontal-horizontal polarization. The RCS signatures are used to train 15 different classification algorithms, each belonging to one of three different categories: statistical learning (SL), machine learning (ML), and deep learning (DL). The study shows that while the classification accuracy of all the algorithms increases with the signal-to-noise ratio (SNR), the ML algorithm achieved better accuracy than the SL and DL algorithms. For example, the classification tree ML achieves an accuracy of 98.66\% at 3 dB SNR using the 15 GHz VV-polarized RCS test data from the UAVs. We investigate the classification accuracy using Monte Carlo analysis with the aid of boxplots, confusion matrices, and classification plots. On average, the accuracy of the classification tree ML model performed better than the other algorithms, followed by the Peter Swerling statistical models and the discriminant analysis ML model. In general, the classification accuracy of the ML and SL algorithms outperformed the DL algorithms (Squeezenet, Googlenet, Nasnet, and Resnet 101) considered in the study. Furthermore, the computational time of each algorithm is analyzed. The study concludes that while the SL algorithms achieved good classification accuracy, the computational time was relatively long when compared to the ML and DL algorithms. Also, the study shows that the classification tree achieved the fastest average classification time of about 0.46~ms.
\end{abstract}


\maketitle
\section{Introduction}
\label{sec:introduction}
\IEEEPARstart{u}nmanned aerial vehicles (UAVs) are aircrafts designed to operate without an onboard human pilot. These vehicles can function autonomously, controlled by a remote controller (ground station), or by a hybrid of both means. Depending on applications and mode of operations, UAVs are sometimes referred to as drones, remotely piloted vehicles (RPV), remotely operated aircraft, remotely piloted aircraft system (RPAS), and micro-aerial vehicles\cite{granshaw2018rpv}. In the past, UAVs were designed for military applications such as reconnaissance, aerial surveillance, and tactile aerial attack on ground targets. However, in recent times, new use cases for UAVs are arising in public safety, government, and civilian/commercial applications~\cite{shakhatreh2019unmanned}. On the other hand, there is an increase in the use of unauthorized UAVs or rogue drones for malicious purposes. Therefore, there is a need to correctly detect and identify UAVs in the airspace. However, in recent times, more worrying is the use of UAVs by non-state actors for acts of terrorism. Also, some hobbyists have used UAVs to encroach in no-fly zones. These new public safety concerns have raised interest in the design of counter-unmanned aircraft systems (C-UAS) technologies and automatic target recognition (ATR) of UAVs.

Radars are valuable for ATR applications. In its active mode of operation, radar can detect a target by sending electromagnetic radiations and receiving the backscattered signal~\cite{richards2010principles}. In comparison to other UAV detection modalities like acoustics and video cameras (computer vision) techniques, radars are valuable because they yield wide coverage in both azimuth and elevation planes, achieve long detection distances, and can effectively operate in harsh weather conditions such as fog where visibility is poor~\cite{richards2010principles,guvenc2018detection}. The all-weather capability attributed to most radars can be explained by the fact that at their operational frequencies, rain, cloud, and other atmospheric attenuation are relatively small in comparison with electro-optical and infrared (IR)-based detection system~\cite{richards2010principles}. Also, some military-grade radar systems exploit the refractive effect of the earth's ionosphere to achieve very long range (over-the-horizon) target detection~\cite{richards2010principles}. Moreover, unlike radio frequency (RF)-based detection techniques (see e.g.~\cite{ezuma2019detection,ezuma2019micro}), radars can detect a target UAV flying in autonomous mode (no communication channel). Once a radar target has been detected, the process of classifying or identifying the target often requires the exploitation of any information imprinted on the scattered signals (target echo) by the target~\cite{lang2020comprehensive}. However, detection of UAVs using radars could be tricky due to the small RCS of these targets as compared to larger aircraft~\cite{guvenc2018detection} or other objects, like cars~\cite{Schipper11}.

In this study, we focus on UAV identification using the backscattered echo power or radar cross-section (RCS) obtained when the target is illuminated by a surveillance radar. We perform a comparative analysis of different RCS-based techniques or methods for UAV classification. The methods investigated include the SL approach (single distribution and Gaussian mixture models), ML technique, and DL approach. The different algorithms will be evaluated using the RCS data measured from six different UAVs at 15~GHz and 25~GHz. The reason we measured the UAV RCS at 15 and 25 GHz is mainly informed by the trend in the commercial counter-UAV radar industry and academic research. For example, Fortem Technologies and Luswave Technology are providing smart counter-UAV radar sensors at 15 and 25 GHz, respectively~\cite{fortem_radar1,LUSWAVE_radar1}. The average classification performances of the different algorithms are evaluated at different noise levels by varying the signal-to-noise ratio (SNR). Further analysis is provided by means of the Monte Carlo analysis, with emphasis on low SNR conditions. Box plots, classification plots, and confusion matrices are generated to analyze the performance of the different classifiers. Also, the complexities of the algorithms are evaluated by means of the average computational time needed to identify/classify an unknown UAV.

\begin{table}[t!]
\setlength{\tabcolsep}{4.8pt}
\centering
\caption{Literature Review of RCS-based detection/Identification of UAVs.}
\label{LITERATURE_REVIEW}
\begin{threeparttable}
 \begin{tabular}{|P{12mm}|P{10mm}|P{12mm}|P{6mm}|P{10mm}|P{15.5mm}|}
\hline
\textbf{Ref.} & \textbf{Freq. (GHz)} & \textbf{Method} & $\textbf{\#}$ \textbf{of UAVs} & \textbf{Accuracy analysis} & \textbf{Computational time analysis}\\

\hline
\cite{ezuma2021radar} & 15, 25 & SL & 6 & \checkmark & \xmark     \\
\hline
\cite{torvik2016classification} & 3.25 & SL & 2 & \checkmark & \xmark    \\

\hline
~\cite{morris2021detection} & 5.725 & SL & 1 & \checkmark & \xmark  \\
\hline

~\cite{de2018drone,de2019drone} & 8.75 & SL & 1 & \checkmark &  \xmark   \\

\hline
~\cite{semkin2021drone, semkin2020analyzing} & 26-40 & SL & 10 & \checkmark & \xmark  \\
\hline


\cite{pieraccini2017rcs} & 8-12 & SL & 2 & \xmark &  \xmark  \\
\hline

\cite{sedivydrone} & 9 & SL & 9 & \xmark &  \xmark  \\
\hline

\cite{yang2019experimental_1} & 8-12 & SL & 1 & \xmark &  \xmark   \\
\hline
\cite{guay2017measurement} & 8-10 & SL & 1 & \checkmark &  \xmark  \\
\hline
\cite{mohajerin2014feature} & 2.5 & ML & N/A & \checkmark &  \xmark   \\
\hline
\cite{lehmann2020simulation} & 10 & ML & 10 & \checkmark &  \xmark  \\
\hline

\cite{samaras2019uav} & 9.35 & DL & 5 & \checkmark &  \xmark  \\
\hline

\cite{roldan2020dopplernet} & 8.75 & DL & 1 & \checkmark &  \xmark  \\
\hline
\cite{diamantidou2021multimodal} & N/A & DL & N/A &  \checkmark &  \checkmark   \\
\hline
\textbf{Our work} & \textbf{15, 25} & \textbf{SL, ML, DL} & \textbf{6} &  \checkmark & \checkmark\\  \hline
\end{tabular}
  \end{threeparttable}
\end{table}

The rest of the paper is organized as follows. Section~\ref{TWO} provides a brief literature review and Section~\ref{THREE} provides a summary description of the UAV RCS data acquisition technique that is used in our analysis. Section~\ref{statistical_techniques}, Section~\ref{machine_learning_techniques}, and Section~\ref{DEEP_LEARNING_FRAMEWORK} describes the various SL, ML, and DL approaches to RCS-based UAV classification, respectively. Section~VII presents our numerical results and a comparative analysis of the considered SL, ML, and DL approaches, while Section~VIII provides some concluding remarks.

\section{Literature Review}\label{TWO}
Radars can classify or identify targets of interest using features such as on-board radar waveform, multidimensional radar images like a micro-Doppler signature, inverse synthetic aperture radar features (ISAR), and one-dimensional radar cross-section (1-D RCS) signature or high-resolution range profiles (HRRPs)~\cite{luo2009micro, chen2014inverse, wang2019feature}. However, since many commercial UAVs do not have on-board surveillance radars like the military-grade UAVs, it is not possible to identify the former by using automatic radar waveform recognition techniques. This has led researchers to focus micro-Doppler, ISAR, and RCS approaches for UAV identification using radars.
 
For example, in recent times, radar micro-Doppler signatures have been investigated for identifying small UAVs~\cite{tahmoush2015review,MartinsDrone}. The micro-doppler technique relies on recognizing the micromotions caused by the rotating propellers of a UAV. The rotating propellers have different motion dynamics from the mainframe of the UAV, producing a different Doppler shift or harmonics that can be processed using time-frequency transform techniques~\cite{tahmoush2015review,chen2019micro}. In~\cite{MartinsDrone}, we showed how the radar micro-Doppler analysis can be used to distinguish a UAV from a walking man and the rotating propeller of a simulated helicopter. Also, in~\cite{li2018numerical}, numerical and experimental analysis of the radar micro-Doppler signature is used for UAV recognition. Also, radar ISAR imaging techniques are used to generate high-resolution radar images for target recognition. Generating ISAR images require the use of 2D or 3D inverse Fourier transform and exploiting the relative motion of the target UAV. In~\cite{lee2017identification,li2016investigation}, the radar ISAR images of 
a multi-rotor UAV platform is exploited for the target recognition. Also, in~\cite{li2018wide}, high-resolution ISAR images are used to distinguish a multi-rotor from a moving car. 

A significant challenge in the use of micro-Doppler and ISAR imaging techniques is a large amount of radar bandwidth required for processing and generating these images~\cite{klaer2020investigation, chen2019micro}. For example, to generate a high-resolution micro-Doppler of a rotating propeller, one needs to generate a pulse repetition frequency that is at least four times the maximum Doppler shift induced by the propeller~\cite{chen2019micro}. For this reason, our focus will be on the 1-D RCS data for UAV recognition. The RCS returns can be processed using a simple radar without a high energy demand (transmit power), resolution (or bandwidth requirement), storage memory, and computational resources~\cite{ehrman2010using,mertens2016ground}. 

Several RCS-based techniques have been employed for UAV radar target recognition. These techniques could be broadly classified as SL, ML, and DL approach. Table~\ref{LITERATURE_REVIEW} provides a brief literature review of RCS-based techniques used for UAV identification. While most of the studies in Table~\ref{LITERATURE_REVIEW}
have focused on X-band radars (8-12 GHz), the UAV identification techniques can be applied to radars with much lower or higher operational frequencies. To the best of our knowledge, the current work is the first to perform a comparative analysis of the accuracy and computational time of the SL, ML, and DL approaches for UAV identification using radars. 
\begin{figure}[t!]
\center{
 \begin{subfigure}[]{\includegraphics[width=0.445\textwidth]{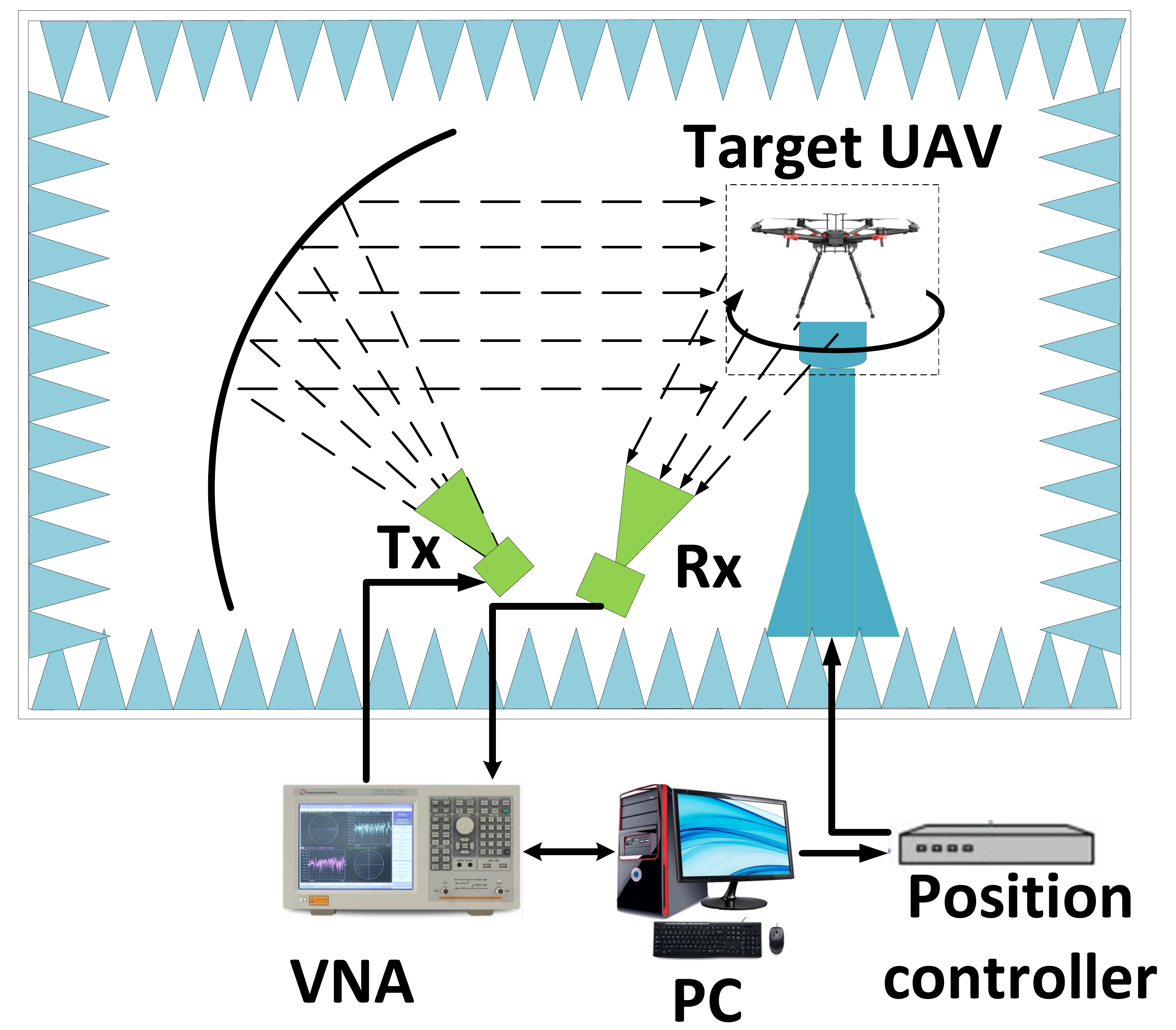}
\label{Fig:setup1}}
\end{subfigure}
 \begin{subfigure}[]{\includegraphics[width=0.45\textwidth]{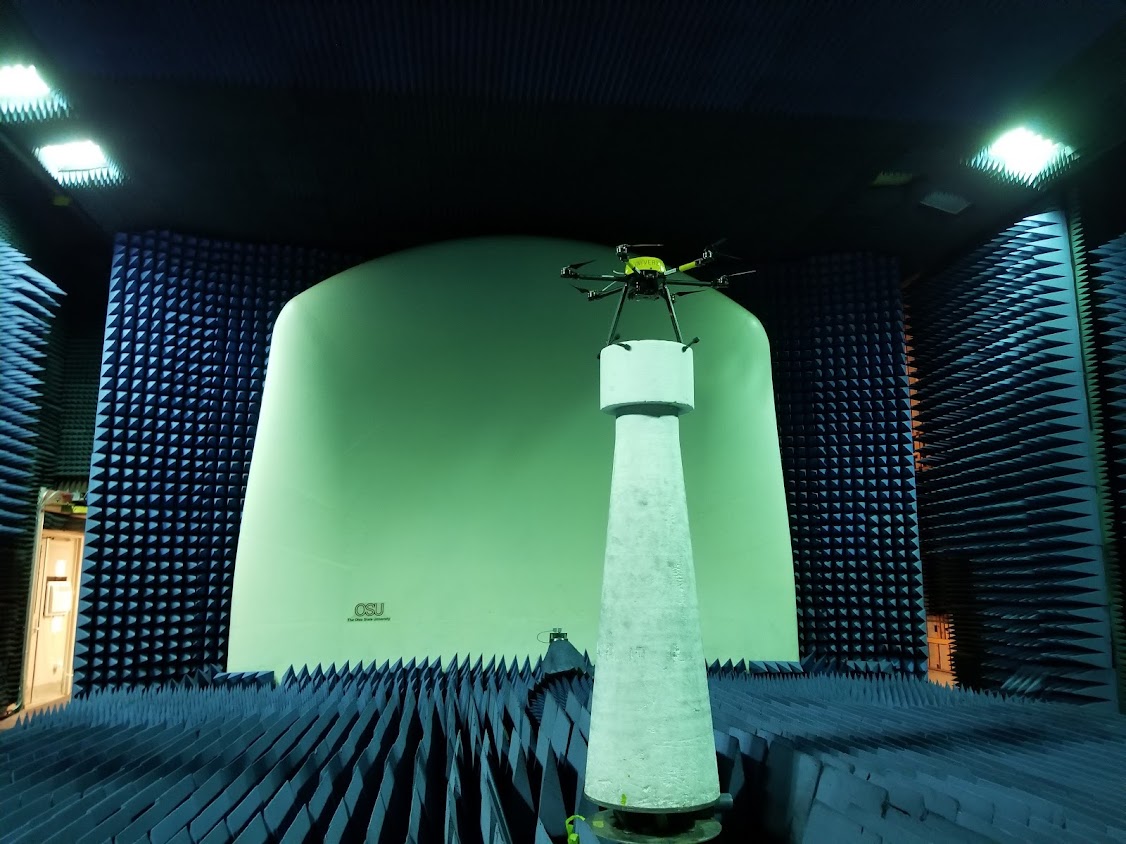}\label{Fig:setup_2}}
\end{subfigure}
\caption{{(a) Schematic of the Indoor RCS measurement of target UAVs. The target UAV is standing on a rotating turntable with low reflectivity, (b) The actual UAV RCS measurement environment shows the target UAV on a low reflective rotating pylon stand (turntable). The walls, floor, and roof of the compact-range anechoic chamber are laced with fidelity RAM. The large white parabolic reflector behind the UAV is used to create plane waves radar illumination to achieve the far-field Fraunhofer condition~\cite{ezuma2021radar}}.}
}
 \end{figure}

\begin{figure}[t]
\center{
\begin{subfigure}[]{\includegraphics[width=0.32\linewidth]{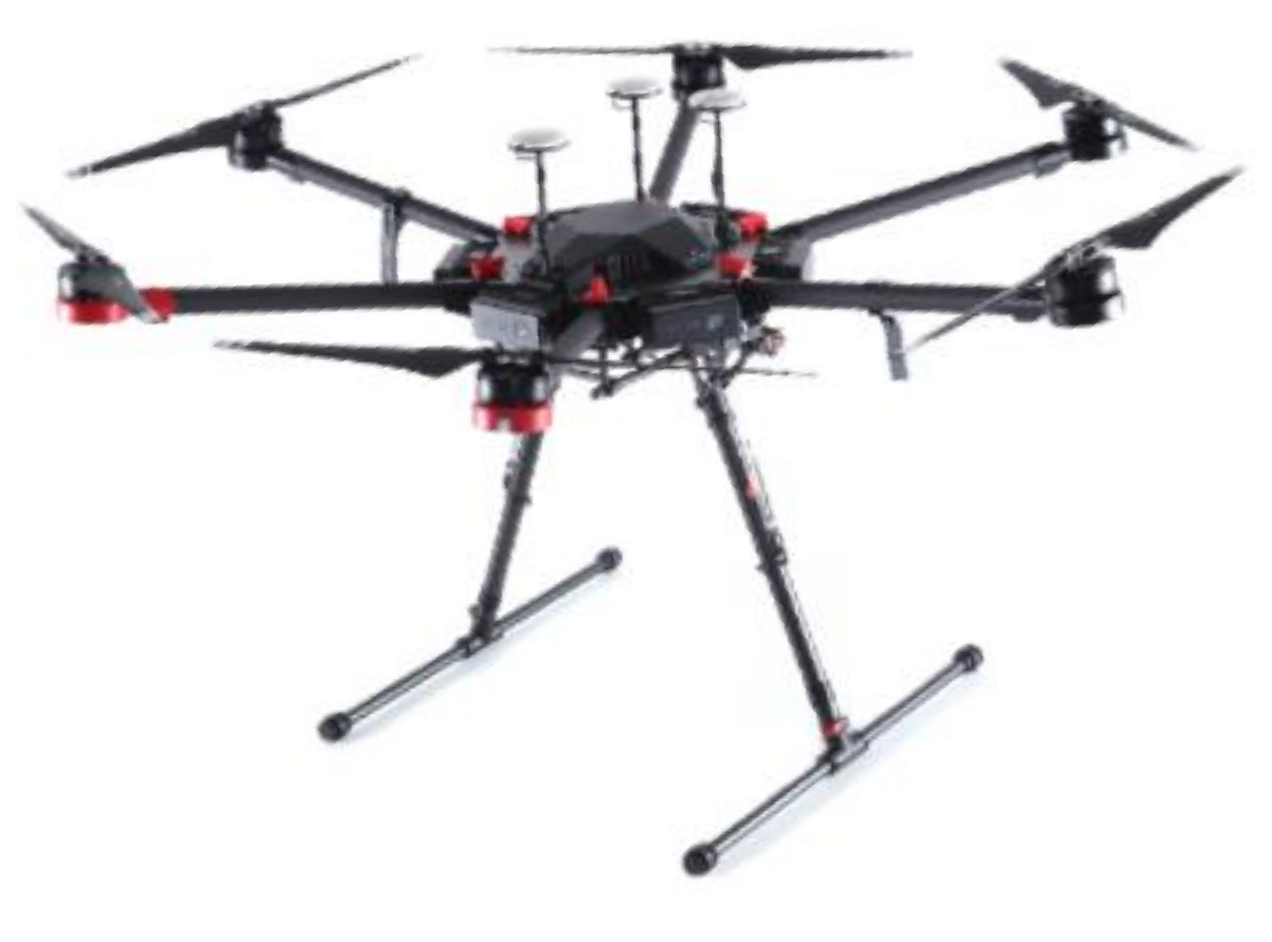}\label{DJI_M600_Pro}}
\end{subfigure}
\begin{subfigure}[]{\includegraphics[width=0.31\linewidth]{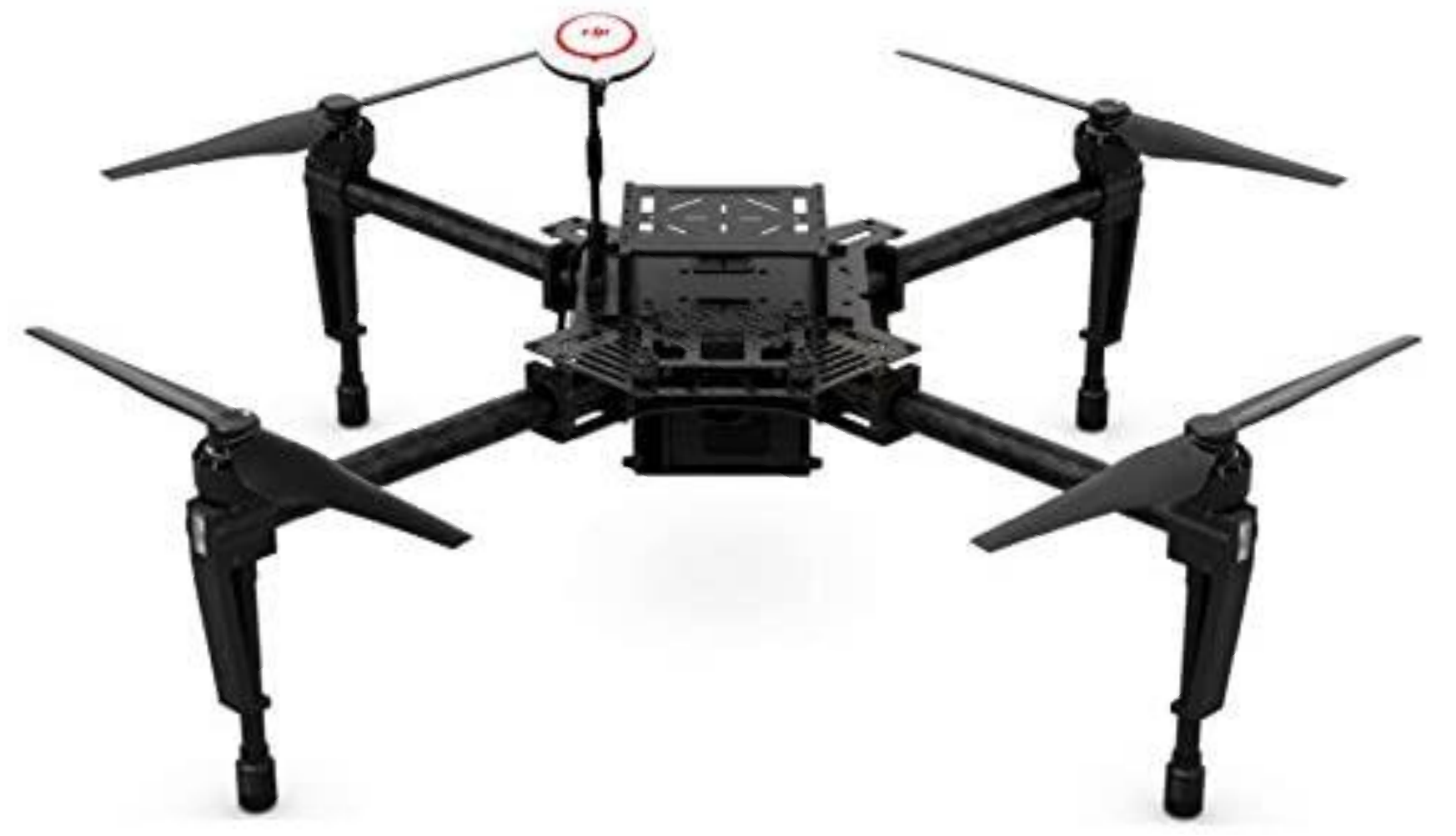}\label{dji_matrice-100}}
\end{subfigure}
\begin{subfigure}[]{\includegraphics[width=0.31\linewidth]{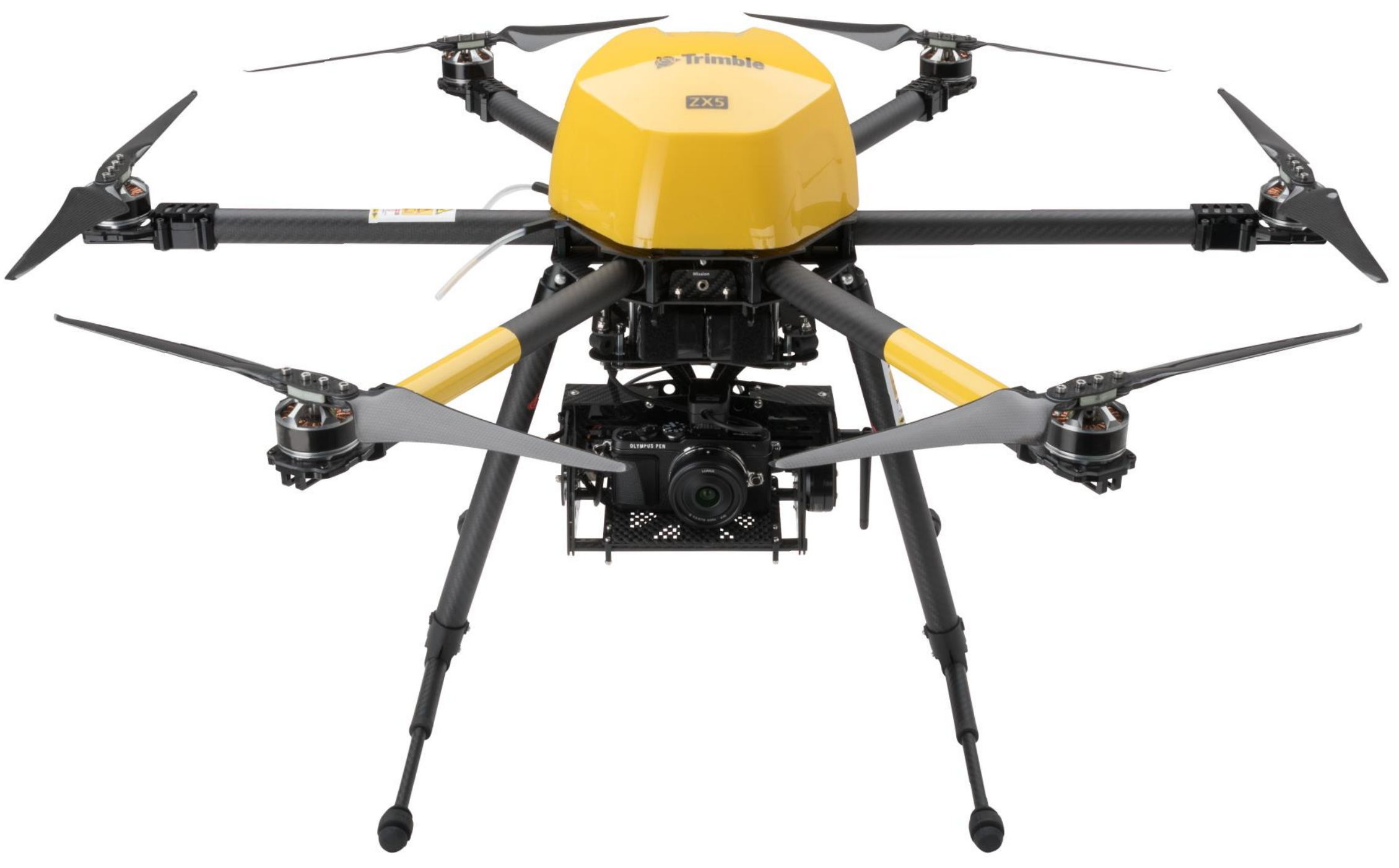}\label{trimble_UAV}}
\end{subfigure}\\
\begin{subfigure}[]{\includegraphics[width=0.3\linewidth]{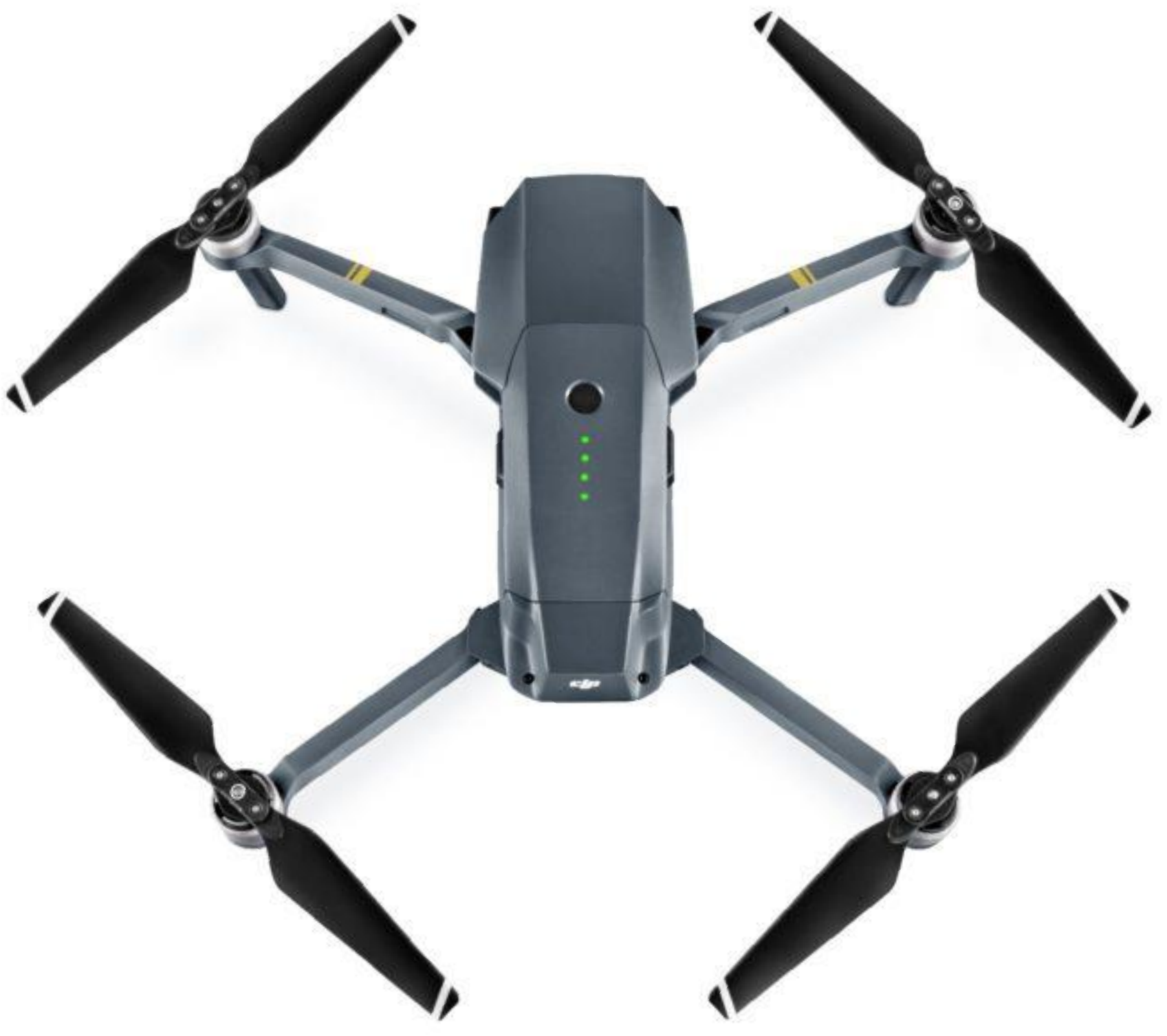}\label{dji_mavic_pro_1}}
\end{subfigure}
\begin{subfigure}[]{\includegraphics[width=0.31\linewidth]{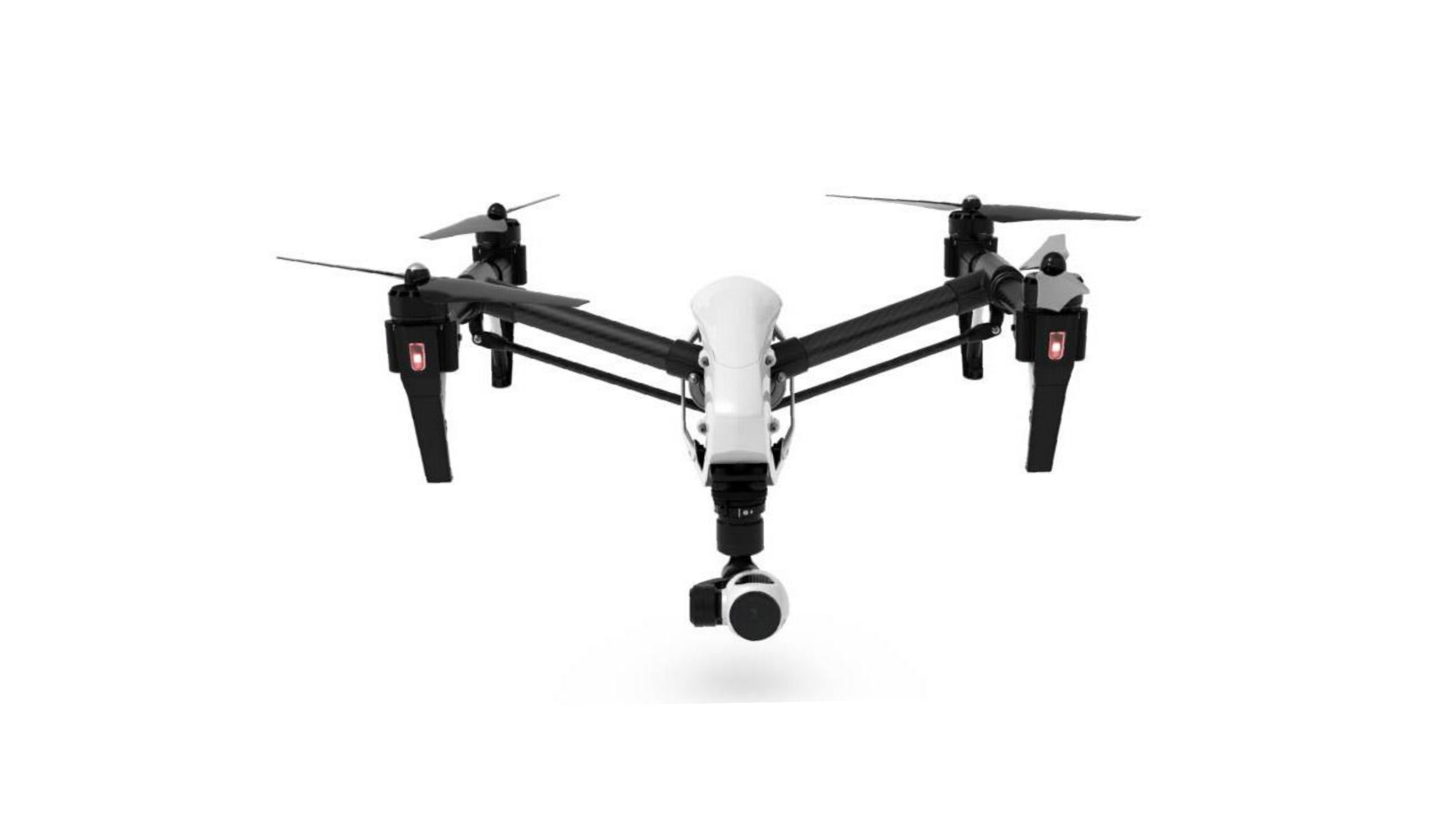}\label{DJI_inspire}}
\end{subfigure}
\begin{subfigure}[]{\includegraphics[width=0.32\linewidth]{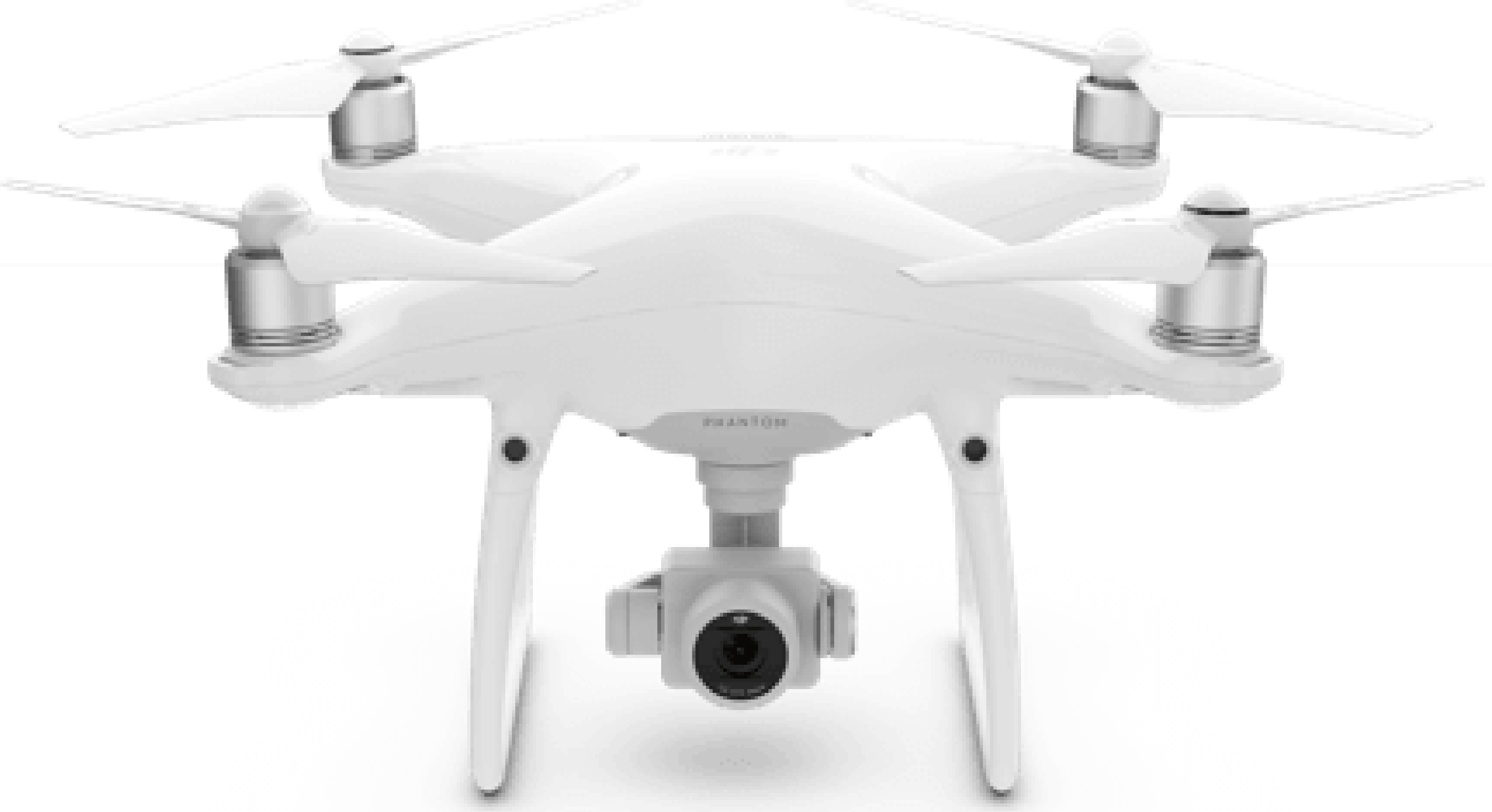}\label{DJI_phantom4pro}}
\end{subfigure}\vspace{-2mm}
 \caption{{The six UAVs considered in this study: (a) DJI Matrice 600 Pro, (b) DJI Matrice 100, (c) Trimble zx5, (d) DJI Mavic Pro 1, (e) DJI Inspire 1 Pro, (f) DJI Phantom 4 Pro.}}
 \label{SMALL_UAVs}}
 \end{figure} 

\begin{figure*}{}
\center{
\begin{subfigure}[]{\includegraphics[width=0.4\linewidth]{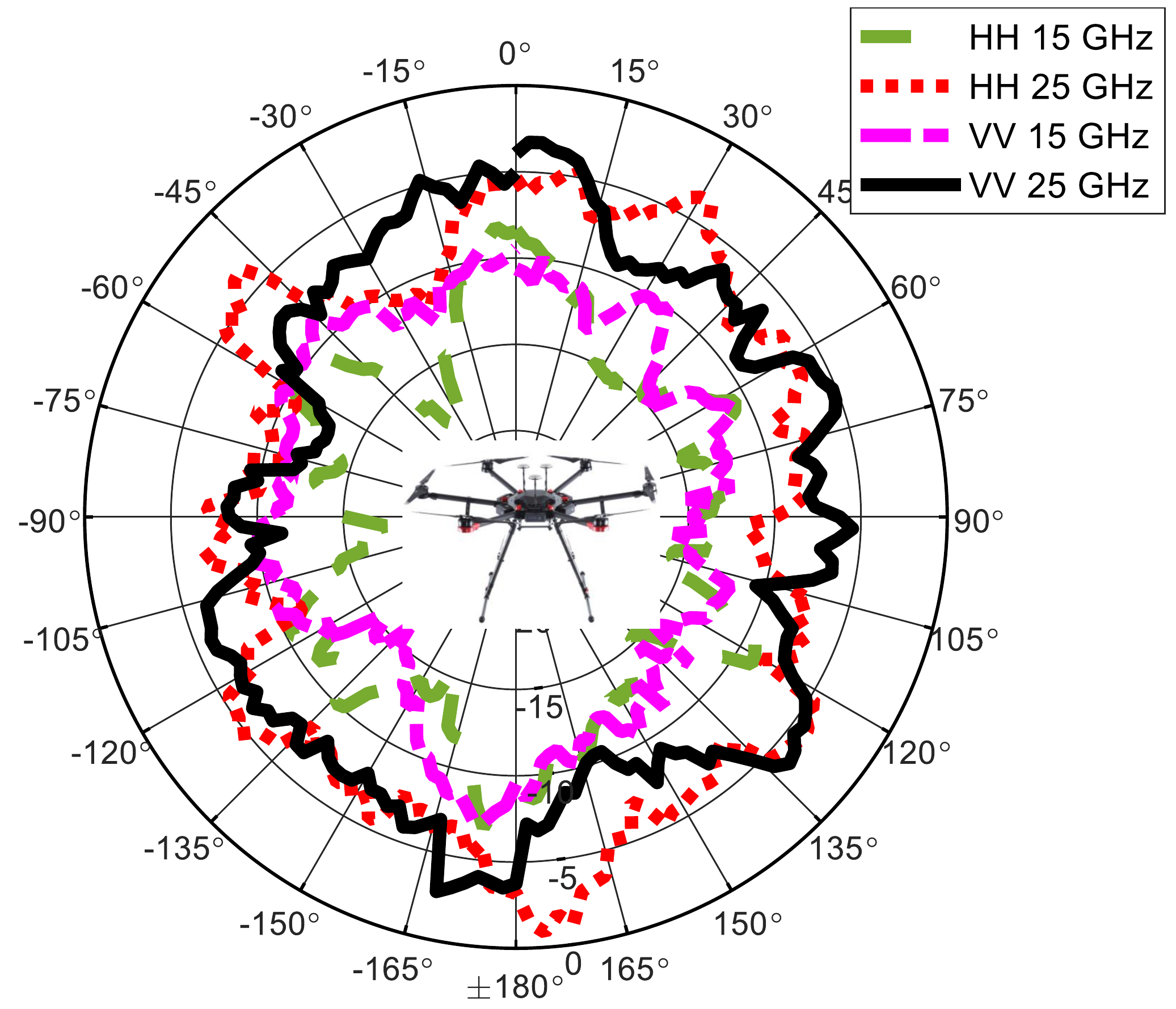}\label{RCS_M600}}
\end{subfigure}
\begin{subfigure}[]{\includegraphics[width=0.4\linewidth]{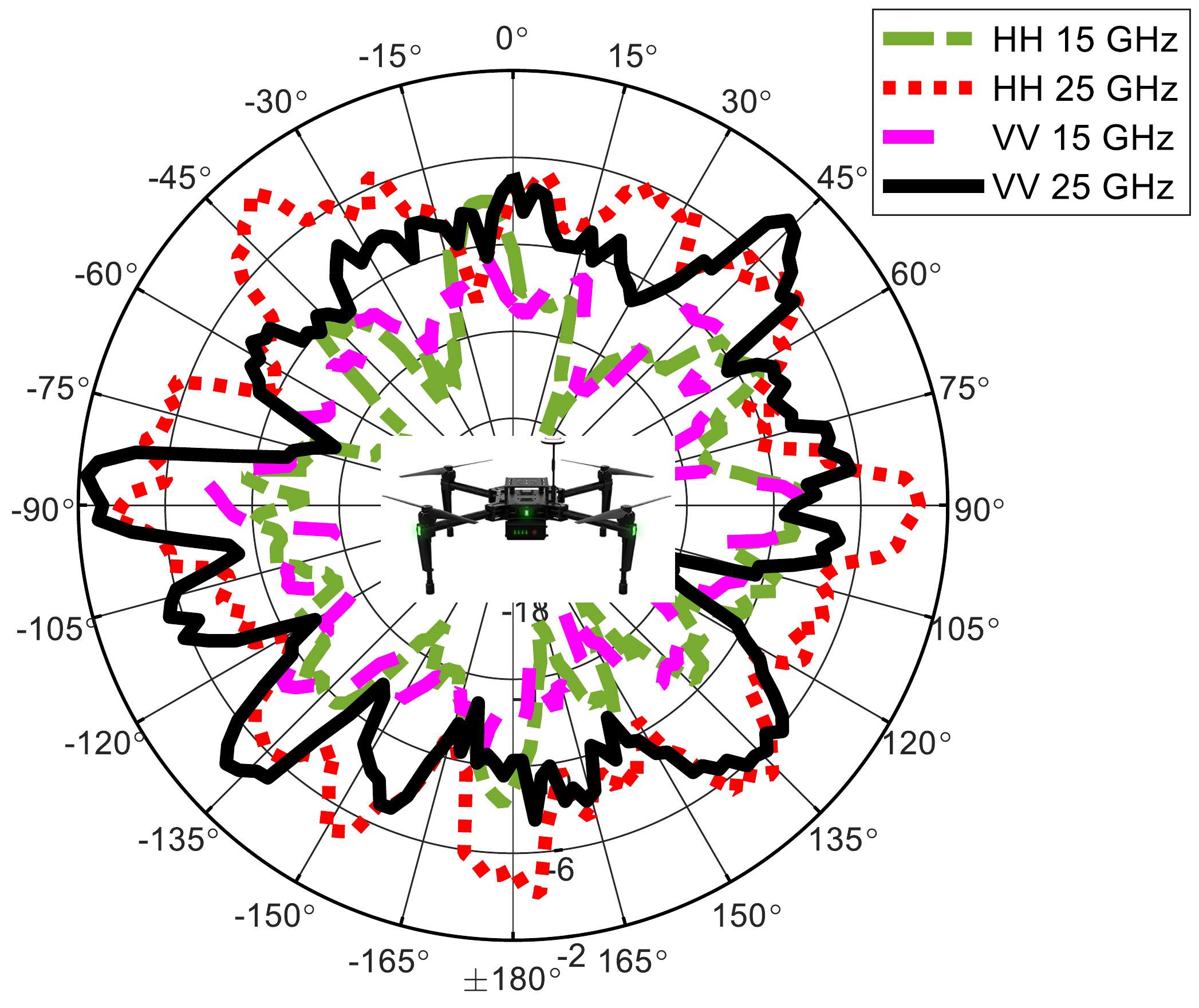}\label{RCS_DJI_M100}}
\end{subfigure}\\
\begin{subfigure}[]{\includegraphics[width=0.4\linewidth]{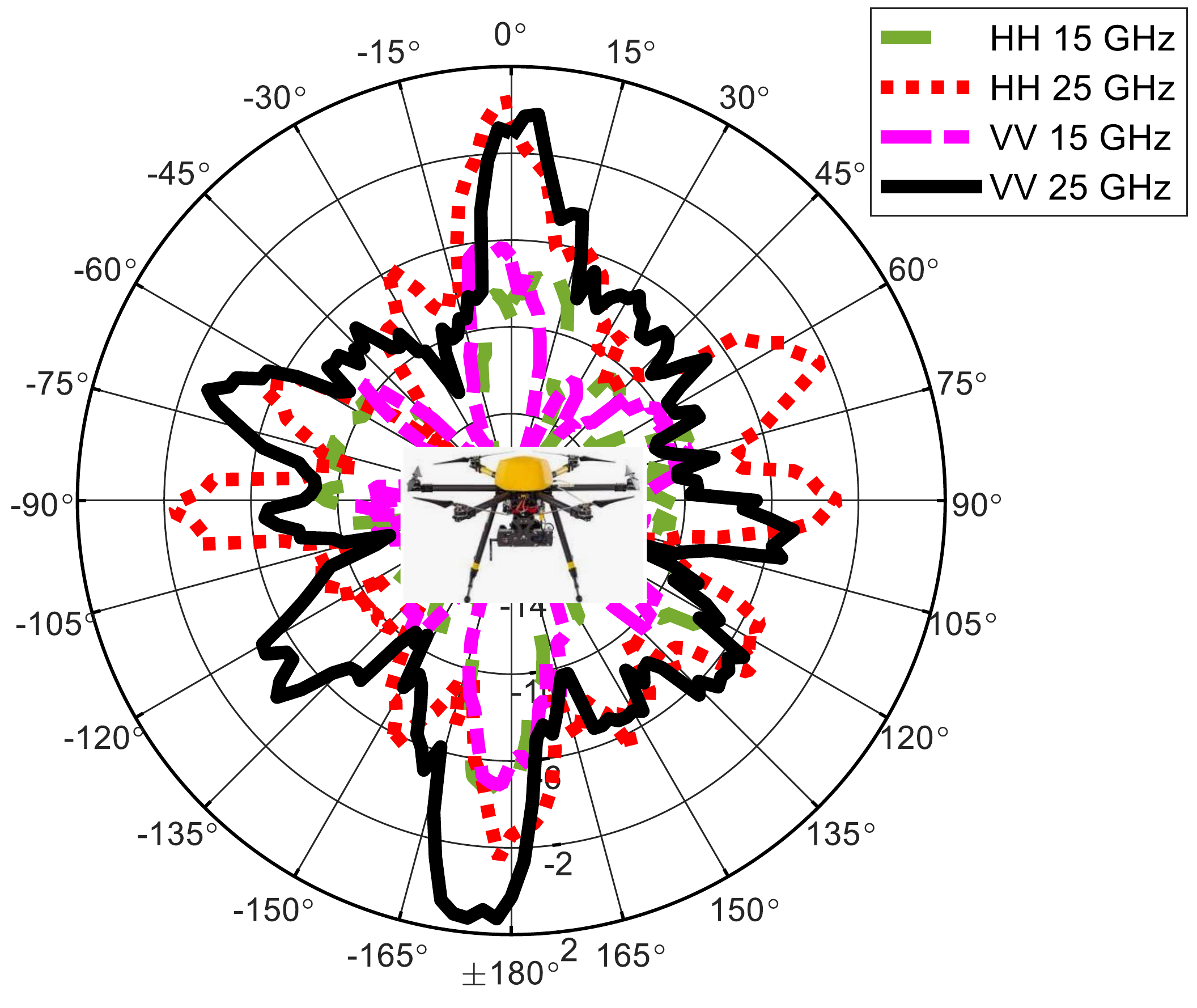}\label{RCS_Trimble}}
\end{subfigure}
 \begin{subfigure}[]{\includegraphics[width=0.4\linewidth]{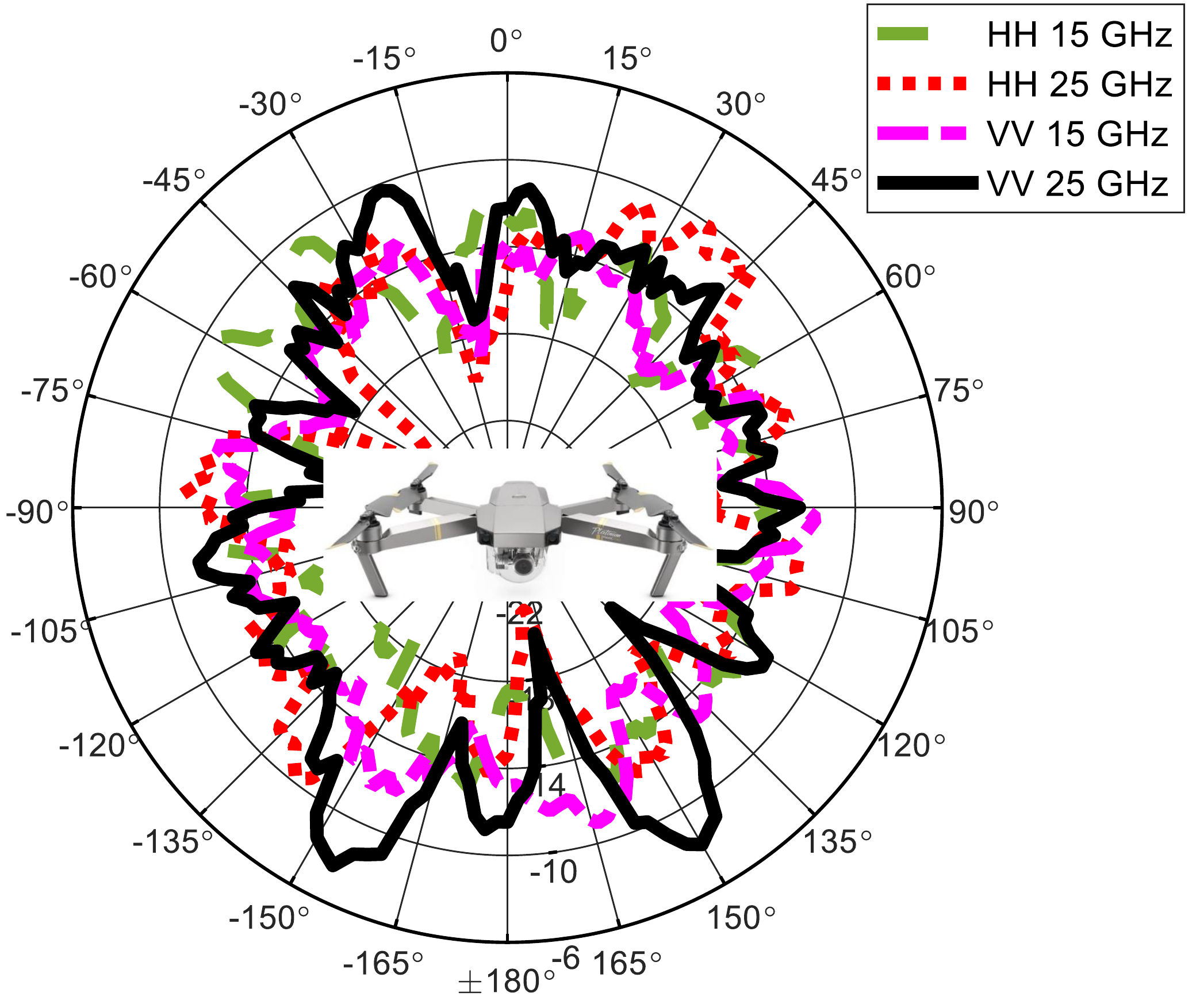}\label{RCS_MAVICPRO}}
\end{subfigure}

\caption{{The measured RCS (dBsm) versus azimuth angles ($\phi\in[0\degree, 360\degree]$) of four different UAVs: (a) DJI Matrice 600 Pro, (b) DJI Matrice 100, (c) Trimble zx5, (d) DJI Mavic Pro 1. The measured RCS signatures are measured at 15 GHz and 25 GHz in both VV and HH polarizations. The RCS signatures highlights the physical features of the target UAVs~\cite{ezuma2021radar}}.}
 \label{RCS_POLAR_PLOT_new}}
 \vspace{-4mm}
 \end{figure*}

\section{RCS Data Acquisition}\label{THREE}
The RCS of a UAV measures its reflectivity when illuminated by a radar. The mathematical definition of RCS is given in~\cite{knott2004radar} as:
\begin{equation} \label{RCSequation_1}
\sigma = \lim_{R\to\infty} 4\pi R^2\frac{|\boldsymbol{\rm{E_s}}|^2}{|\boldsymbol{\rm{E_i}}|^2},
\end{equation}
which states that the RCS ($\sigma$) of a target is the ratio of the incident electric field ($\boldsymbol{\rm{E_i}}$) and the scattered electric field ($\boldsymbol{\rm{E_s}}$) as seen at distance $R$. The RCS of a UAV is a function of the polarization, azimuth, and elevation angles, and operational frequency of the radar as well as the geometry and material properties of the UAV. 

The first step in designing an RCS-based UAV identification system is measuring the RCS of the target of interest as a function of the aspect angle. The UAV RCS measurements are carried out at the compact-range anechoic chamber, Electroscience Laboratory, Ohio State University. Fig.~\ref{Fig:setup1} is a schematic of a compact-range anechoic chamber for measuring the RCS of the target UAVs. Fig.~\ref{SMALL_UAVs} shows the six UAVs considered in the study. The UAVs are DJI Matrice 600 Pro, DJI Matrice 100, Trimble zx5, DJI Mavic Pro, DJI Inspire 1 Pro, and DJI Phantom 4 Pro. Besides the onboard camera, no other payload is placed on the UAV during measurement. The measurements are performed at 15~GHz and 25~GHz. In the schematic shown in Fig.~\ref{Fig:setup1}, radar signals are generated by a vector network analyzer (VNA) and transmitted towards the UAV using a horn antenna (TX) positioned at the focus of a 20-foot parabolic reflector. The parabolic reflector ensures that the target UAV is illuminated by plane waves, thus satisfying the Fraunhofer far-field condition~\cite{ezuma2021radar}. The scattered or reflected signals from the target UAV are captured by the receive antenna (RX). After each azimuth measurement, the target is rotated for the next azimuth measurement. The rotation is done in steps of~2$\degree$ until the entire azimuth plane of the target is covered. Fig.~\ref{Fig:setup_2} shows the actual UAV RCS measurement environment. To minimize or eliminate multipath reflections from the walls, pylon turntable stand, ceiling, and floor, low reflective radar-absorbing materials and pyramidal-shaped radar-absorbing structures are used. After the measurement is completed, the captured data is transformed from the frequency domain to the time domain for further post-processing using a time gating technique. Also, the measurement results are calibrated using standard perfectly electrical conducting spheres with known theoretical RCS values. In~\cite{ezuma2021radar}, we provide additional details of the UAV RCS measurement procedure and the data acquisition process.

For each UAV, RCS measurement is carried out in both the vertical-vertical (VV) and horizontal-horizontal (HH) polarization. Fig.~\ref{RCS_POLAR_PLOT_new} shows the RCS signature of four of the UAVs as a function of the aspect angles, frequency, and polarization. The RCS signature of the UAVs shows the underlying physical features of the target. Therefore, the UAV RCS data captured by radar can be exploited by the SL, ML, and DL algorithms for the target identification. The next section describes the RCS-based UAV identification using SL techniques.

\section{UAV Classification using SL Technique}\label{statistical_techniques}
Mathematically, a complex radar target such as a UAV can be modeled as having many scattering centers which are randomly distributed. Suppose there are $N$ scattering centers on a given UAV, each identified by its own RCS $\sigma_i$ (point reflectivity) and located at a radial distance $R_i$ from the radar. The complex voltage of the backscattered signal (radar echo) from the UAV is given in~\cite{richards2010principles} as:
\begin{eqnarray}
\begin{aligned}
  v(t)  &= \sum_{i=1}^{N}\sqrt\sigma_i e^{j2\pi f(t-\frac{2R_i}{c})}, \\
  &=e^{j2\pi f t}\sum_{i=1}^{N}\sqrt\sigma_i e^{j4\pi \frac{R_i}{\lambda}},
\end{aligned}
\end{eqnarray}
where $c$ is the speed of light while
$f$ and $\lambda$ are the frequency and wavelength of the illuminating radar. Therefore, the echo amplitude $\zeta$ of the backscattered signal from the UAV is proportional to  $\left| v \right|$, and the UAV RCS (echo power) is given~as:
\begin{equation}
  \sigma=  \left| \zeta \right|^2= \left| e^{j2\pi f t}\sum_{i=1}^{N}\sqrt\sigma_i e^{j4\pi \frac{R_i}{\lambda}} \right|^2.
   \label{RCS_distribution}
\end{equation}

Suppose a field surveillance radar captures the RCS data $\boldsymbol{\sigma}=(\sigma_1,\cdots,\sigma_n)$ from an unknown UAV. Given that the field surveillance radar have been trained to recognize $M$ different UAV classes, then the UAV classification problem becomes an $M$-ary Bayesian hypothesis testing problem,
with the corresponding hypothesis $\{H_1, H_2, \cdots H_M\}$. Using Bayes theorem, the class membership of the unknown UAV is the one that maximizes the posterior probabilities $P(C=j|\boldsymbol{\sigma})$ given in~\cite{ezuma2021radar} as:
\begin{equation}
 P(C=j|\boldsymbol{\sigma})= \frac{P(\boldsymbol{\sigma}|C=j) P(C=j)}{\sum_{j=1}^{M} P(\boldsymbol{\sigma}|C=j) P(C=j)}~,
\end{equation}
where $P(C=j|\boldsymbol{\sigma})$ is the posterior probability (class membership probability), $P(\boldsymbol{\sigma}|C=j)$ is the conditional class density or the likelihood function, and $P(C=j)$ is the class prior of the $j^{\text{th}}$ class. If we assume the priors $P(C=j)$ are known, say equal priors $P(C=j)=\frac{1}{M}$ for all the classes, then the decision of the SL classifier depends on estimating the class-conditional density $P(\boldsymbol{\sigma}|C=j)$ given the training data~\cite{ezuma2021radar,  jain2000statistical}. The statistical decision rule for the UAV classification problem can then be written as:
\begin{eqnarray}
\begin{aligned}
\widehat{C}  &=  \arg\max_{C=1,2,\cdots, M}{\ln P({C}=j|\boldsymbol{\sigma})},\\
   &=  \arg\max_{C=1,2,\cdots, M} {\ln P(\boldsymbol{\sigma}|C=j)}.
   \label{decision_rule}
\end{aligned}
\end{eqnarray}

Therefore, when an unknown UAV RCS signature is presented to the SL classifier, the classifier will predict the class label by comparing the class likelihood probabilities $p(\sigma_1,\cdots,\sigma_n|C=j)$ estimated from the conditional class densities given the unknown test data $\boldsymbol{\sigma}=(\sigma_1,\cdots,\sigma_n)$. Consequently, it is important to model and estimate the parameters of the conditional class densities $P(\boldsymbol{\sigma}|C=j)$ using either parametric or non-parametric (kernel-based) techniques~\cite{webb2003statistical} to approximate the distribution of the UAV scattered signal power $\sigma$ given in (\ref{RCS_distribution}). 

A major drawback of the kernel density estimation is that it is always biased around the boundaries of bounded data~\cite{zambom2013review}. That is, kernel density estimators have a weakness when estimating densities with bounded support due to boundary/edge effects~\cite{karunamuni2005boundary}. This is because kernel density estimators are not consistent when estimating a density near a finite or the end points of the bounded support. Thus, there is a need for boundary correction, which may be computationally expensive, when using kernel density estimation~\cite{karunamuni2005boundary}. Also, spurious noise appears in the tail of the estimates especially when the underlying distribution is long tailed~\cite{zambom2013review}. These drawbacks of kernel-based target estimation could undermine the accuracy of modeling fluctuating radar targets like UAVs. Therefore, we are going to focus on parametric density estimation for $P(\boldsymbol{\sigma}|C=j)$.


Two major approaches to the parametric estimation of $P(\boldsymbol{\sigma}|C=j)$ will be explored: unimodal parametric density and the multi-modal parametric density model. The estimation of $P(\boldsymbol{\sigma}|C=j)$ is achieved using the UAV RCS measurement data obtained from Section~\ref{THREE}. 




\subsection{Unimodal UAV RCS Statistical Distribution}
In this study, the unimodal parametric densities considered are 1) \textbf{Peter Swerlings statistical models}, 2) \textbf{gamma distribution}, and 3) \textbf{generalized Pareto distribution (GPD)}. The Peter Swerlings statistical models have been widely used to describe the radar scattering characteristics of airborne military aircraft and ground vehicles/tanks~\cite{persson2017radar,fang2020stochastic}. The Peter Swerling RCS density is described by the generalized Chi-Squared ($\chi^2$ or CS) distribution with $N=2m$ degrees of freedom given as:
\begin{equation}
  P(\boldsymbol{\sigma})=\begin{cases}
   \frac{m}{\Gamma(m)\overline{\sigma}}\left[ \frac{m\sigma}{\overline{\sigma}}\right]^{m-1}e^{-\frac{m\sigma}{\overline{\sigma}}}, & \text{if $\sigma\geqslant0$},\\
    0, & \text{otherwise},
  \end{cases}
  \label{chi_square}
\end{equation}
where $\Gamma(a)=\int_{0}^{\infty}t^{a-1}e^{-t}{\rm d}t~$ is the Euler gamma function, $\overline{\sigma}$ is the sample mean of the RCS, and $m$ is a positive number. For Swerling cases 1 and 2, $m=1$ and the Chi-squared model in~(\ref{chi_square}) reduces to the $2\textsuperscript{nd}$ degree distribution given in (\ref{CS-2}). For the Swerling cases 3 and 4, $m=2$ and the Chi-squared model in~(\ref{chi_square}) becomes a $\nth{4}$ degree Chi-squared distribution given in (\ref{CS-4}), which is a special case of the gamma distribution with $N=4$, shape parameter $\alpha=2$ and scale parameter $\beta=2$. 
 \begin{eqnarray}
 \begin{aligned}
f_{\rm{CS-2}}  =\frac{1}{\overline{\sigma}}e{^{-\frac{\sigma}{\overline{\sigma}}}}\label{CS-2}, \\
f_{\rm{CS-4}}  =\frac{4\sigma}{\overline{\sigma}^2}e{^{-\frac{2\sigma}{\overline{\sigma}}}}\label{CS-4}
\end{aligned}
 \end{eqnarray}

Another important parametric density for RCS modeling is the two-parameter gamma density~\cite{du2006two}. In fact, the Chi-square density is a special case of the gamma density. Mathematicaly, the two parameter gamma density is defined as:
\begin{equation}
\scriptstyle 
P(\boldsymbol{\sigma}|\alpha,\beta)=\begin{cases}
   \frac{\beta^\alpha \sigma^{\alpha-1}}{\Gamma(\alpha)}e^{-\beta \sigma}, & \text{if $\sigma\geqslant0$},\\
    0, & \text{otherwise}.
  \end{cases}
  \label{Gamma}
\end{equation}
where $\alpha$ and $\beta$ are the shape and scale parameters respectively.

Also, we will consider the GPD RCS density which is a power-law distribution with a negative exponent. The GPD is a family of related Pareto distributions (Pareto Type I, Type~II, Lomax Type, Type III, and Type IV Pareto) that can be used to describe extreme measurement data. Mathematically, in terms of the shape parameter $\kappa$ and scale parameter $\xi$, the GPD density has the general form:
\begin{equation}
{\scriptstyle 
  P(\boldsymbol{\sigma}|\kappa,\xi)=\begin{cases}\frac{1}{\xi}\left( 1+\kappa\frac{\sigma}{\xi}\right)^{-\frac{1}{\kappa}-1},& \text{if $\sigma\geqslant0$},\\
    0, & \text{otherwise}.
  \end{cases}}
  \label{GPD_Distr}
\end{equation}
When the scale parameter $\xi>0$, the GPD becomes a heavy-tailed Pareto distribution of type II (or Lomax distribution), and if $\xi=0$ the GPD becomes a two-parameter exponential distribution with medium tail~\cite{lee2019exponentiated}. As a result, the GPD density is flexible to fit various RCS data structures. Several studies have shown that radar targets like aircraft and clutters can be modeled using the GPD and Pareto-based densities~\cite{persson2017radar, rosenberg2015radar, gali2020glrt}.

Whenever a unimodal parametric density is used to model the class density~$P(\boldsymbol{\sigma}|C=j)$ of a family of UAVs,  there is need to estimate the model parameters from the RCS measured data obtained in Section~\ref{THREE}. The parameter estimation can be done using the maximum likelihood estimation (MLE) technique~\cite{van2004detection}. The MLE parameter estimation method entails maximizing a log-likelihood function, $\mathcal{L}(\boldsymbol{\theta},\boldsymbol{\sigma})$, such that the assumed class density model $P(\boldsymbol{\sigma}|C=j)$, is most probable given the observed or measured RCS data. For example, suppose a UAV RCS class density, $P(\boldsymbol{\sigma}|C=j)$, is described by the Lomax type GPD distribution with shape parameter $\alpha=\frac{1}{\kappa}$ and scale parameter $\lambda=\xi\alpha$, then (\ref{GPD_Distr}) reduces to: 
\begin{equation}
  P(\boldsymbol{\sigma}|\alpha,\lambda)=  \frac{\alpha}{\lambda}\big[1+\frac{\sigma}{\lambda} \big]^{-(\alpha+1)}
  \label{}
\end{equation}
If we assume the measured RCS from a UAV $\boldsymbol{\sigma}=(\sigma_1,\cdots,\sigma_n)$ are independent and identically distributed (i.i.d) and the unknown parameter vector $\boldsymbol{\theta}=[\alpha~\lambda]$, then the log-likelihood function,~$\mathcal{L}(\boldsymbol{\theta},\boldsymbol{\sigma})$, is given as:
\begin{eqnarray}
\begin{aligned}
\mathcal{L}(\boldsymbol{\theta},\boldsymbol{\sigma}) &= \ln \prod_{i=1}^{n}  P(\sigma_i|\alpha,\lambda) =\sum_{i=1}^{n}\ln \frac{\alpha}{\lambda}\big[1+\frac{\sigma}{\lambda} \big]^{-(\alpha+1)} \\
  &= n\ln{\alpha}-n\ln{\lambda}-(1+\alpha)\sum_{i=1}^{n}\ln (1+\frac{\sigma_i}{\lambda})~.
\end{aligned}
\end{eqnarray}
Finding the MLE for $\alpha$ and $\lambda$ requires the computation of the partial differential equations $\frac{\partial \ln \mathcal{L}({\alpha,\lambda}|\sigma)}{\partial \alpha}=0$ and $\frac{\partial \ln \mathcal{L}({\alpha,\lambda}|\sigma)}{\partial \lambda}=0$. However, in the case of GPD-based densities, there is no closed-form solution to the partial differential equations. Thus, numerical search algorithms such as the Newton-
Raphson numerical method and the expectation-maximization (EM) algorithm can be used to determine the MLEs of the GPD-based density parameters~\cite{giles2013bias,pak2018estimating}. In MATLAB, these numerical search methods for MLE parameter estimation are provided as in-built functions.

\subsection{Multimodal UAV RCS Statistical Distribution}
Thus, far we assumed that the UAV RCS data can be modeled completely using a unimodal parametric density. That is, each observation of the target UAV RCS is assumed to come from one specific parametric density. However, in some cases, this assumption is too restrictive. Sometimes, RCS measurement data may be very complex and multimodal in structure (containing multiple peak regions with high probability mass). In such cases, mixture parametric densities can serve as good models for the UAV class. Mixture models are a class of densities that can represent arbitrarily complex class-conditional densities in supervised learning problems~\cite{jain2000statistical,theodoridis2010introduction}.


Several mixture models have been investigated for radar target and clutter modeling. In~\cite{du2006two} a statistical model comprising two distribution forms, i.e., Gamma
and Gaussian mixture distribution (GMM), are used to model the radar echoes from a target. In~\cite{seng2012gaussian}, Gaussian-Rayleigh mixture is applied to radar image segmentation which enables radar target detection over a wall. In~\cite{copsey2003bayesian, webb2000gamma} gamma mixture models are used to model radar reflections from targets of various types. However, in the current study, our focus will be on the GMM model for identifying UAVs using the measured RCS data. 

The GMM is a density function that consists of a weighted sum or linear mixture of a finite number of Gaussian densities. The GMM densities have the advantage that they can better describe the distribution shape of various multi-modal measurement data. In~\cite{bilik2006gmm,yessad2011svm}, GMM are used for radar target identification using echo data. Adopting a similar approach, we will model the UAV RCS class densities using GMM. In terms of $K$ Gaussian components, the GMM density model for the UAV class conditional density $P(\boldsymbol{\sigma}|C=j)$ where the measured RCS samples are i.i.d, is given by (\ref{GMM_2}):

\begin{equation}
  P(\boldsymbol{\sigma}|\Theta_{(K)})=\prod_{i=1}^{n}\sum_{m=1}^{K}\pi_{m}P_{m}({{\sigma_{i}}|\theta_{m}}),
  \label{GMM_2}
\end{equation}
where each $P_{m}(\boldsymbol{\sigma}|\theta_{m})$ is a Gaussian component parameterized by ${\theta_{m}} = (\mu_m, \Sigma_m)$, with $\mu_m$ and $\Sigma_m$ as the mean and covariance parameters. Also, $m=1,\dots,K$ is an integer that identifies a particular Gaussian component, and $\pi_{m}$ is the component weight with  $\sum_{m=1}^{K}\pi_{m}=1$. Therefore, to model or fit a given UAV RCS observation data $\boldsymbol{\sigma}=(\sigma_1,\cdots,\sigma_n)$ using the GMM statistical approach, we need to estimate the mixture parameter vector $\boldsymbol{\Theta}_{(K)}=\{\theta_1,\dots, \theta_K, \pi_1,\dots,\pi_{K}\}$ and the number of Gaussian components $K$~\cite{jain2000statistical}. The basic MLE parameter estimator used for unimodal statistical model is not suitable for estimating the parameters of the GMM densities. This is because of the basic limitation of the MLE technique: the assumption that the training dataset is complete~\cite{jain2000statistical}. That is, the MLE technique assumes that all the relevant variables needed to describe the dataset are present in the density model~\cite{jain2000statistical}. Thus, the MLE estimator becomes intractable if some relevant variables are hidden or not observed (latent variables). 


The parameters of the GMM density can be estimated using the expectation-maximization (EM) algorithm. The EM algorithm is an iterative solution for computing the local MLE or maximum a posterior (MAP) estimate of mixture models, such as GMM, in the presence of latent or hidden variables. The EM algorithm considers the UAV RCS observation data $\boldsymbol{\sigma}=(\sigma_1,\cdots,\sigma_n)$ as incomplete data, either explicitly or by construct~\cite{jain2000statistical}. Thus, we can express the hidden or missing random variables as $\boldsymbol{z}=\{\boldsymbol{z}^{(1)},\dots, \boldsymbol{z}^{(K)} \}$, where $\boldsymbol{z^{(i)}}=\{z_{1}^{(i)},\dots, z_{K}^{(i)} \}$ indicates which of the $K$ Gaussian components generated the $i^{\text{th}}$ observation in the RCS data dataset (i.e. $\sigma_{i}$ data point). If it was the $m^{\text{th}}$ Gaussian component that generated $\sigma_{i}$, then ${z_{m}^{(i)}}=1$ otherwise ${z_{m}^{(i)}}=0$. Then, the complete log-likelihood function obtained from (\ref{GMM_2}), which includes the latent variables, is given as:
\begin{equation}
  \mathcal{L}(\boldsymbol{\Theta}_{(K)},\boldsymbol{z},\boldsymbol{\sigma} )= \sum_{i=1}^{n} \sum_{m=1}^{K} z_{m}^{(i)}\ln{[ \pi_{m}P_{m}({\sigma_{i}}|\theta_{m})]}.
  \label{GMM_complete_likelihood}
\end{equation}

Assume that at time $t=0$, the initial guess (initialization) for the mixture parameter is $\boldsymbol{\Theta}_{(K)}^{(0)}$. Then at $(t+1)^{\text{th}}$ time, the parameter estimate $\hat{\boldsymbol{\Theta}}_{(K)}^{(t+1)}$ is obtained by iterating through two steps until the convergence criteria is satisfied: the expectation step (E-step) and the maximization step (M-step). 

\begin{itemize}
  \item 
  ${\textbf{E-step}}$: Given the training UAV RCS training data or observation $\boldsymbol{\sigma}=(\sigma_1,\cdots,\sigma_n)$ and the current parameter estimate ${\boldsymbol{\hat\Theta}_{(K)}^{(t)}}$, compute the conditional expectation of the complete log-likelihood given in (\ref{GMM_complete_likelihood}). This gives:
\begin{eqnarray}
\begin{aligned}
& Q({\boldsymbol{\Theta}_{(K)}|\boldsymbol{\Theta}_{(K)}^{(t)}},\boldsymbol{\sigma}) = E \Big[ \mathcal{L}(\boldsymbol{\Theta}_{(K)},\boldsymbol{z},\boldsymbol{\sigma} ) \Big]
\end{aligned}
 \label{conditional_Expec_1}
\end{eqnarray}
\begin{equation*}
    =\sum_{i=1}^{n} \sum_{m=1}^{K}\Big[ E(z_{m}^{(i)}|\hat{\Theta}_{(K)}^{(t)},\sigma_{i})\Big[\ln(\pi_{m}) +  \ln(P_{m}({\sigma_{i}}|\theta_{m}) \Big] \Big] 
\end{equation*}
with the conditional expectation of the missing variable, computed from Bayes theory as follows:
\begin{eqnarray}
\begin{aligned}
 E(z_{m}^{(i)}|\hat{\Theta}_{(K)}^{(t)},\sigma_{i})&= 1 \cdot P(z_{m}^{(i)}=1|\hat{\Theta}_{(K)}^{(t)},\sigma_{i}) + \\ 0 \cdot P(z_{m}^{(i)}=0|\hat{\Theta}_{(K)}^{(t)},\sigma_{i})
&=\frac{{\pi}_{m}^{t} P_{m}({\sigma_{i}}|{\theta}_{m}^{(t)})}{ \sum_{m=1}^{K} {\pi}_{m}^{t} P_{m}({\sigma_{i}}|{\theta}_{m}^{(t)})}\\
&=\gamma_{z_{m}^{(i)}}^{(t)}~,
\end{aligned}
 \label{conditional_Expec_2}
\end{eqnarray}
where $ P(z_{m}^{(i)}=1|\hat{\Theta}_{(K)}^{(t)},\sigma_{i})= \gamma_{z_{m}^{(i)}}^{(t)}$, which can be easily computed, is the posterior probability of the latent variable, at the $t^\text{th}$ iteration, conditioned on the UAV RCS training data. 
 
    \item ${\textbf{M-step}}$: Update or estimate the new parameter $\hat{\Theta}_{(K)}^{(t+1)}$ using the current values of the posterior probability $\gamma_{z_{m}^{(i)}}^{(t)}$. That is, the new parameter estimate $\hat{\Theta}_{(K)}^{(t+1)}$ is obtained by maximizing the expected log-likelihood computed in the ${\textbf{E-step}}$. Often times, this optimization problem is solved using the Lagrange method. That is:
\begin{equation}
\hat{\Theta}_{(K)}^{(t+1)}= \arg\max_{\boldsymbol{\Theta}_{(K)}}Q({\boldsymbol{\Theta}_{(K)}|\boldsymbol{\hat{\Theta}}_{(K)}^{(t)}},\boldsymbol{\sigma})~.
  \label{M-step}
\end{equation}
From~\cite{du2006two,seng2012gaussian}, the result of the optimization are:
\begin{align*} 
\hat{\pi}_{m}^{t+1} &=  \frac{1}{n}\sum_{i=1}^{n}\gamma_{z_{m}^{(i)}}^{(t)} , \quad  
\hat{\mu}_{m}^{t+1} =  \frac{\sum_{i=1}^{n}\gamma_{z_{m}^{(i)}}^{(t)}\cdot \sigma_{i}}{ \sum_{i=1}^{n}\gamma_{z_{m}^{(i)}}^{(t)}}~,\\
 \hat{\Sigma}_{m}^{t+1} &= \frac{\sum_{i=1}^{n}\gamma_{z_{m}^{(i)}}^{(t)}\cdot (\sigma_{i}-\hat{\mu}_{m}^{t+1})(\sigma_{i}-\hat{\mu}_{m}^{t+1})^T}{ \sum_{i=1}^{n}\gamma_{z_{m}^{(i)}}^{(t)}}~.
\end{align*}

\end{itemize}

 \begin{figure}[t!]
\center{
 \begin{subfigure}[]{\includegraphics[width=0.9\linewidth]{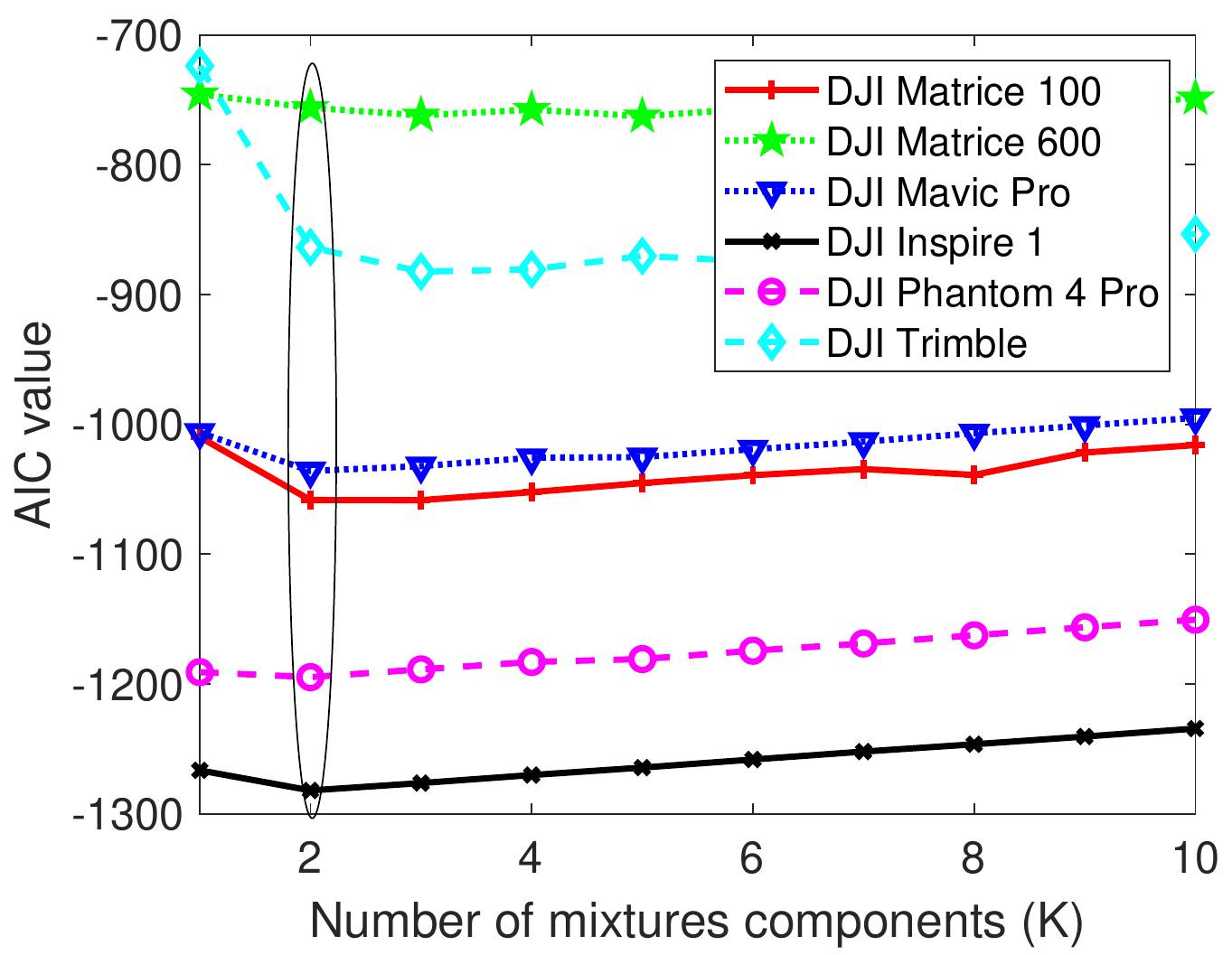}
 \label{Confusion_100_DL_Training}}
\end{subfigure}\\
 \begin{subfigure}[]{\includegraphics[width=0.9\linewidth]{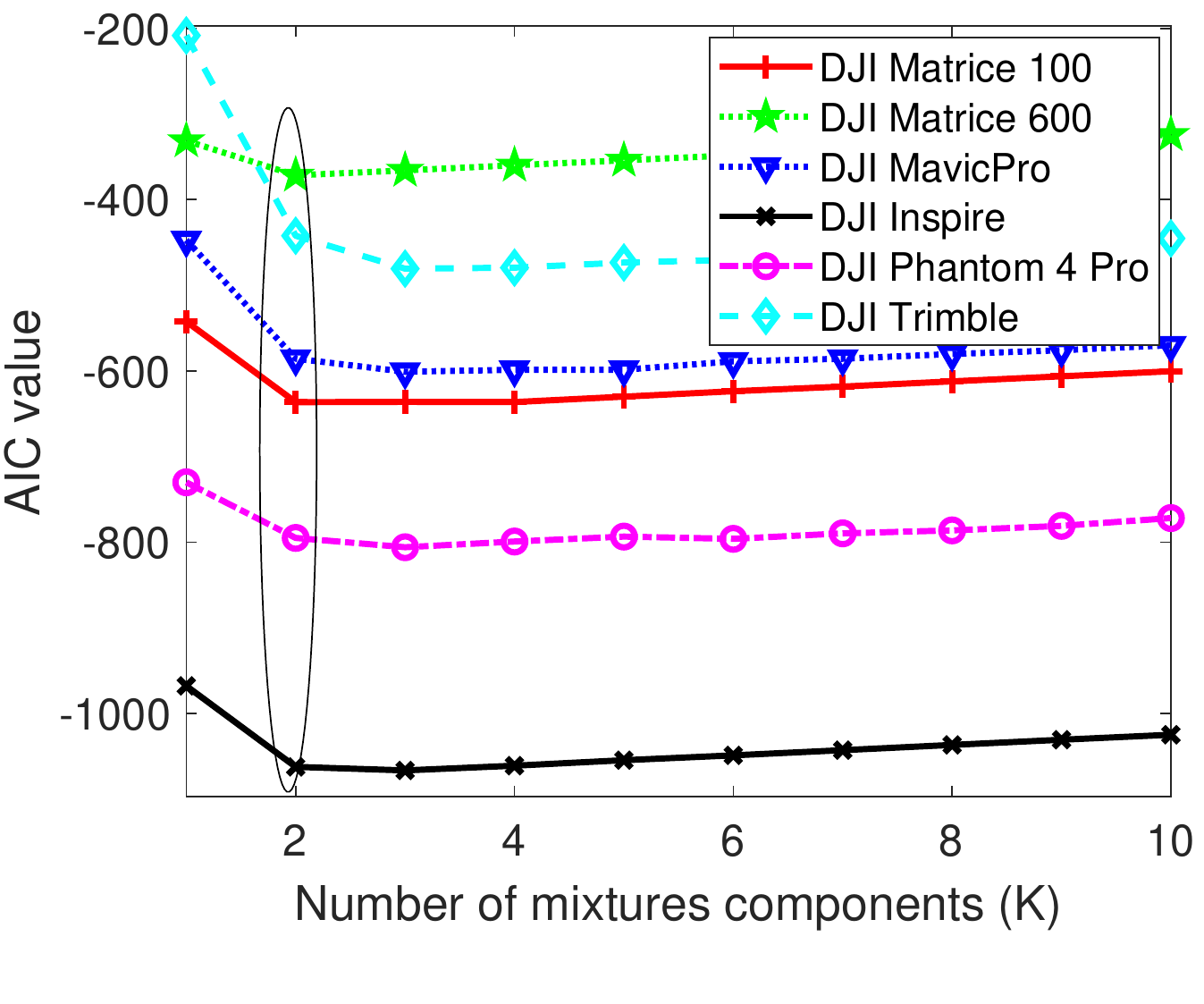}
 \label{AIC_15GH_25GH_VV}}
\end{subfigure}}
\caption{The AIC value as a function of the number of GMM mixture components for each UAV in the training set (a) 15 GHz VV-polarized UAV RCS Data, (b) 25 GHz VV-polarized UAV RCS Data.}
\label{AIC_plots} 
\end{figure}
The EM algorithm runs the iteration (the E and M steps above) until the convergence condition: $\parallel \Theta_{(K)}^{(t+1)} - \Theta_{(K)}^{(t)} \parallel < \varepsilon$, is satisfied, where $ \varepsilon$ is a very small positive number of the order of $10^{-5}$~\cite{seng2012gaussian,vegas2014gamma}. In summary, using EM algorithm, if we knew the current parameters of the GMM model, we can compute an estimate of the posterior probability $\gamma_{z_{m}^{(i)}}^{(t)}$ and vice-versa. Once the GMM-based class conditional density $P(\boldsymbol{\sigma}|C=j)$ for each UAV class has been estimated using the UAV RCS training data obtained from Section~\ref{THREE}, the classification decision for an unknown UAV target follows the Bayesian decision rule given in (\ref{decision_rule}).

An important task when using the GMM to model the UAV RCS data is estimating the number of mixture components~$K$ that can parsimoniously describe the data~\cite{jain2000statistical}. When $K$ is too small, the GMM model may not satisfactorily fit the RCS training data. On the other hand, if $K$ is large, the GMM model will overfit the RCS training data~\cite{seng2012gaussian,jain2000statistical}. This can lead to errors in prediction. The choice of $K$ can be determined using techniques such as the minimum description length (MDL) criterion, Akaike's
information criterion (AIC), and Schwarz's Bayesian inference criterion (BIC)~\cite{jain2000statistical}. In this study, our focus is on the AIC technique.

Given several candidate GMM models, each with a different number of mixtures, the AIC criterion selects the optimal GMM model as the one with the lowest AIC score. Suppose that the UAV RCS training data is given by $\boldsymbol{\sigma}=(\sigma_1,\cdots,\sigma_n)$, then the AIC score is computed as follows:
\begin{equation}
   {\rm AIC(\boldsymbol{\sigma})}=-2\ln P(\boldsymbol{\sigma}|\boldsymbol{{\Theta_{(K)}}}) + 2K~.   \label{AIC}
\end{equation}





Fig.~\ref{AIC_plots} shows plots of the AIC value versus the number of mixture components ($K$) of several GMM models fitted to the VV-polarized UAV RCS measured at 15 GHz and 25 GHz. From Fig.~\ref{AIC_plots}, it is obvious that the lowest AIC score occurs around $K=2$. Therefore, the two-component GMM model is considered optimized for modeling the UAV class conditional density from the training dataset. 


\section{UAV Classification Using ML Technique}\label{machine_learning_techniques}
The ML is an emerging tool in 
radar signal processing. It relies on developing a mathematical model that learns high-dimensional data (e.g. target RCS, ISAR, SAR, or micro-Doppler images) and makes a decision based on the learned information~\cite{abdu2021application,lang2020comprehensive}. Three main types of ML classifiers have been investigated for radar target recognition: supervised, semi-supervised, and unsupervised learning~\cite{majumder2020deep,lang2020comprehensive}. However, the focus of this section is on supervised learning algorithms and RCS-based target features for UAV identification.


\begin{figure}[t]
 \center
 \includegraphics[width=0.5\textwidth]{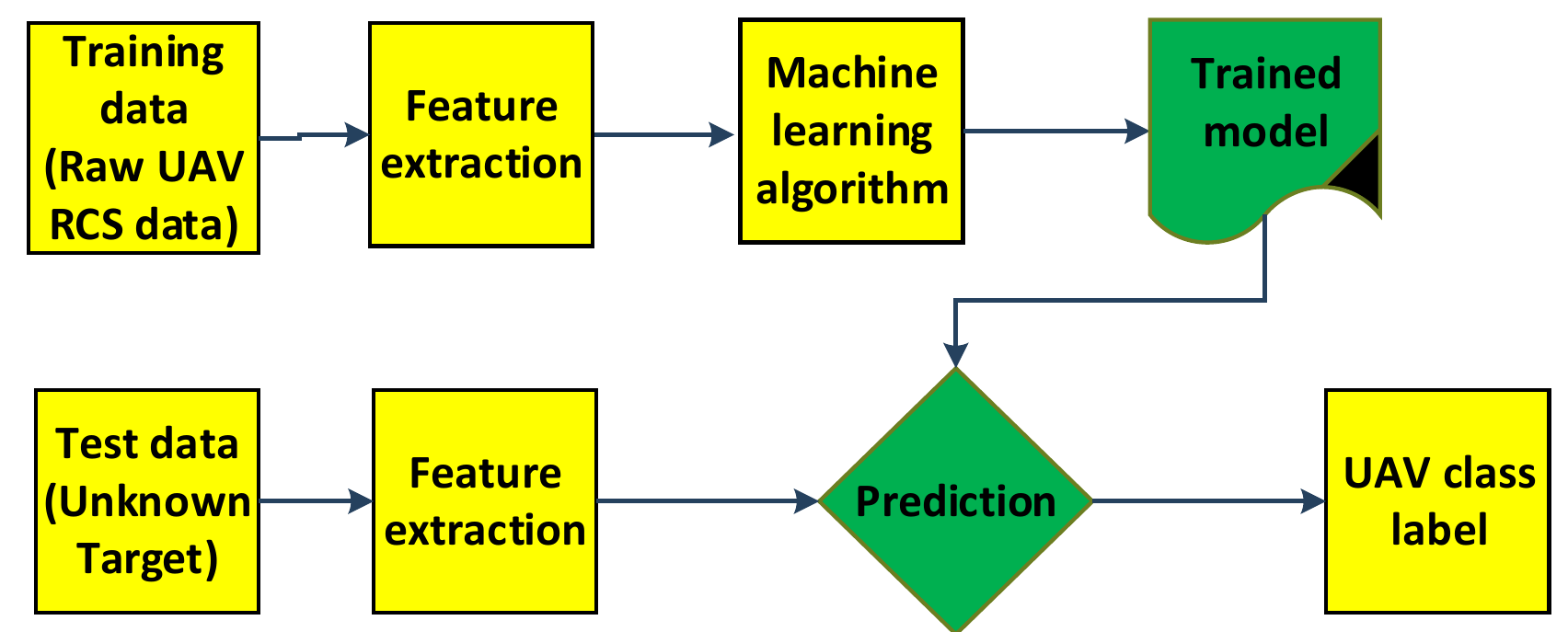}
\caption{A typical flowchart for the ML based UAV classification. Feature selection is vital for achieving good accuracy while reducing the dimensionality of the UAV RCS training and test data.}
\label{ML_Flowchart}
\end{figure}

\begin{table}[t!]
\centering
\caption{RCS Statistical Features for ML based UAV classification.}
\label{stat_feature}
\begin{tabular}{|p{2.5cm}| m{3cm}|m{2cm}|m{2cm}|}
\hline
\textbf{Features} & \textbf{Formula} & \textbf{Measures}\\
\hline
Peak value ($x_{\rm{pv}}$)& $\rm{max(\sigma_{i})}$ & Amplitude \\
Minimum ($x_{\rm{rms}}$)~& $\left[\frac{1}{N}\sum_{i=1}^N \sigma_{i}^2\right]^\frac{1}{2}$ & Minimum value\\
Mean ($\mu$)& $\frac{1}{N}\sum_{i=1}^N \sigma_{i}$ & Central tendency \\
Standard deviation ($\sigma_{T}$)~& $\left[\frac{1}{N-1}\sum_{i=1}^N (\sigma_{i}-\mu)^2\right]^{\frac{1}{2}}$ & Dispersion \\
Variance & $\frac{1}{N}\sum_{i=1}^N (\sigma_{i}-\mu)^2$ & Dispersion \\
Median &  $\rm{median(\sigma_{i})}$ &Central tendency~\\
Mode  & $\rm{mode(\sigma_{i})}$ & Central tendency\\
\hline
\end{tabular}\label{Table_feature}
\end{table}

\begin{figure*}[t!]
\center{
 \begin{subfigure}[]{\includegraphics[width=0.45\linewidth]{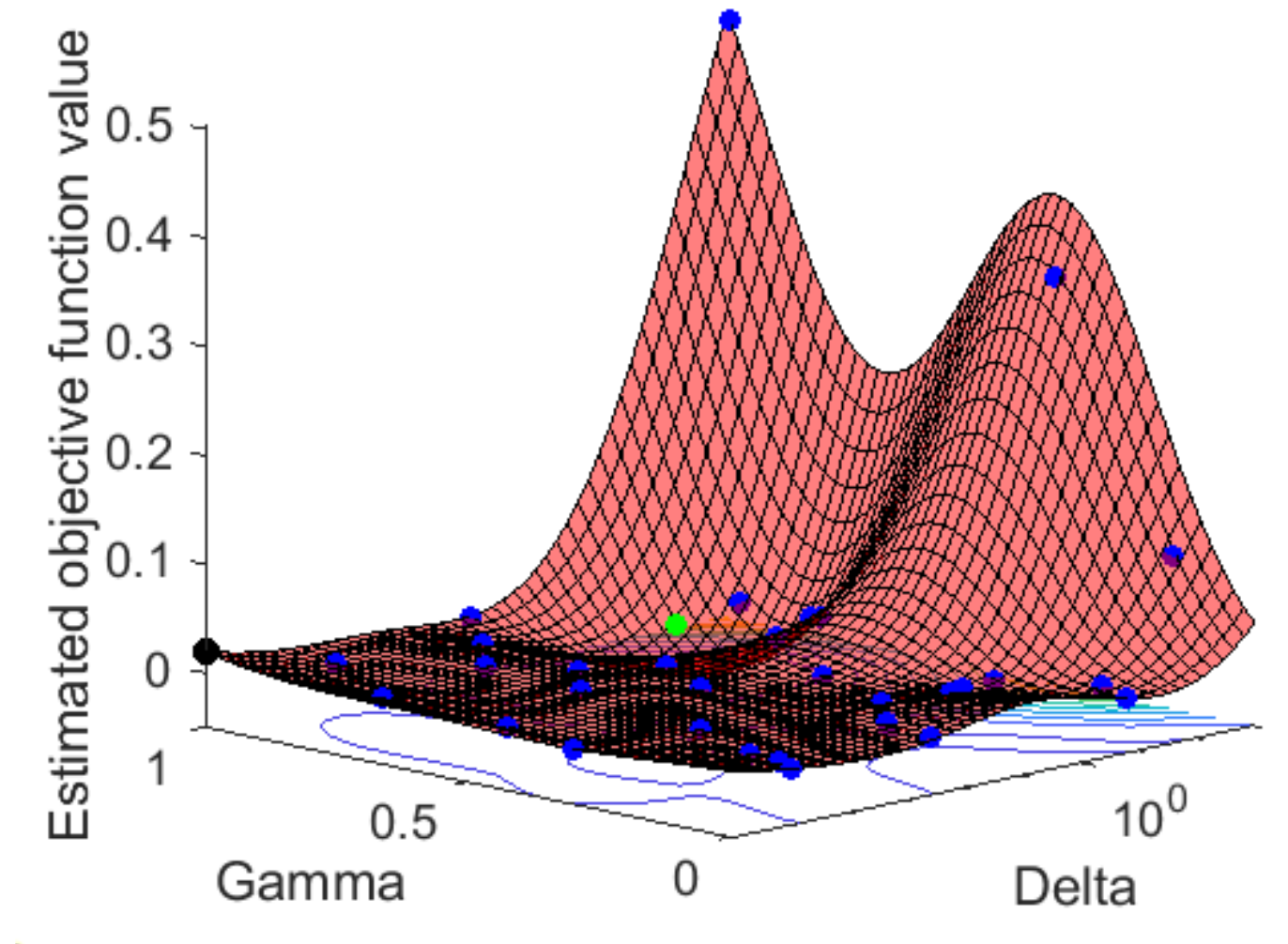}
 \label{discriminant_opt_1}}
\end{subfigure}
 \begin{subfigure}[]{\includegraphics[width=0.45\linewidth]{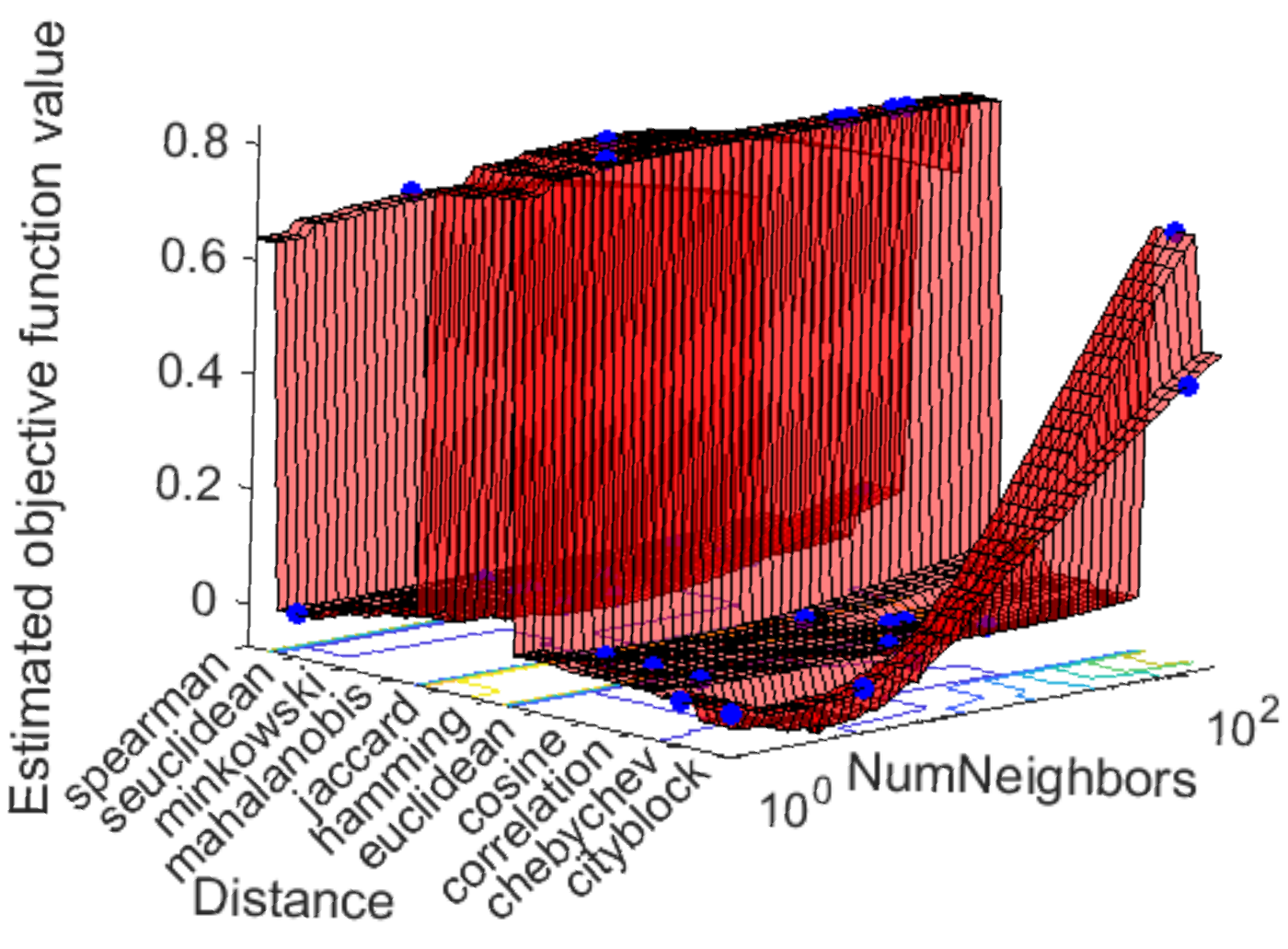}\label{knn_opt_1}}
\end{subfigure}
\caption{{(a) The estimated 3-dimensional objective function versus the ML model hyperparameters was obtained using the Bayesian optimization technique for DA ML Classifier. The optimal hyperparameters correspond to points where the objective function is minimized.  (b) Hyperparameter optimization for the kNN ML classifier showing the effects of the distance metric (Distance) and the number of nearest neighbors (NumNeighbors).  }\label{objective_knn_plots}}
}
 \end{figure*}

Fig.~\ref{ML_Flowchart} shows a typical flowchart for the supervised ML-based classification of an unknown UAV. The choice of the ML algorithm/ML classifier and the extracted features are two important decisions that determine the classification accuracy. If the feature vector is too large, often a feature selector may be required. For the UAV classification problem discussed herein, seven statistical moments of the measured or observed UAV RCS are extracted as features for training the ML classifiers. Table~\ref{stat_feature} provides the list of the statistical features extracted from UAV RCS data which are used to train five different ML classifiers: k-nearest neighbors (kNN), support vector machine (SVM), Ensemble, Naive Bayes, classification tree, and discriminant analysis (DA). The ML classifiers learn the patterns from the feature set extracted from the UAV RCS training dataset obtained in Section~\ref{THREE}~\cite{lang2020comprehensive}. The learned pattern is exploited when classifying an unknown UAV target in a supervised framework. 

To improve the ability of our ML classifiers to learn the UAV RCS training dataset, it is necessary to carefully tune the model hyperparameters. Besides, it is desirable to automate such a procedure. For this reason, we apply the Bayesian optimization method to automatically tune the hyperparameters of our ML models. The Bayesian optimization is a robust technique that models the objective function used to train an ML classifier as a Gaussian process (GP)~\cite{snoek2012practical}. Just like other
optimization methods, in Bayesian optimization we are interested in finding the optimum model parameters that minimize an objective function (such as the holdout cross-validation loss of the ML classifiers) on some bounded set.

Fig.~\ref{objective_knn_plots} shows the result of the Bayesian optimization applied to tune the hyperparameters of the DA and kNN ML classifiers trained with the 25 GHz HH UAV RCS training data obtained from Section~\ref{THREE}. The optimization was carried out using the MATLAB bayesopt() function. Fig.~\ref{objective_knn_plots} presents the 3-dimensional plot of the estimated objective function versus the hyperparameters of the classifiers. The optimal set of hyperparameters correspond to the minimum estimated objective function. In Fig.~\ref{discriminant_opt_1}, the hyperparameters tuned for the DA ML classifier are the two regularization parameters: Delta and Gamma, which are used to remove redundant predictors from the classifier. On the other hand, in Fig.~\ref{knn_opt_1}, the hyperparameters tuned for the kNN ML classifier are the number of neighbors (NumNeighbors) and distance metric (Distance).

The values of the critical ML model hyperparameters for different classifiers, obtained from the Bayesian optimization using the 25 GHz HH-polarized UAV RCS training data presented in Section~\ref{THREE}, are as follows:



\begin{itemize}
    \item \textbf{kNN:} Number of neighbors (NumNeighbors) = 1, Distance metric (Distance) = chebychev
    \item \textbf{DA:} Delta = $7.9588\times 10^{-5}$, Gamma = 0.2689
    \item \textbf{SVM:} Coding: onevsall, Box constraint (BoxConstraint)  = 473.16, Kernel scale parameter (KernelScale) = 0.0014583
    \item \textbf{Ensemble:} Method= Bag,  Number of ensemble learning cycles (NumLearningCycles) = 67, Minimum number of leaf node observations (MinLeafSize)  = 86
    \item \textbf{Naive Bayes:} Distribution name:~kernel, Width: 0.15096
     \item \textbf{Classification Tree:} MinLeafSize = 26
\end{itemize}
After training and optimizing the ML classifiers on the UAV RCS data, the models are ready for testing or classification.

\section{UAV Classification Using DL Technique}\label{DEEP_LEARNING_FRAMEWORK}
DL algorithms are a subset of ML techniques and can be viewed as an extension of an artificial neural network (ANN) with representation learning~\cite{abdu2021application}. ANN are bio-inspired networks and DL algorithms are structured on multiple layers of ANN designed to mimic the human brain functions~\cite{abdu2021application}. Recently, several DL architectures such as deep neural networks, deep recurrent networks, and convolutional neural networks (CNN) have been applied to the problem of radar target recognition~\cite{lang2020comprehensive,majumder2020deep,abdu2021application,geng2021deep,santra2020deep,wan2019convolutional}. For some applications, DL algorithms are preferred over traditional ML algorithms because the former do not require the manual extraction and selection of handcrafted features. For example, radar image classification using manually extracted features and traditional ML algorithms will require the skills of a subject-matter expert with strong domain knowledge of the target and flight terrain. Often, this is a time-consuming task.

Furthermore, most traditional ML techniques can only learn approximately limited types of decision boundaries. For example, the logistic regression ML algorithm can only learn linear decision boundaries. Thus, if the decision boundary is non-linear, such an algorithm will not do well in recognizing the specific radar targets. On the other hand, while the SVM algorithm can learn non-linear decision boundaries, it is sometimes difficult to choose the right kernel to achieve good accuracy performance. The DL frameworks are not without their challenges as they require large training data and more computational resources. Without sufficiently large training data, DL algorithms will overfit. Therefore, due to the unavailability of high-quality large or sufficient training data, pre-trained DL architectures are preferred to overtraining a network from scratch. The process of using pre-trained network to solve another classification problem is called transfer learning~\cite{xiao2019radar, santra2020deep}. That is, transfer learning algorithms leverage or seed valuable knowledge and experience from a previous domain to enhance the learning performance of the current domain.

In this section, we will investigate the performance of transfer learning DL frameworks for UAV classification using RCS-based input data. However, since the RCS-data is one-dimensional, it needs to be transformed into a two-dimensional image before feeding into the transfer learning DL classifier. This is because most transfer learning frameworks have internal structures that are designed to operate on 2-D image data feed~\cite{wan2019convolutional, abdu2021application,lang2020comprehensive}. For this reason, time-frequency (T-F) transform of the 1-D RCS data or high-resolution radar (HRR) returns can be used to create two dimensional radar images for the DL-based UAV target classification. As an example, in~\cite{wan2019convolutional}, HRR returns scattering from an aircraft are transformed to two-dimensional Short-time Fourier transform (STFT) images (spectrogram images) that are fed into a CNN network for target identification.

In general, the use of wavelet-based T-F transforms for the detection and classification of nonstationary signals (e.g. radar and sonar signals) has been well investigated~\cite{chen2002time}. The nonstationarity of the signal implies that the frequency of such a signal is time-dependent. Thus, the nonstationary signal is better analyzed using a joint T-F analysis. The T-F transforms can be grouped into two main types: linear and bilinear  transforms~\cite{chen2002time}. Linear T-F includes the STFT, continuous wavelet transform (CWT), and adaptive T-F representation~\cite{chen2002time}. On the other hand, bilinear transforms include Wigner-Ville distribution (WVD), Cohen's class distribution functions, and T-F
distribution series (TFDS)~\cite{chen2002time}. Due to space limitations, the focus of this section will be the CWT transform for the transfer learning-based classification of UAVs. Next, we briefly discuss the CWT transform of the UAV RCS data.


\begin{figure}[t!]
\center{
 \begin{subfigure}[]{\includegraphics[width=0.65\linewidth]{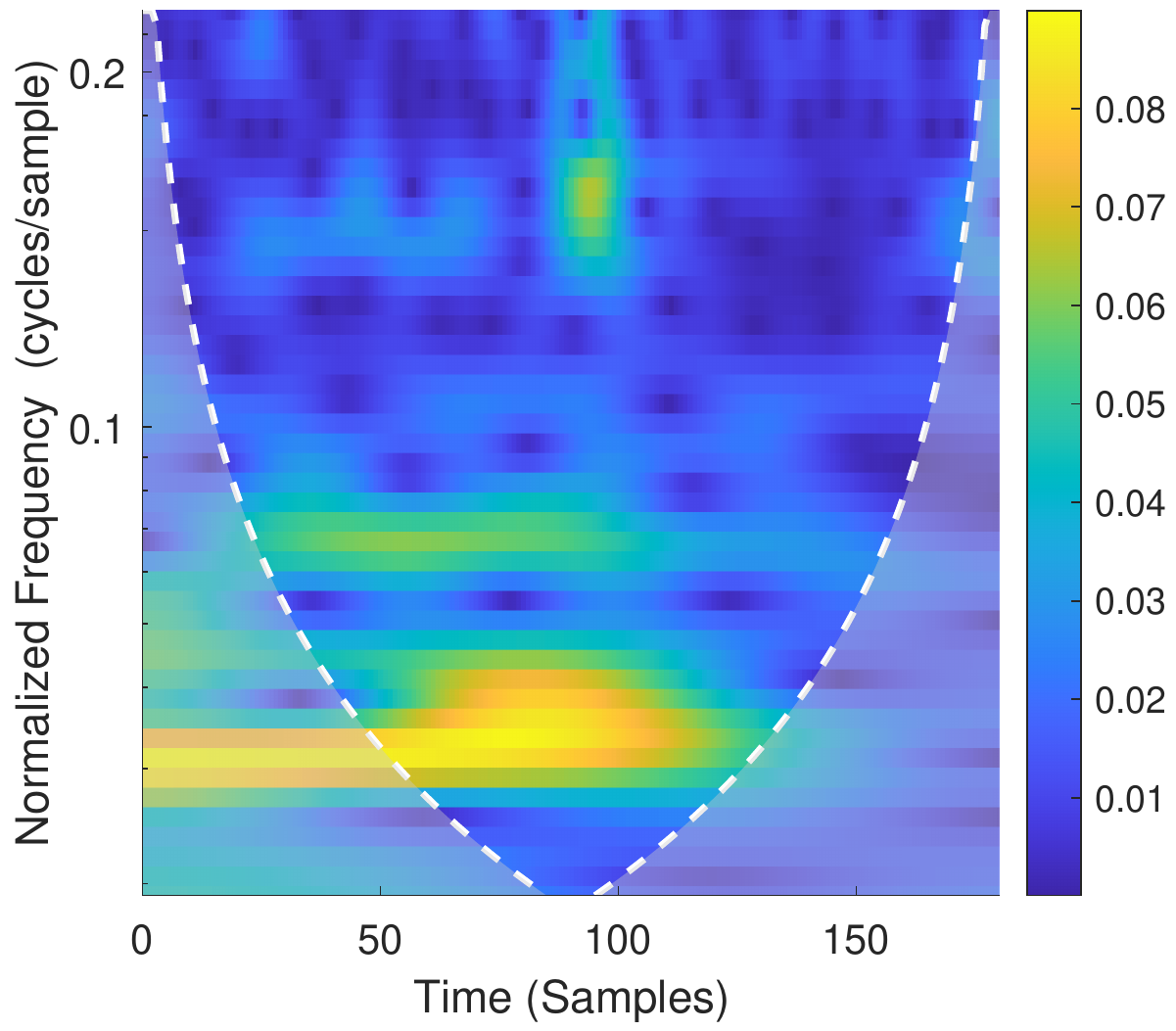}
 \label{discriminant_opt}}
\end{subfigure}
 \begin{subfigure}[]{\includegraphics[width=0.65\linewidth]{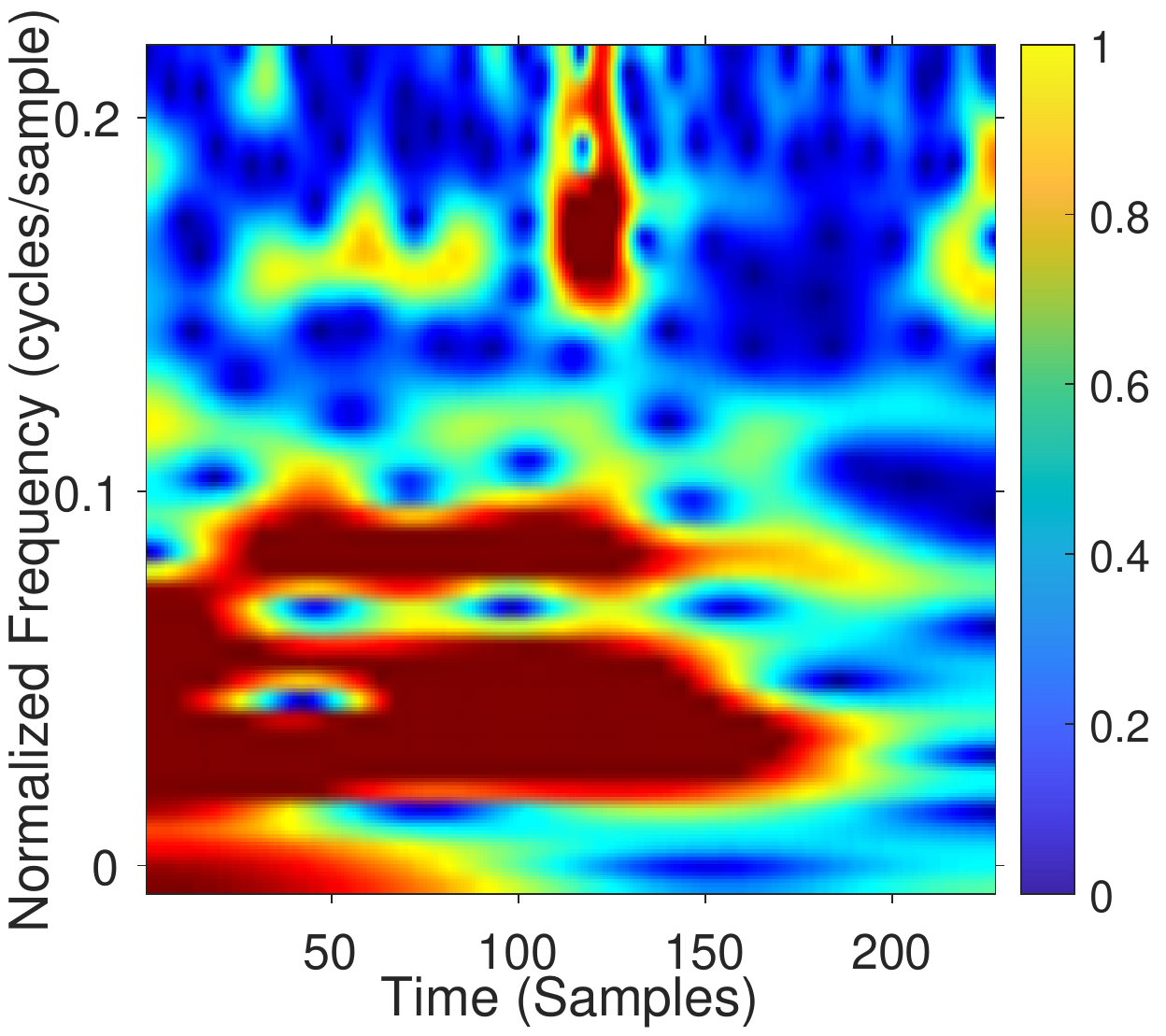}\label{knn_opt}}
\end{subfigure}
\caption{{(a) CWT scalogram of the 25~GHz VV-polarized RCS data measured from the DJI Matrice 600 Pro UAV. (b) Enhanced image that is fed into the SqueezeNet for UAV classification. The enhanced portions of the CWT image indicate the RCS samples with a higher magnitude as compared with the blurry parts of the image. The overall color distribution of the CWT scalogram image indicates the RCS signature of the UAV in the T-F domain.}\label{cwt scalogram}}
}
 \end{figure}

 \begin{figure*}[t!]
\centering
\begin{subfigure}[15 GHz RCS DJI M600 Pro.]{\includegraphics[width=0.235\linewidth]{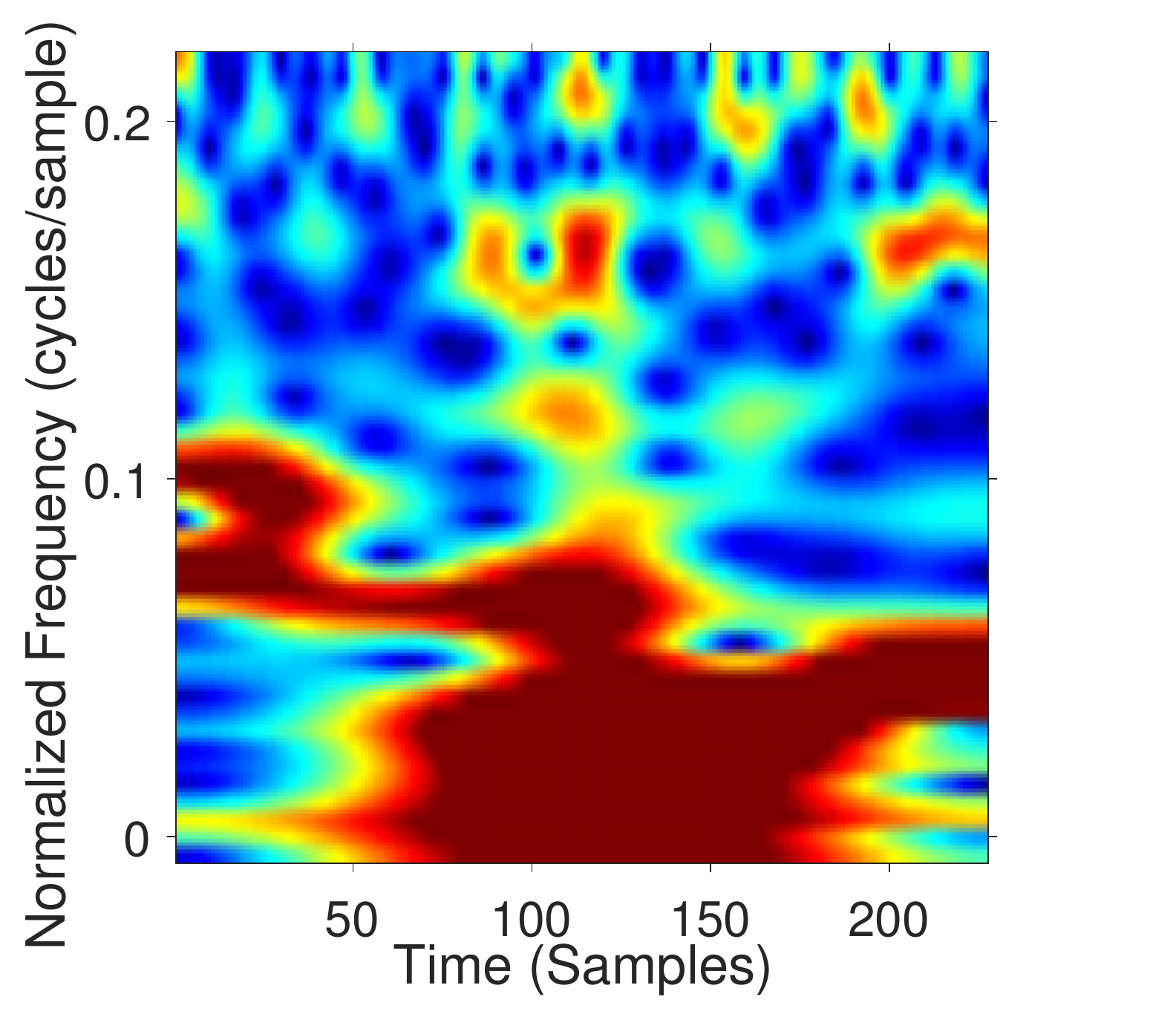}
 \label{DJI_M600_CWT_VV15GHz_processed_image}}
 \end{subfigure}
\begin{subfigure}[25 GHz RCS DJI M600 Pro.]{\includegraphics[width=0.235\linewidth]{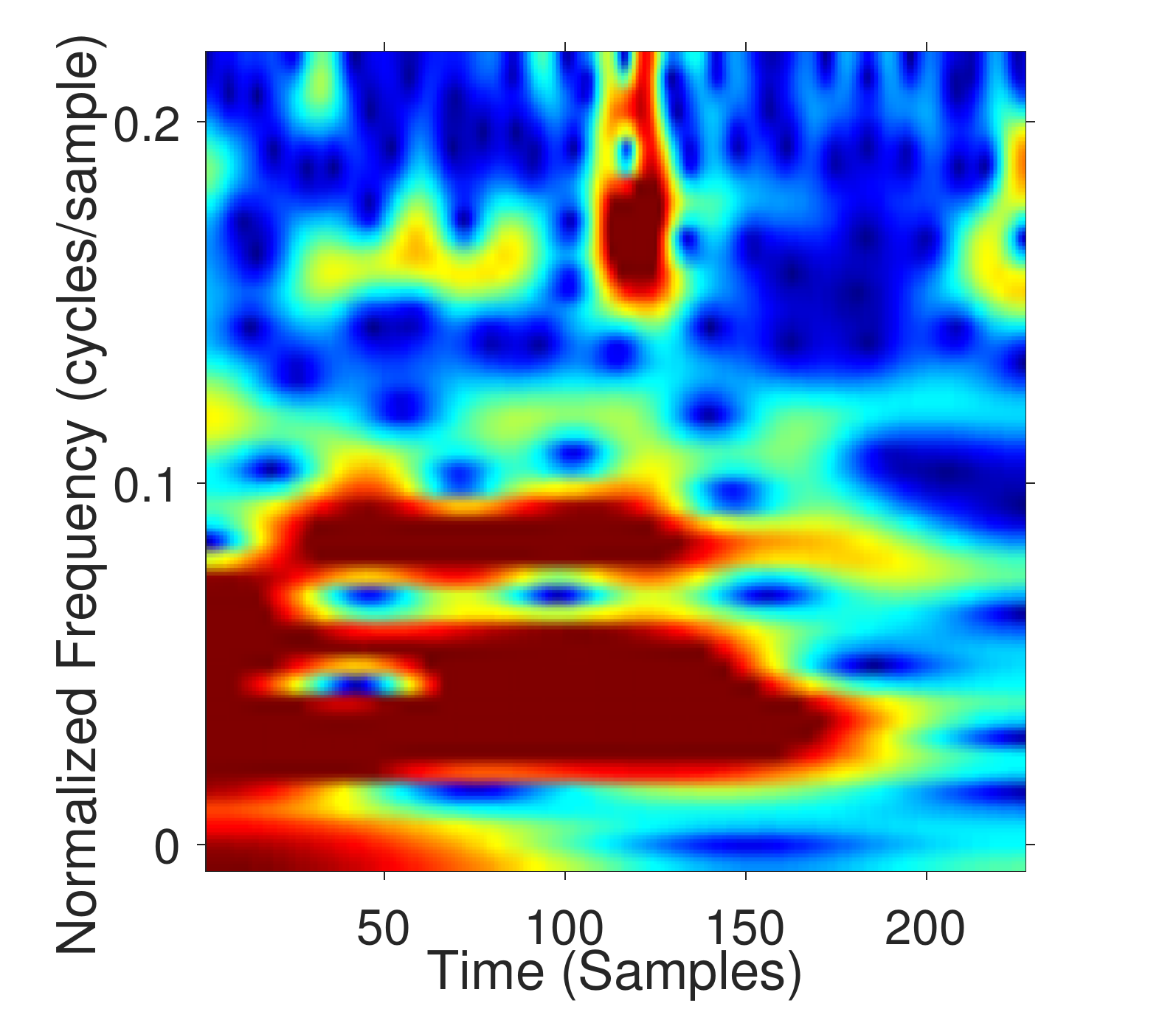}
 \label{DJI_M600_CWT_VV25GHz_processed_image_VV25GHz}}
 \end{subfigure}
 \begin{subfigure}[15 GHz RCS DJI M100.]{\includegraphics[width=0.235\linewidth]{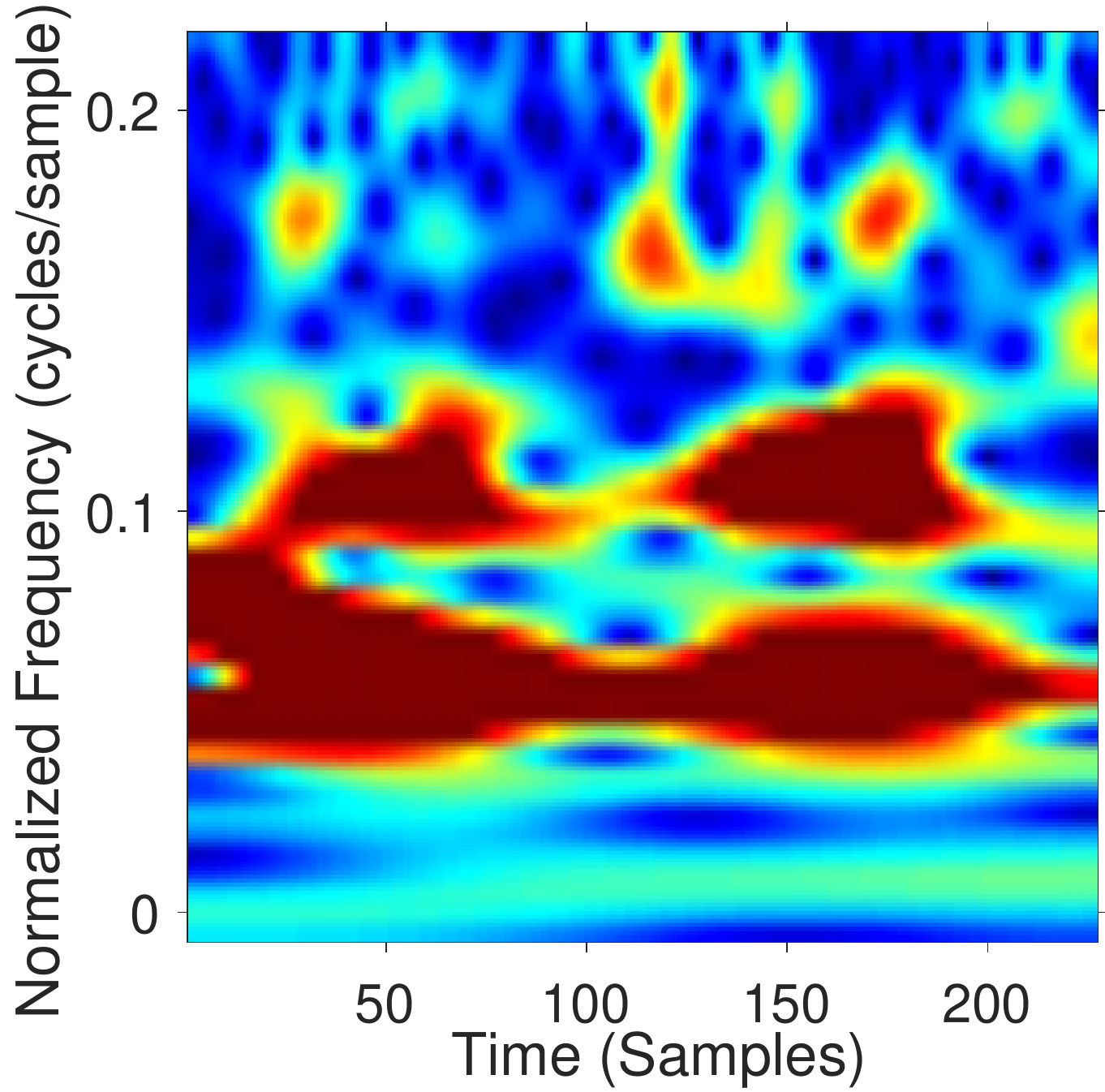}\label{DJI_M100_CWT_VV15GHz_processedImage}}
 \end{subfigure}
 \begin{subfigure}[25 GHz RCS DJI M100.]{\includegraphics[width=0.235\linewidth]{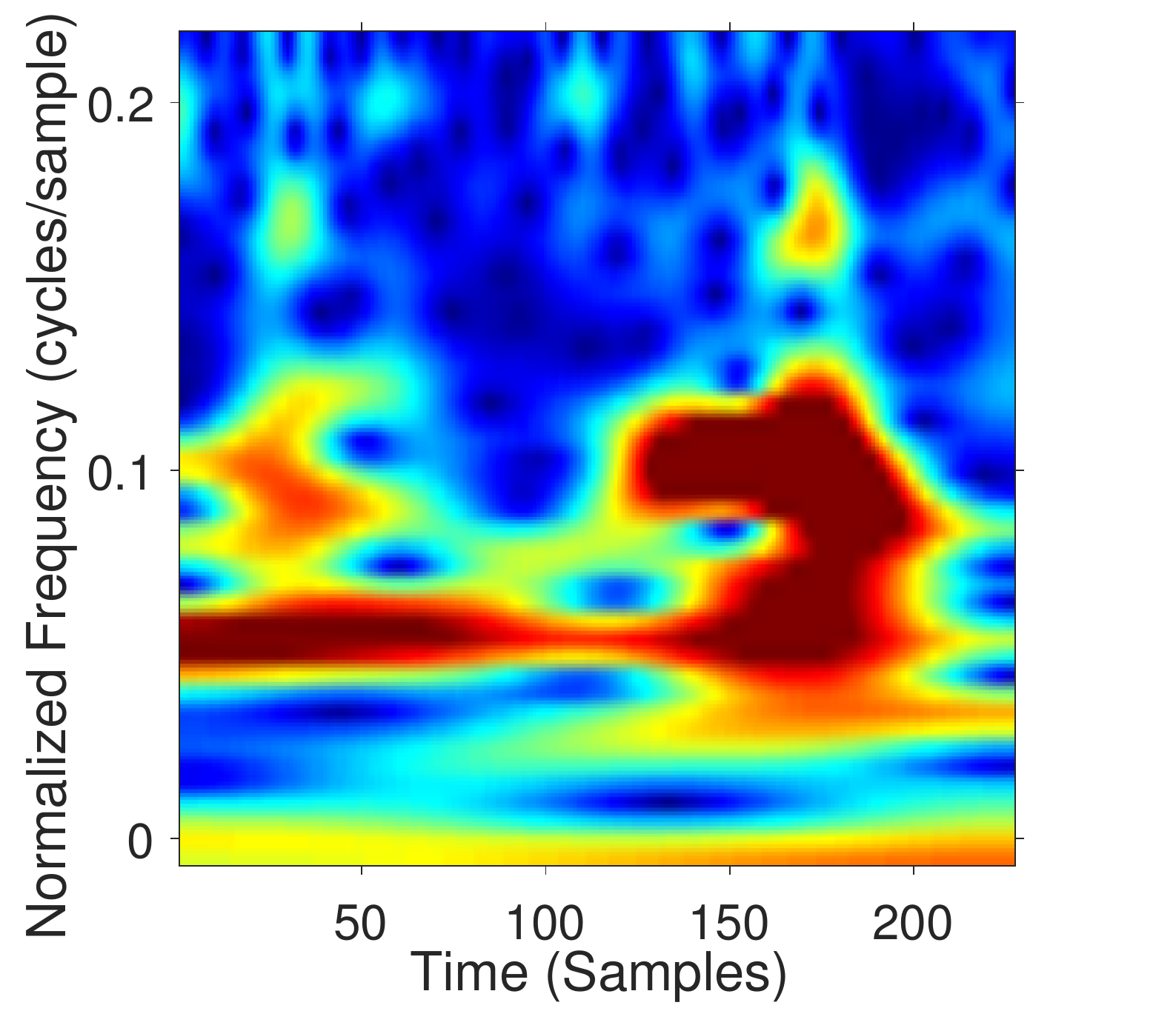}\label{DJI_M100_CWT_VV25GHz_processedImage}}
\end{subfigure}\\
 \begin{subfigure}[15 GHz RCS DJI Inspire 1 Pro.]{\includegraphics[width=0.235\linewidth]{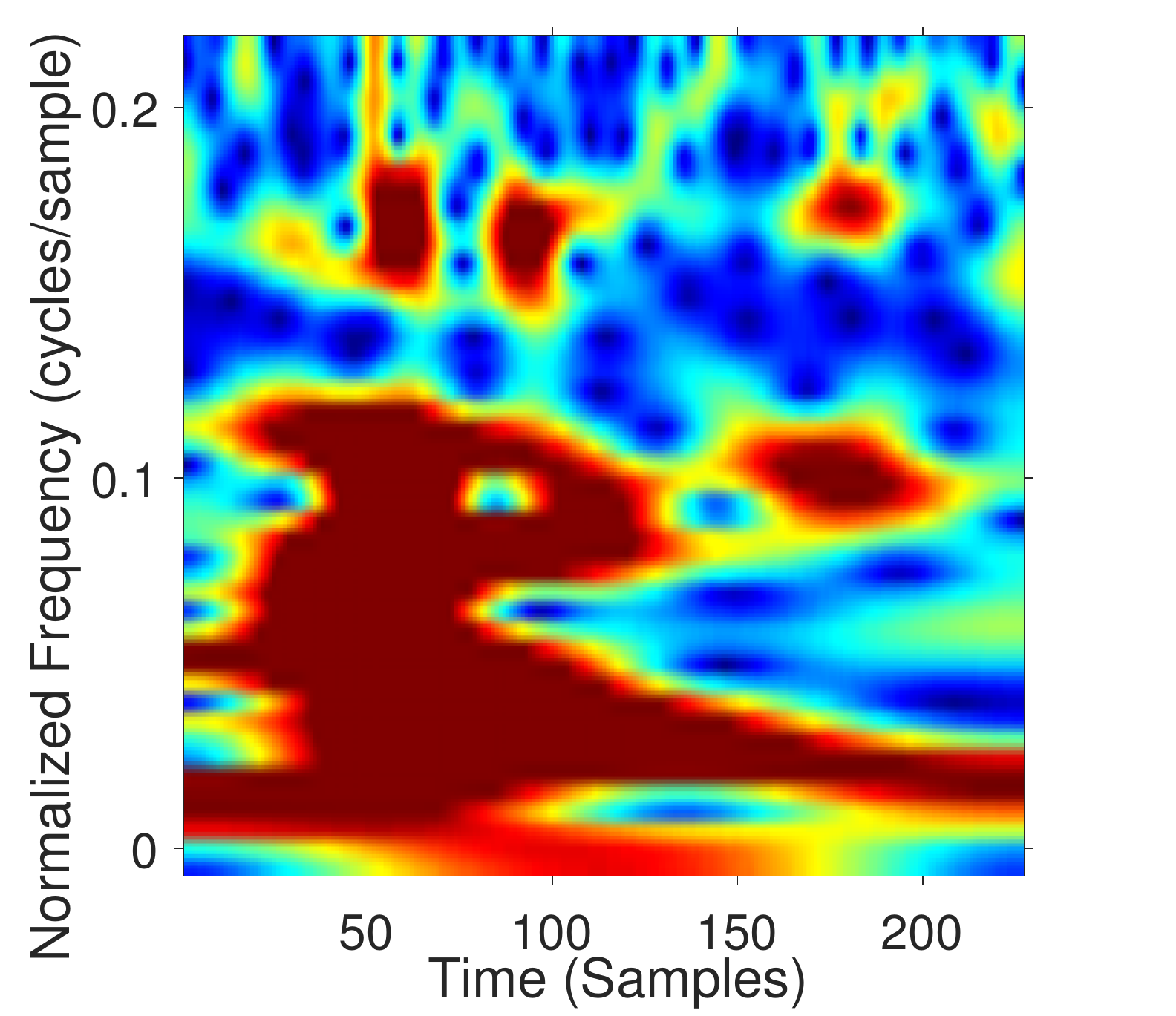}}\label{DJI_INSPIRE_CWT_VV15GHz_processsedImage}
\end{subfigure}
\begin{subfigure}[25 GHz RCS DJI Inspire 1 Pro.]{\includegraphics[width=0.235\linewidth]{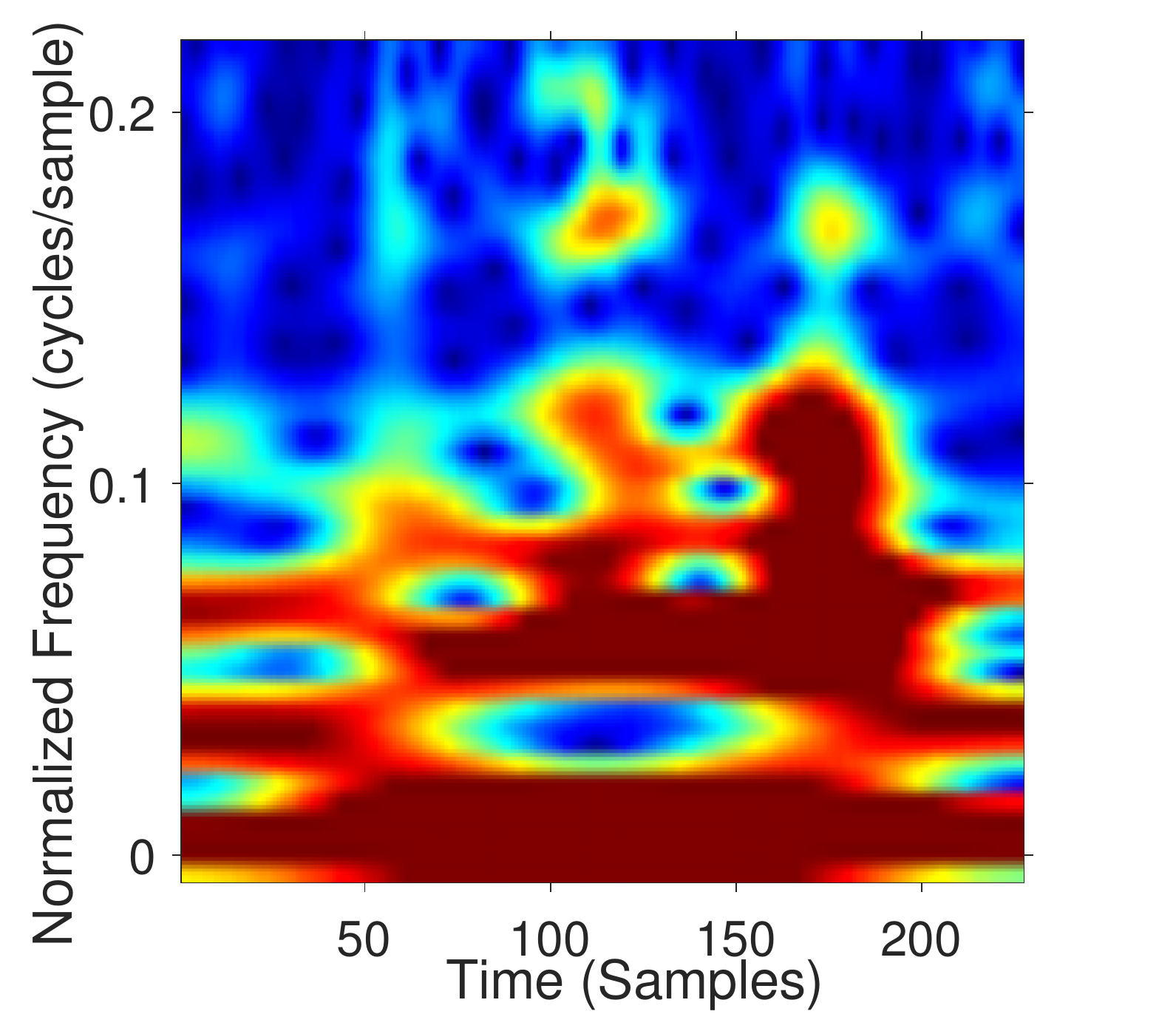}}\label{DJI_INSPIRE_CWT_VV25GHz_processsedImage}
\end{subfigure}
 \begin{subfigure}[15 GHz RCS DJI Phantom 4 Pro.]{\includegraphics[width=0.235\linewidth]{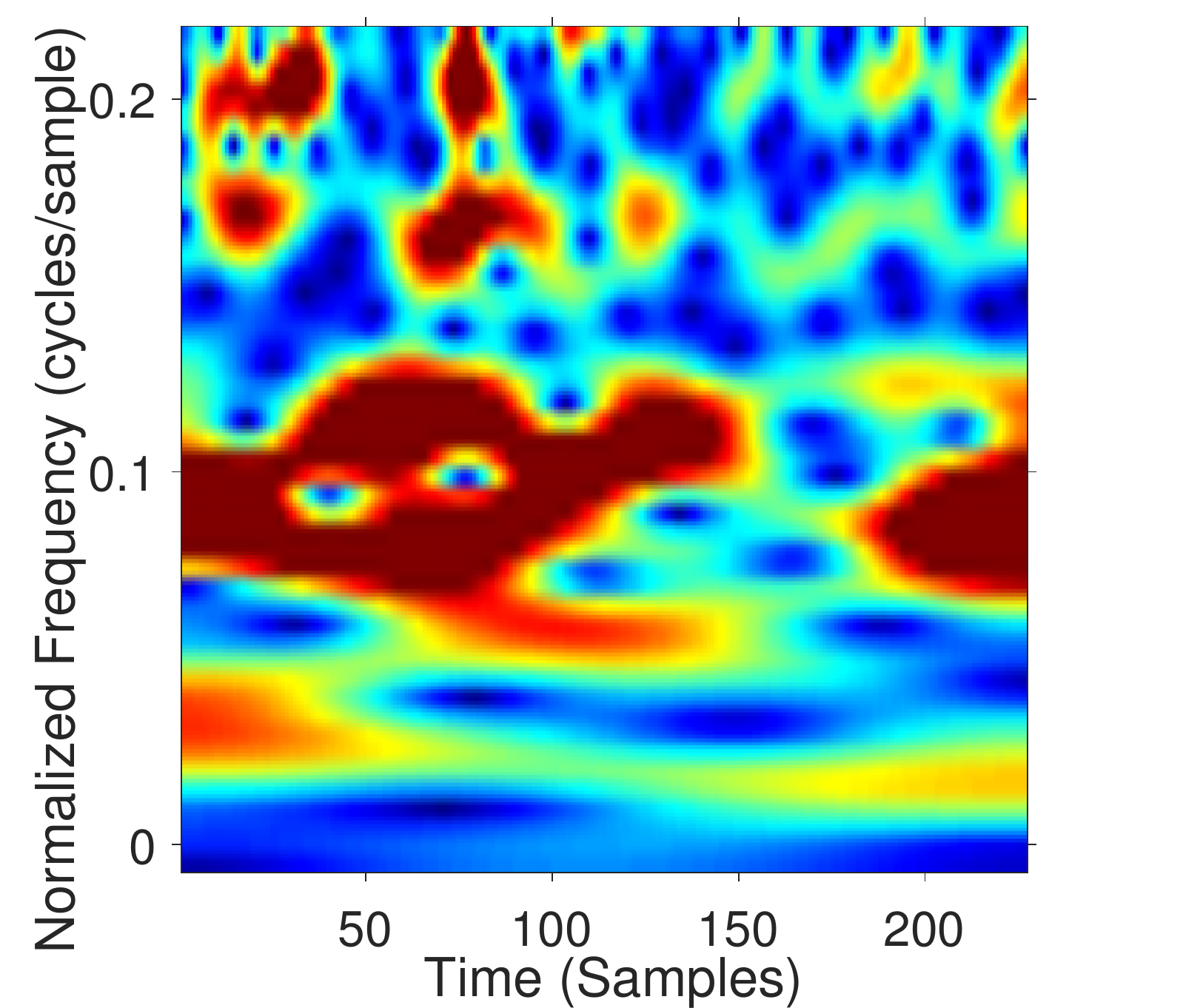}}\label{DJI_Phantom4Pro_CWT_VV15GHz_ProcessedImage}
\end{subfigure}
 \begin{subfigure}[25 GHz RCS DJI Phantom 4 Pro.]{\includegraphics[width=0.235\linewidth]{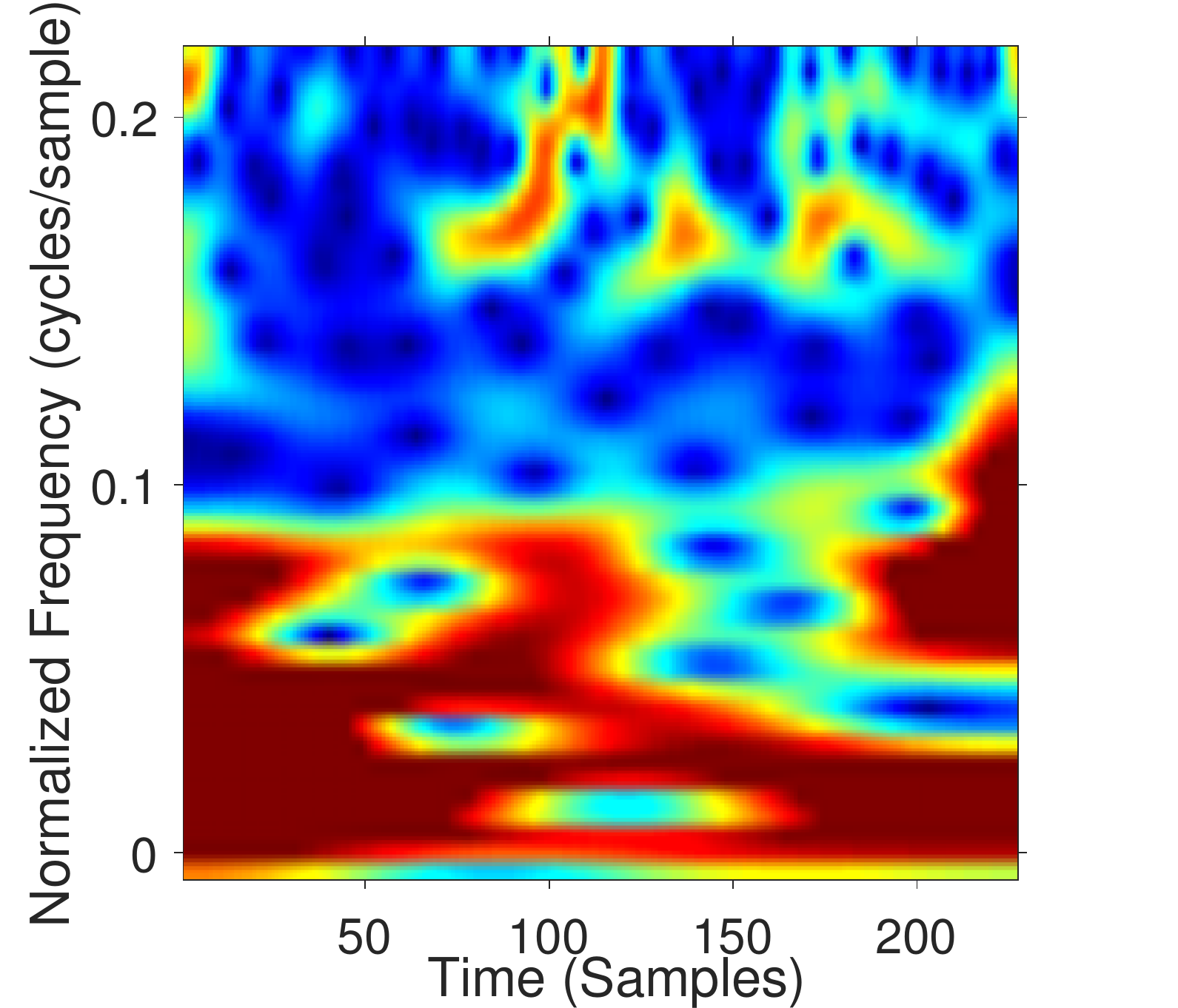}\label{Sample_DJI_Phantom4Pro_CWT_VV}}
\end{subfigure}\\
\begin{subfigure}[ 15 GHz RCS DJI Mavic Pro.]{\includegraphics[width=0.235\linewidth]{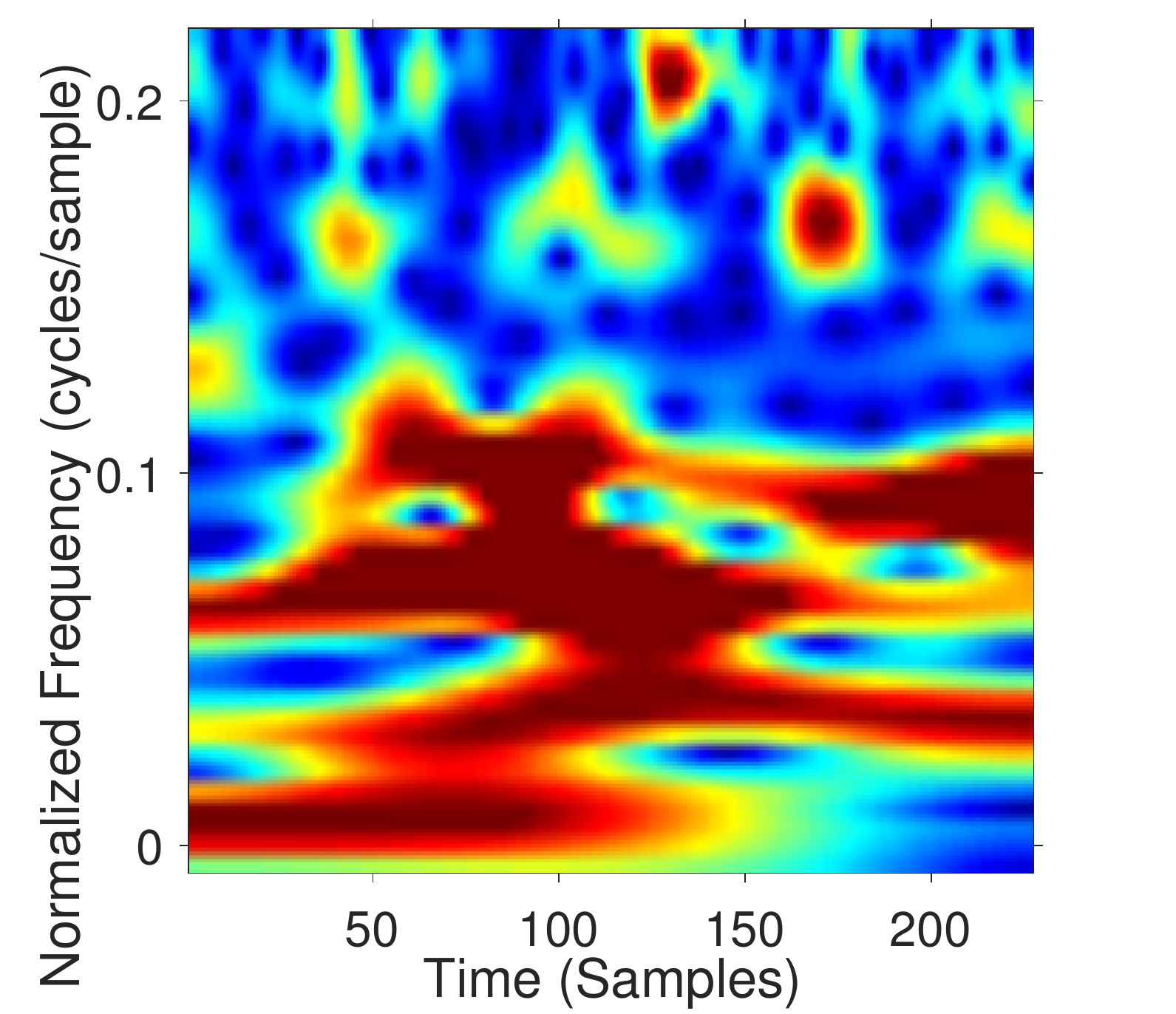}\label{DJI_MAVIC_CWT_VV15GHz_processedImage}}
\end{subfigure}
\begin{subfigure}[ 25 GHz RCS DJI Mavic Pro.]{\includegraphics[width=0.235\linewidth]{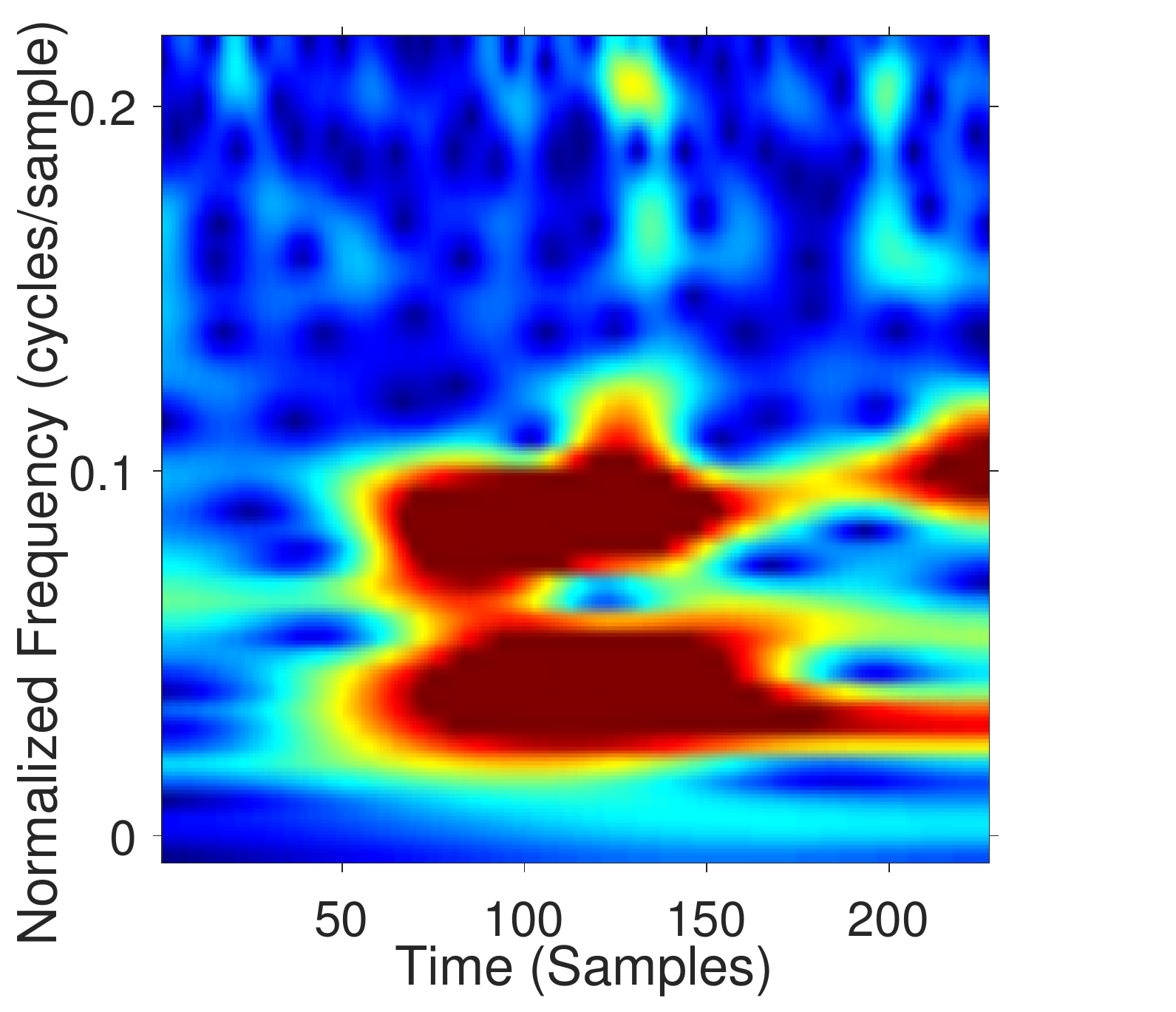}\label{Sample_DJI_MAVIC_CWT_VV25GHz}}
\end{subfigure}
 \begin{subfigure}[15 GHz RCS Trimble zx5.]{\includegraphics[width=0.235\linewidth]{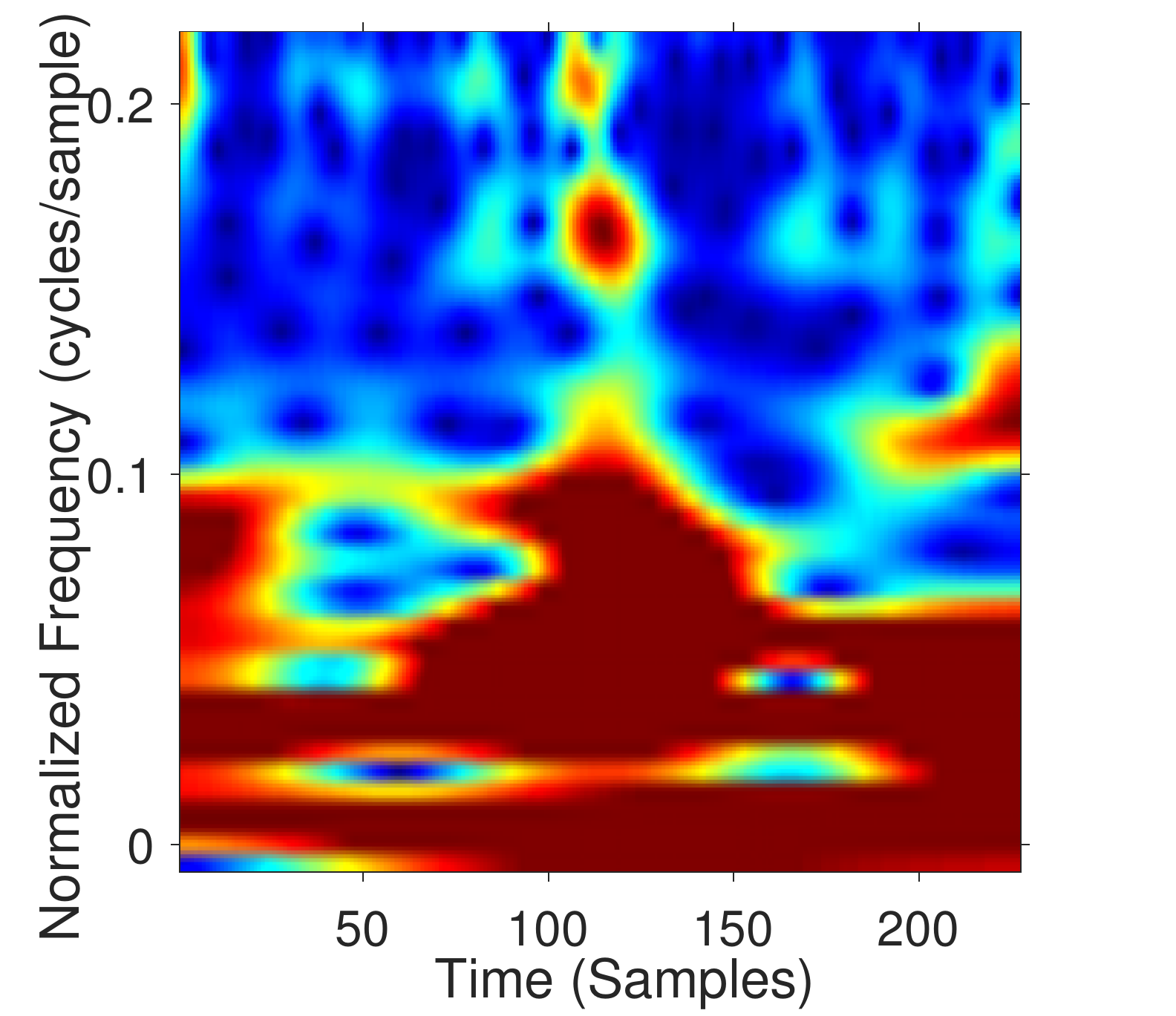}\label{Trimblezx5_CWT_VV15GHz_ProcessedImage}}
\end{subfigure}
\begin{subfigure}[25 GHz RCS Trimble zx5.]{\includegraphics[width=0.235\linewidth]{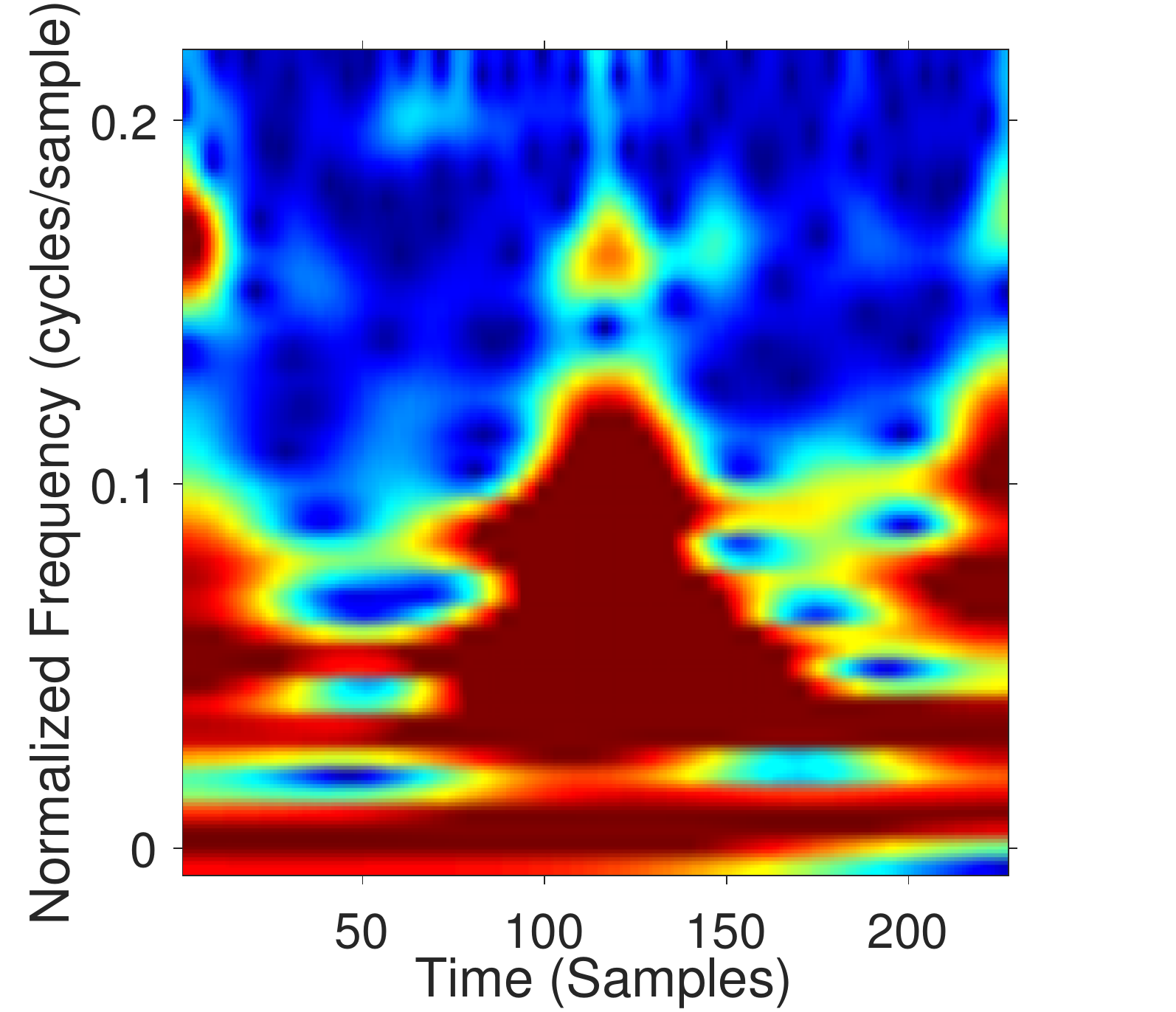}\label{Sample_Trimblezx5}}
\end{subfigure}\\
\caption{The processed CWT scalogram of the six UAVs obtained from their VV-polarized RCS data measured at 15 GHZ and 25 GHz.}\label{scalogram_plots}
\end{figure*}
 
\begin{figure*}[t!]
\center{
 \begin{subfigure}[15~GHz VV-polarized RCS data]{\includegraphics[width=0.45\linewidth]{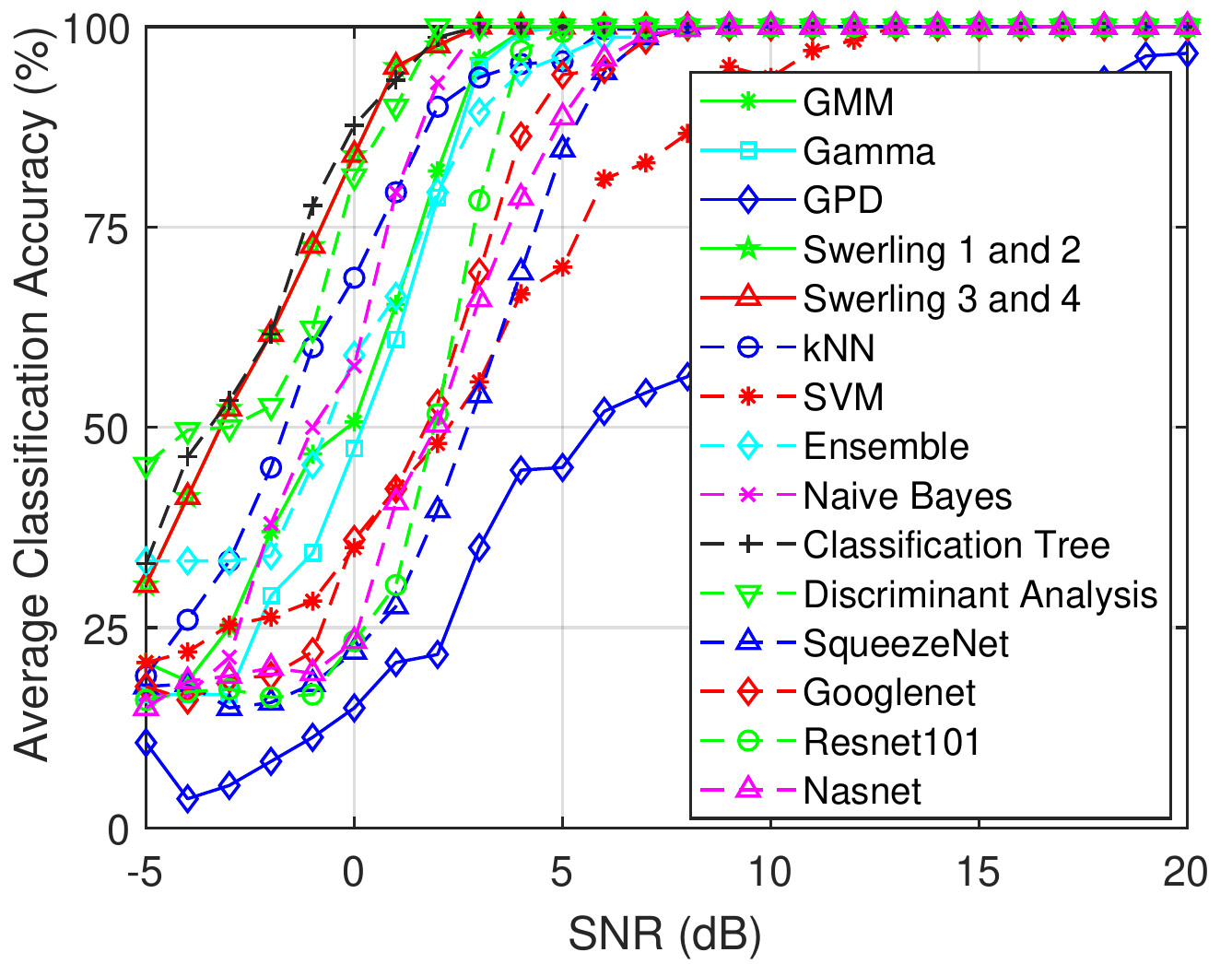}
 \label{VV_15GHZ}}
\end{subfigure}
 \begin{subfigure}[25~GHz VV-polarized RCS data]{\includegraphics[width=0.45\linewidth]{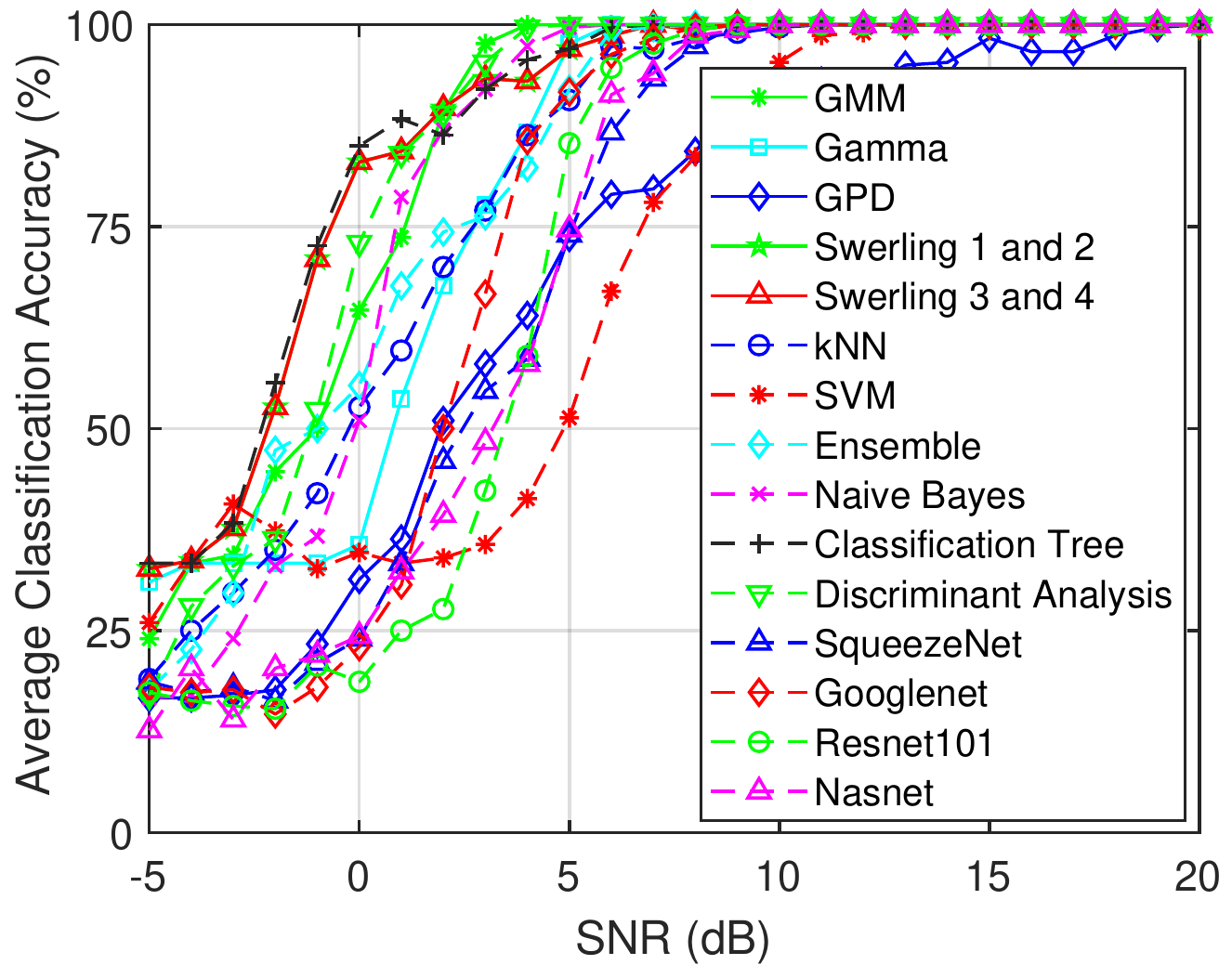}\label{VV_25GHZ}}
\end{subfigure}
 \begin{subfigure}[15~GHz HH-polarized RCS data]{\includegraphics[width=0.45\linewidth]{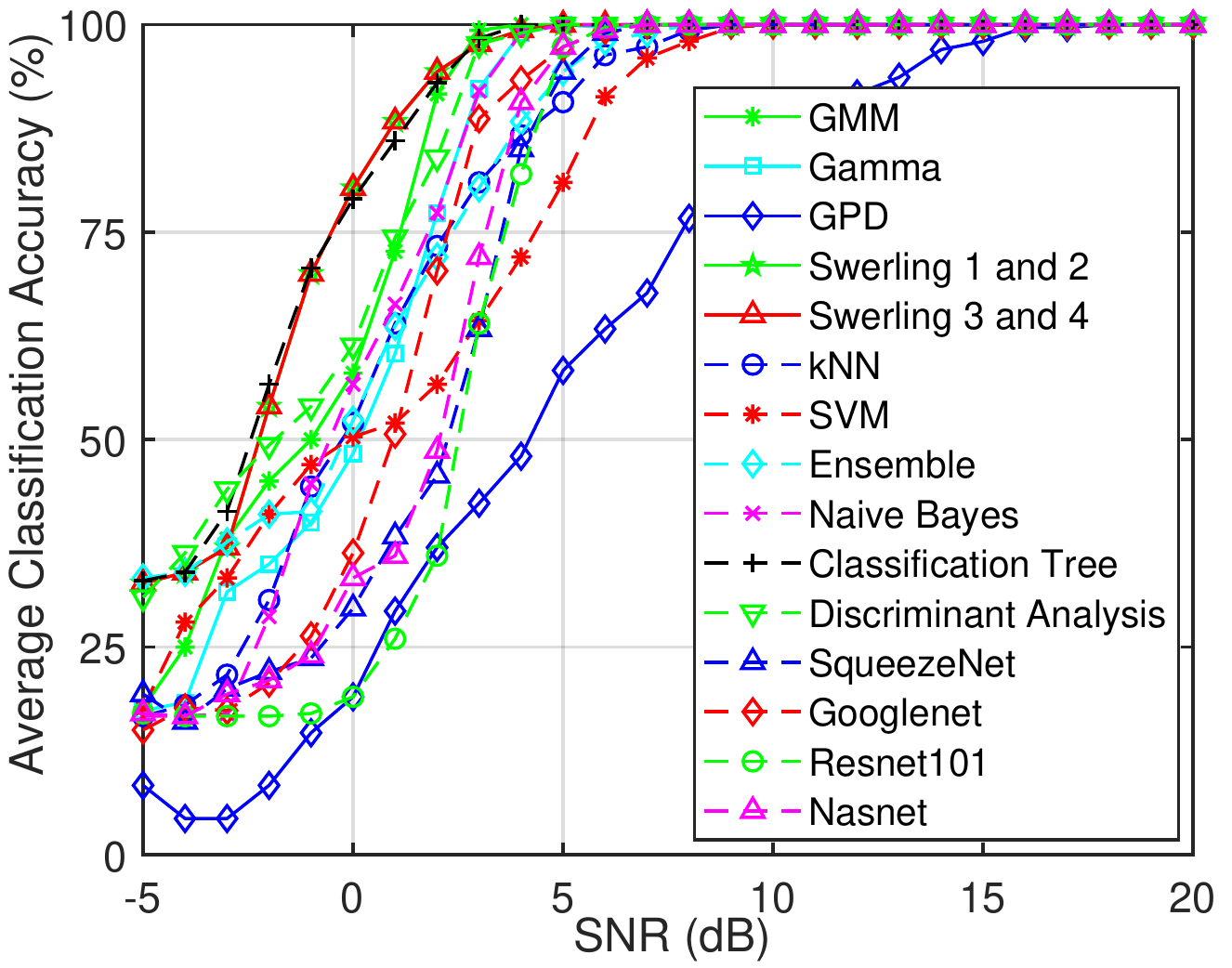}\label{HH_15GHZ}}
\end{subfigure}
 \begin{subfigure}[25~GHz HH-polarized RCS data]{\includegraphics[width=0.45\linewidth]{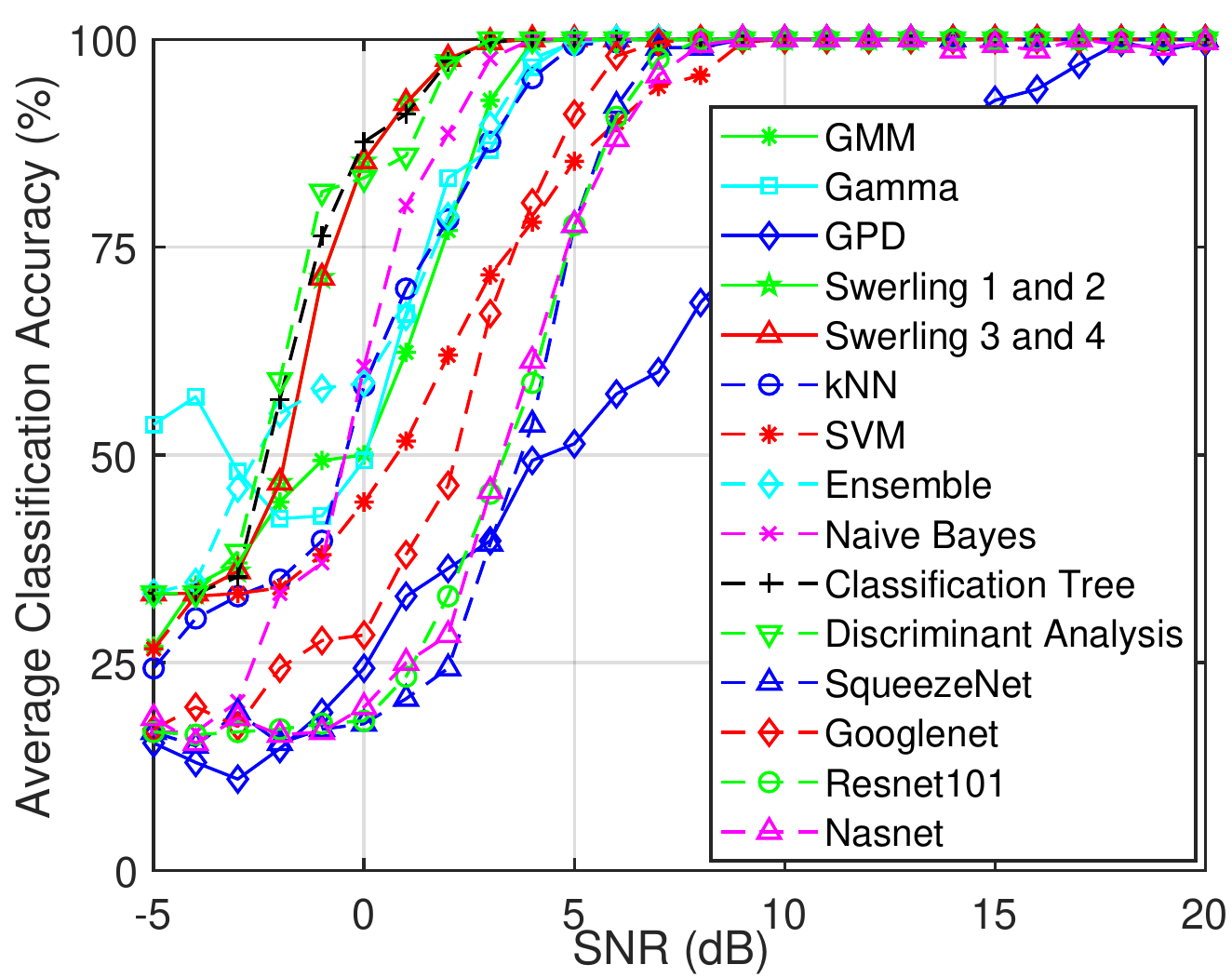}\label{HH_25GHZ}}
\end{subfigure}
\caption{{The average classification accuracy versus SNR for the 15 UAV classification algorithms.}\label{ALL_AZIMUTH_Accuracy_vs_SNR_plots}}
}
 \end{figure*}
 
  \begin{figure*}[t!]
\center{
 \begin{subfigure}[15~GHz VV-polarized RCS data]{\includegraphics[width=0.44\linewidth]{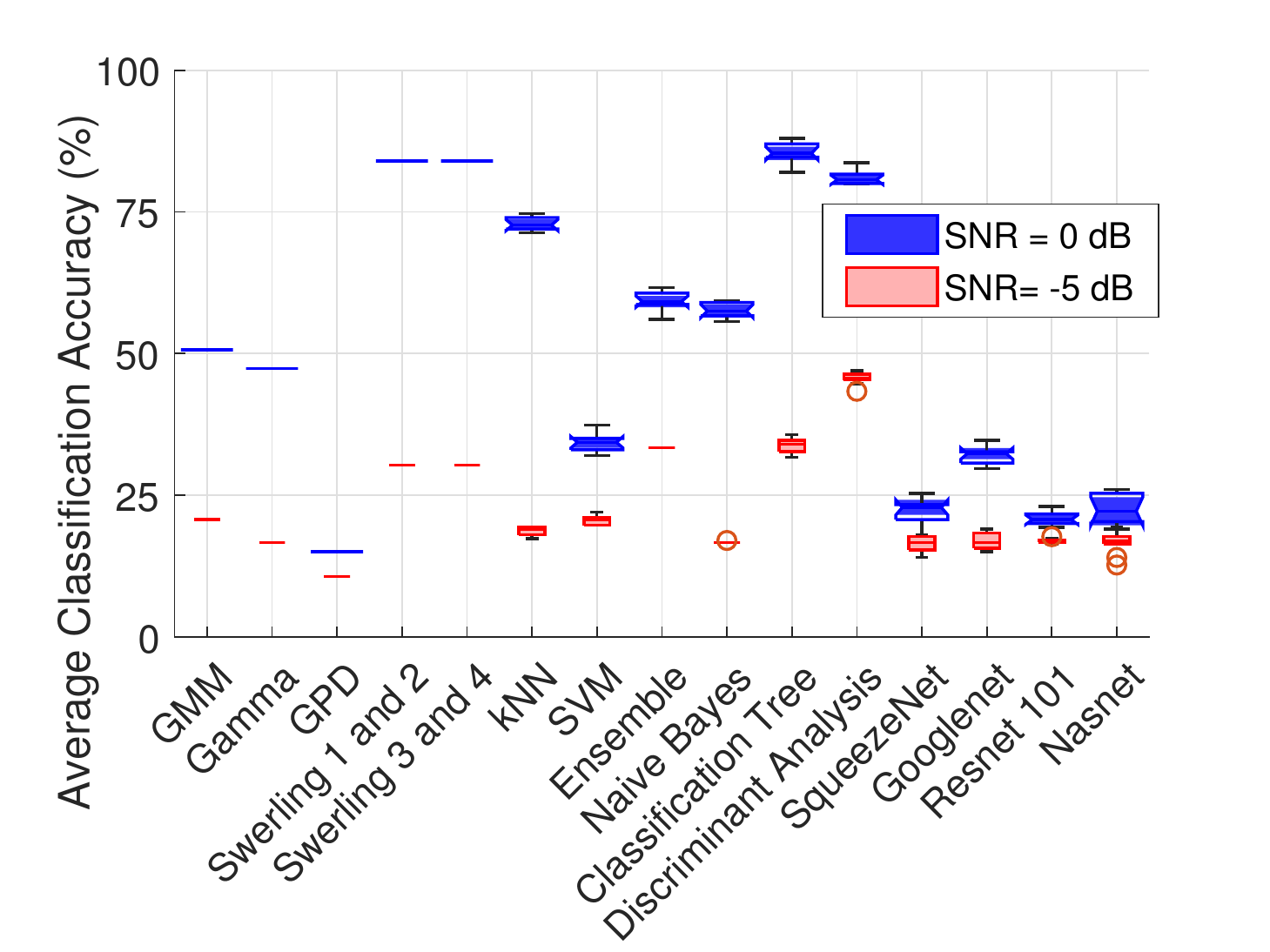}
 \label{VV_15GHz_BOXPLOT}}
\end{subfigure}
 \begin{subfigure}[25~GHz VV-polarized RCS data]{\includegraphics[width=0.44\linewidth]{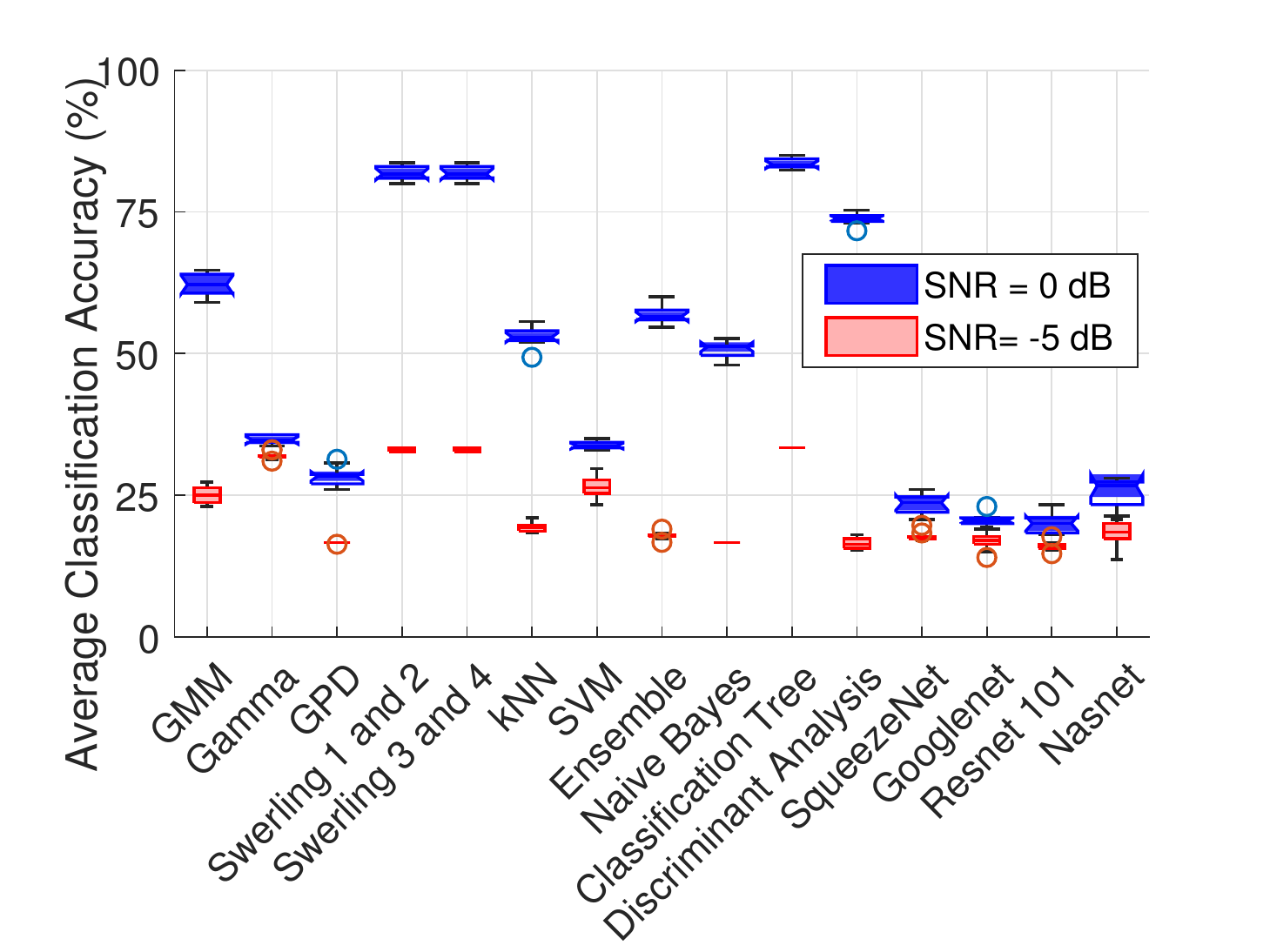}\label{VV_25GHz_BOXPLOT}}
\end{subfigure}
 \begin{subfigure}[15~GHz HH-polarized RCS data]{\includegraphics[width=0.44\linewidth]{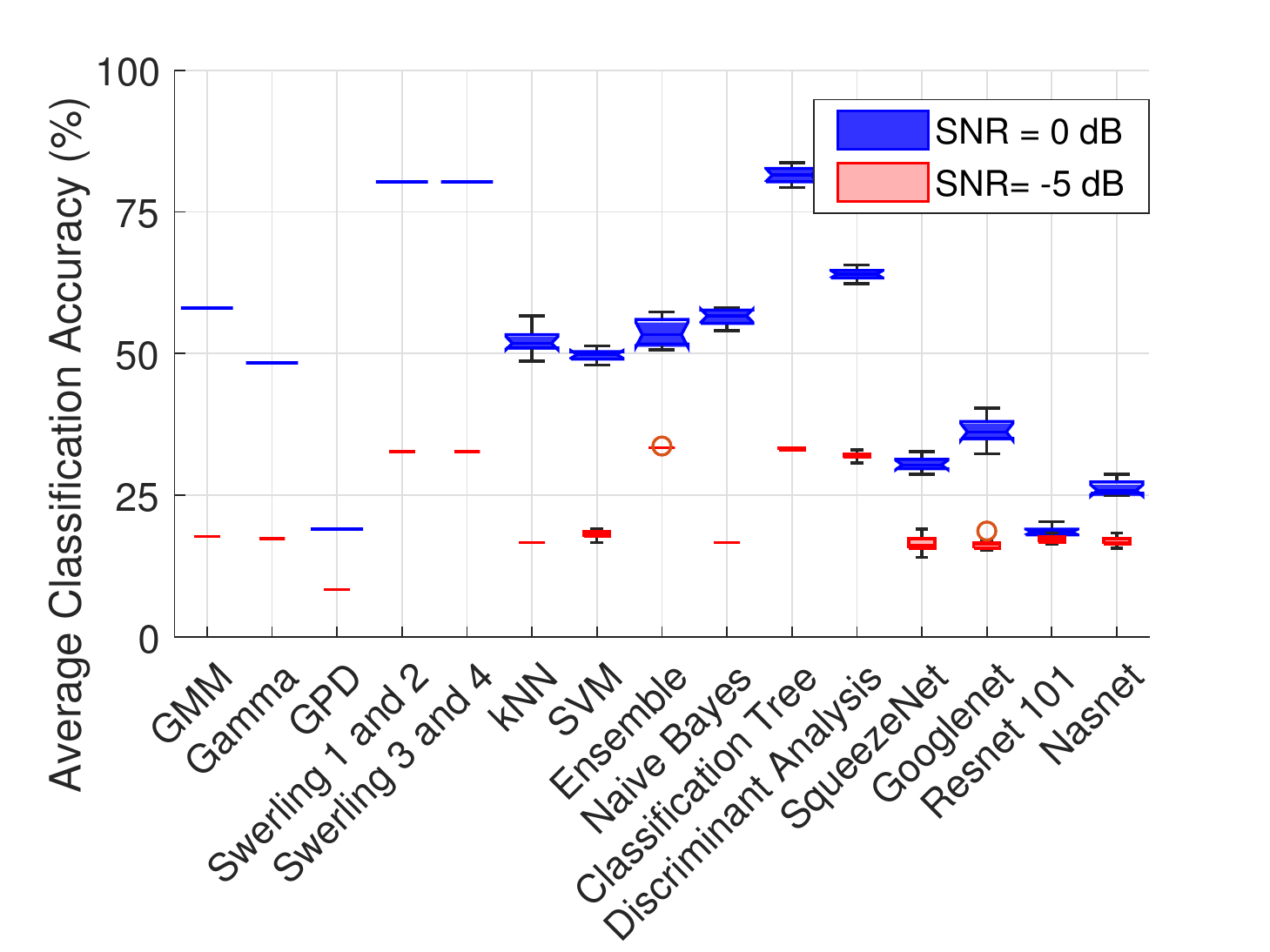}
 \label{HH_15GHz_BOXPLOT}}
\end{subfigure}
 \begin{subfigure}[25~GHz HH-polarized RCS data]{\includegraphics[width=0.44\linewidth]{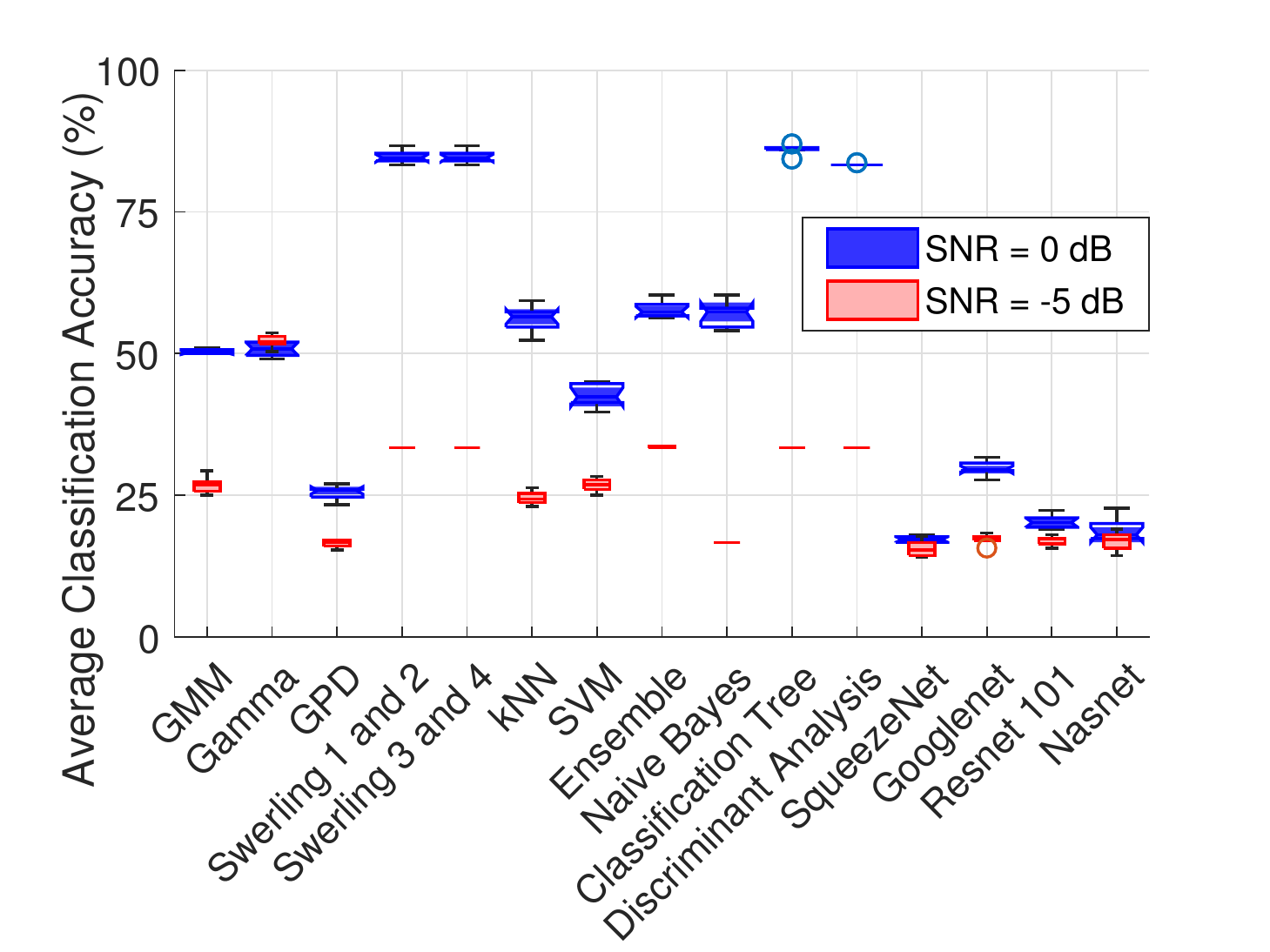}\label{HH_25GHz_BOXPLOT}}
\end{subfigure}
\caption{Box plot analysis of the average classification at low SNR of -5 dB and 0 dB). The results are obtained after 10 Monte Carlo experiments.}\label{Box_charts}
}
 \end{figure*}

\subsection{CWT Time-Frequency Transform of UAV RCS Data}\label{CWT_images}
The CWT provides a T-F representation of a function or signal by windowing the signal with a wavelet function that is continuously scaled and shifted in time. Through the inclusion of all shift and scales, the CWT is able to represent the signal in two-dimension. Mathematically, the CWT of a signal $x(t)$ captured by a radar receiver is defined as:
\begin{equation} \label{CWT_DEFINITION}
CWT(t,\omega;x(t),\psi(t)) = \left(\frac{\omega}{\omega_0}\right) \int x(t') \psi^{\ast}\left( \frac{\omega}{\omega_0} (t'-t)dt'\right)
\end{equation}
where $\psi(\cdot)$ is called the mother wavelet, is a continuous function in both time and the frequency domain. The ratio $\left(\frac{\omega}{\omega_0}\right)$ is called the scale parameter. Basically, (\ref{CWT_DEFINITION}) is a decomposition of $x(t)$ into a family of shifted and dilated wavelet basis functions $\psi\left[\frac{\omega}{\omega_0} (t'-t)\right]$, whose width depends on the value of $\omega$ at time $t$~\cite{chen2002time}. Therefore, by shifting $\psi(\cdot)$ at a fixed $\omega$ or dilating $\psi(\cdot)$ at a fixed $t$, the multiscale events of the scale parameter can be extracted or localized~\cite{chen2002time}. The resulting two-dimensional magnitude display of the CWT in (\ref{CWT_DEFINITION}) is called the scalogram. 

Fig.~\ref{discriminant_opt} shows the CWT scalogram of the 25~GHz VV-polarized UAV RCS data measured from the DJI Matrice 600 Pro UAV. The CWT scalogram images contain the important features that can be learned by a transfer learning DL classifier. This is because the scalogram can simultaneously capture the slowly varying (low frequency) RCS fluctuations of the target UAV as well as the transient (time) characteristics. Thus, the enhanced color on the image indicates RCS returns with higher magnitude while the blurry portions indicate low magnitude RCS samples. The color distribution will indicate the RCS signature of the UAV in the T-F domain. To use a CWT scalogram as the input into a transfer learning DL classifier, the image has to be resized and processed to be compatible with the input size requirement of the network. For this reason, we further process the CWT scalogram image by taking the magnitude of the scalogram and rescaling it to the MATLAB Jet (128) colormap interval. Afterward, the image is resized to fit the input requirement of the transfer learning DL classifier.

Fig.~\ref{knn_opt} shows the processed CWT scalogram image obtained from the 25~GHz VV-polarized UAV RCS data measured from the DJI Matrice 600 Pro UAV. We can see the processed radar image enhances the features of the CWT scalogram in Fig.~\ref{discriminant_opt}. With enhanced image features, it may be easier for the DL classifier to perform UAV identification. Fig.~\ref{scalogram_plots} shows the processed CWT images of VV-polarized RCS data of the six UAVs measured at 15 GHz and 25 GHz. Comparing the processed CWT images, we observe the distinctive features that can aid the identification of the different UAVs. Therefore, to classify or identify unknown UAVs using radar, we feed the processed CWT image of the captured RCS into the transfer learning DL classifier.

\subsection{Transfer Learning DL Algorithms for UAV Identification}
The objective of the transfer learning DL approach to RCS-based UAV classification is to apply knowledge or features learned from a different problem to a new problem rather than starting from scratch to train a DL model. That is, we take the layers from a pre-trained DL network and fine-tune it for use as the starting point for the RCS-based UAV recognition system. This approach is desirable since it is much faster and easier to fine-tune a pre-trained network than training a new DL network from scratch with initialized randomized weights. More importantly, it is desirable to have a transfer learning network with fewer parameters as it reduces the memory footprints and computational cost of training and testing transfer learning-based ATR systems. For these reasons, we adopt the  SqueezeNet~\cite{iandola2016squeezenet} for the RCS-based UAV classification problem.

SqueezeNet, a CNN-like architecture, is reported to achieve a classification accuracy and performance comparable to AlexNet on the ImageNet database challenge, with 50$\times$ fewer parameters (reduced model size) than  AlexNet~\cite{iandola2016squeezenet}. In addition to SqueezeNet, we also included other transfer learning DL networks such as GoogleNet, ResNet~101, and NasNet-Mobile (NasNet) which are relatively heavier in their memory requirement and deeper in their complexity~\cite{mathwork_transferLearning}. Table~\ref{Transfer_Learning_DL} provides a brief comparison of the structure of all the transfer learning DL classifiers used in the study. In comparison to the ML classifiers, the transfer learning DL classifiers do not need handcrafted (manually generated) features for training and testing the UAV recognition system. Instead, transfer learning DL classifiers use automatically generated features that are learned from the processed CWT scalogram images (input data) shown in Fig.~\ref{scalogram_plots}. 


\begin{table}[t!]
\setlength{\tabcolsep}{4.8pt}
\centering
\caption{The Transfer Learning DL Classifiers used for the RCS-based UAV classification.}
\label{Transfer_Learning_DL}
\begin{threeparttable}
 \begin{tabular}{|P{16mm}|P{10mm}|P{11mm}|P{12mm}|P{15mm}|}
\hline
\textbf{Transfer Learning DL} & \textbf{Network Depth} & \textbf{Memory Size} & \textbf{Parameter (Millions)} & \textbf{Image Input Size} \\

\hline
SqueezeNet & 18 & 5.2 MB & 1.24 & 227$\times $227 \\
\hline

GoogleNet & 22 & 27 MB & 7.0 & 224$\times$224\\
\hline

ResNet 101 & 101 & 167 MB & 44.6 & 224$\times$224 \\
\hline

NasNet & \text{*} & 20 MB & 5.3 & 224$\times$224 \\
\hline
\end{tabular}
  \end{threeparttable}
  \begin{tablenotes}
  \item \text{*} NasNet-Mobile (NasNet) does not consist of a linear sequence of modules and as such its complexity is independent of depth~\cite{mathwork_transferLearning}.
  \end{tablenotes}
\end{table}



\begin{figure}[t!]
\label{Confusion_matrices}
 \begin{subfigure}[Classification tree]{\includegraphics[width=0.365\textwidth]{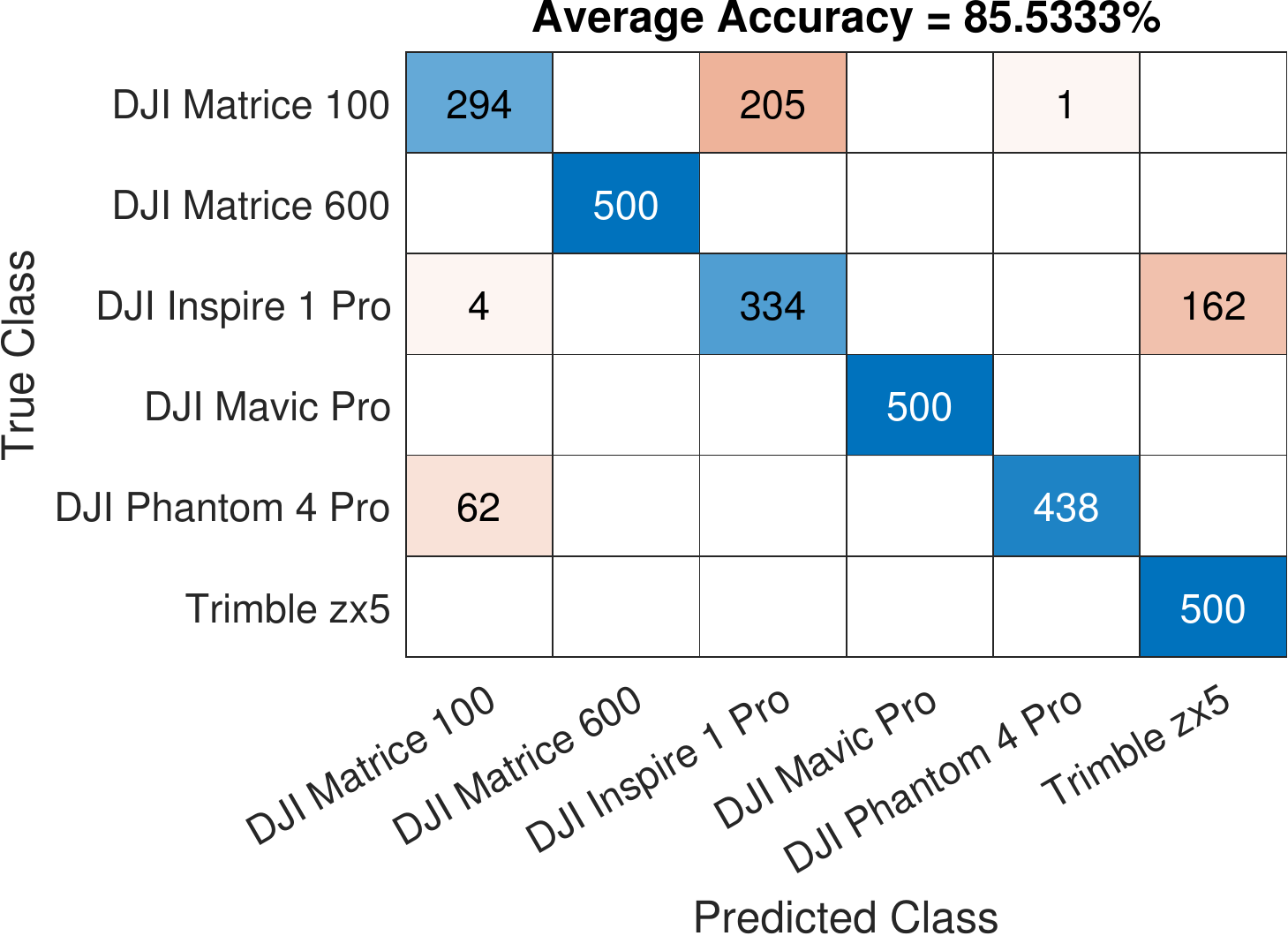}
 \label{Confusion_matrix_CTree}}
\end{subfigure}
 \begin{subfigure}[Swerling 1 and 2 model]{\includegraphics[width=0.365\textwidth]{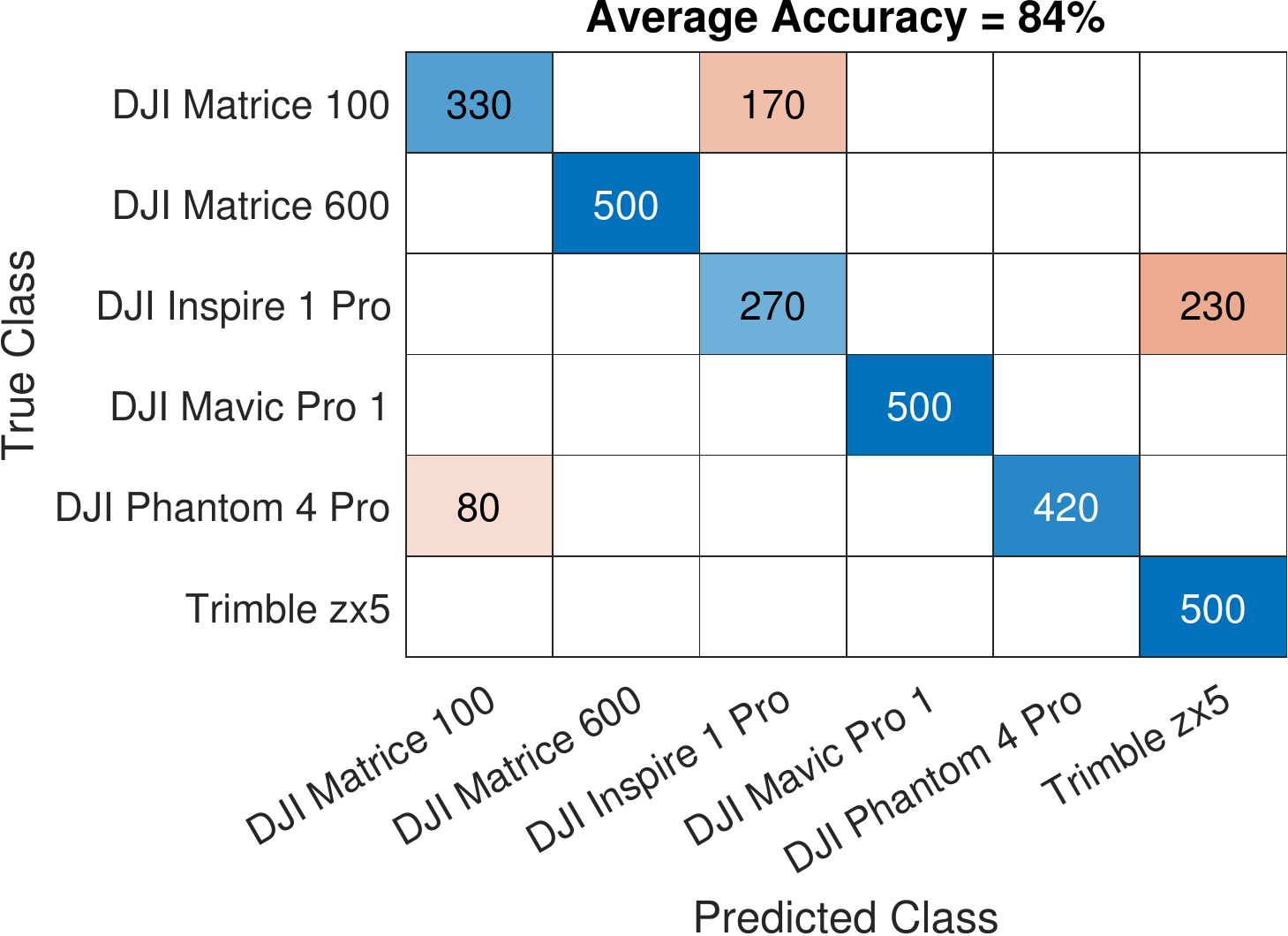}\label{Confusion_matrix_Swerling12}}
\end{subfigure}
\begin{subfigure}[SVM]{\includegraphics[width=0.365\textwidth]{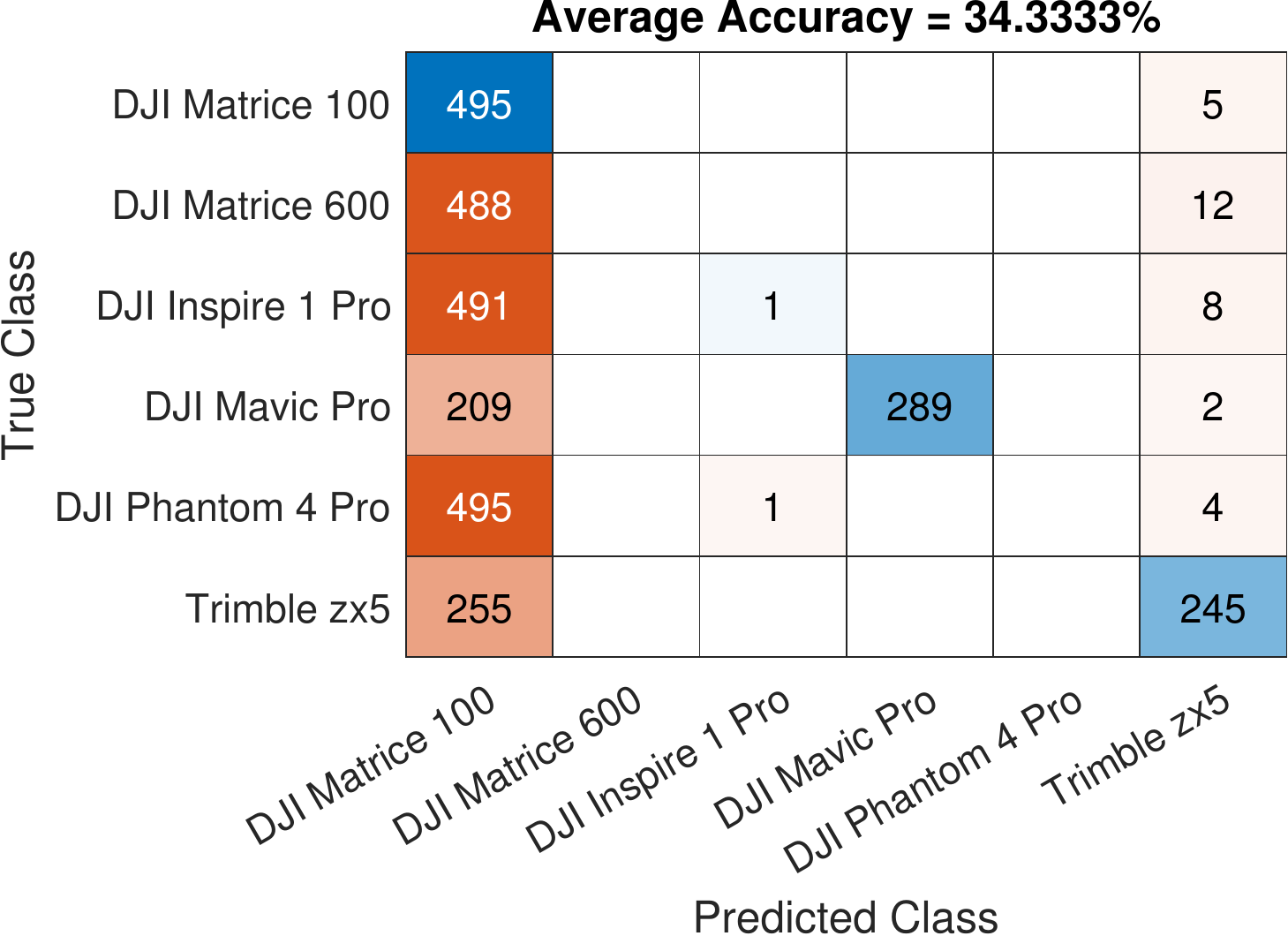}\label{Confusion_matrix_SVM}}
\end{subfigure}
 \begin{subfigure}[GPD]{\includegraphics[width=0.365\textwidth]{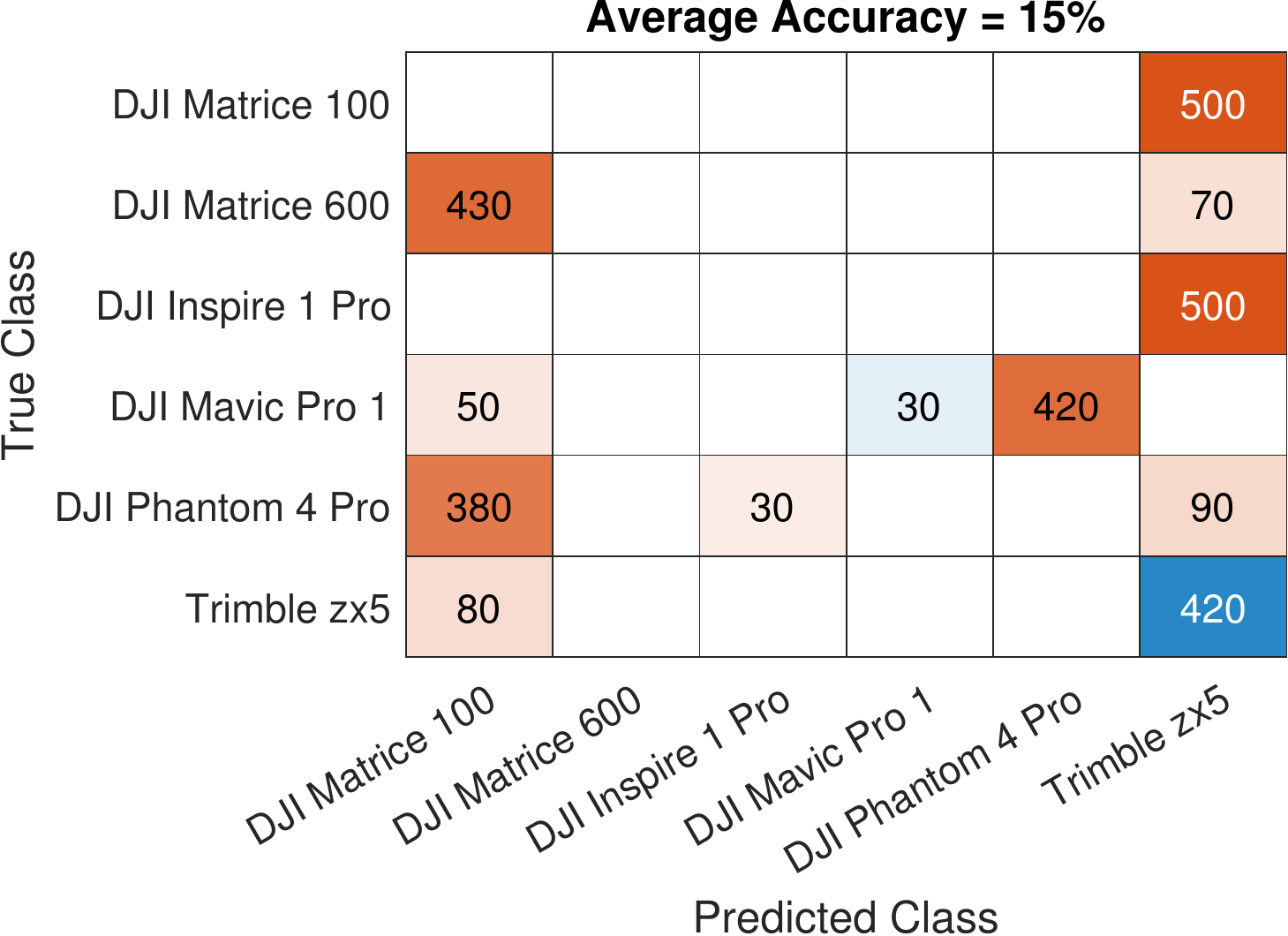}
 \label{Confusion_matrix_GPD}}
\end{subfigure}
 \caption{The confusion matrices of the best and worst classification algorithms at 0 dB SNR using the VV 15 GHz test samples and running 10 Monte Carlo experiments. The confusion matrices depict the strength and weaknesses of the selected classifiers.}
\end{figure}

\section{Results and Discussion}
In this section, we describe and compare the results of the UAV classification/recognition using all the 15 different SL, ML, and DL approaches presented in the last three sections. We investigate the effect of varying the SNR of the backscattered RCS data from a target UAV. All the analysis described in this section is carried out using a 64 bit, 16 GB RAM, 3.4 GHz Intel i7-3770 personal computer~(PC).

{\subsection{Comparison of UAV Recognition Algorithms Using the Complete RCS Data}\label{complete_azimuth}}

\color{black}
Here, we analyze the performance of the 15 classifiers at different SNR using all azimuth RCS data from an unknown UAV. For example, given a  time-domain RCS data $v_{k}(t)$ backscattered from an unknown $k$-th UAV that is illuminated by a monostatic radar, we estimate the noise power ($\sigma_N^2$) of the additive white Gaussian noise (AWGN) as \cite{ezuma2021radar}:

\begin{align}
\label{SNR_EQU}
  \sigma_N^2 =P_{k}10^{-{\rm SNR}/10},
\end{align}
where
\begin{equation}
\label{power_measurement}
P_{k}  =\frac{1}{T}\int_{0}^{T} v_{k} (t)^2 {\rm d}t =\\
  \begin{cases}
    \frac{\sum_{i=1}^{N} \big|\sqrt{\sigma_{\rm VV}}\big|^2}{N},~\text{VV-polarization} \\
    \frac{\sum_{i=1}^{N} \big|\sqrt{\sigma_{\rm HH}}\big|^2}{N},~\text{HH-polarization}\\
  \end{cases}
\end{equation}
is the average power of the backscattered RCS return $v_{k}$, $T$ is the time interval of the data collection, and $N$ is the number of discrete-time samples of the frequency domain RCS return. After estimating $\sigma_N^2$, we add the appropriate AWGN noise to the experimental RCS data to generate noisy RCS samples which are used to evaluate the algorithms. Recall, for the SL algorithms, the process of training the classifiers involves fitting the experimental UAV RCS data to the models and estimating the parameters of the models using either MLE or the EM algorithm as the case may be. For the ML and DL algorithms, we generate a training dataset, consisting of 100 RCS data of each UAV, by adding appropriate AWGN noise to the experimental UAV RCS data. The ML classifiers are trained by extracting and using as inputs a feature vector that consists of the seven handcrafted features, given in Table~\ref{Table_feature}. The ML classifiers are optimized by using the Bayesian hyperparameter optimization technique described in Section~\ref{machine_learning_techniques}. On the other hand, the DL algorithms are trained by converting the training dataset to CWT scalogram images as described in Section~\ref{CWT_images}.

Fig.~\ref{ALL_AZIMUTH_Accuracy_vs_SNR_plots} shows the average accuracy versus SNR plot for the 15 GHz and 25 GHz RCS test data. In Fig.~\ref{VV_15GHZ}, at -5 dB SNR and using the 15 GHz VV-polarized UAV RCS returns, the discriminant analysis algorithm achieved the highest average accuracy of 45.33\%, followed by the classification tree algorithm at 33.00\% and the Swerling 3 and 4 models at 30.33\%. The least performing algorithm at -5 dB SNR is the GPD with an average accuracy of 10.67\%. In general, as the SNR increases the average accuracy of all the algorithms improves steadily. For example, at an SNR of 3 dB, the accuracy of discriminant analysis, classification tree, and Swerling case 3 and 4 models are 100\%, 98.66\%, and 97.66\% respectively. Except for the SVM and GPD models, all other models achieve an average accuracy of 100\% accuracy at 8 dB SNR.

From Fig.~\ref{VV_15GHZ}, we also see that the GPD model is the worst performing even at higher SNR. This is probably because GPD statistical models are mainly suited for modeling heavy-tail distributions and may not correctly fit the peak of the UAV RCS test dataset. Similar trend is observed in Fig.~\ref{VV_25GHZ}, Fig.~\ref{HH_15GHZ}, and Fig.~\ref{HH_25GHZ}. That is, at low SNR, -5 dB to 3 dB, the discriminant analysis, classification tree, and Swerling case 3 and 4 models are relatively better than the other classifiers. On the other hand, in Fig.~\ref{HH_25GHZ}, the performance of the Gamma SL model was superior at -5 dB SNR with an average accuracy of 53.67\%. This is followed by the classification tree and the Swerling models. The average performance of all 15 classifiers can be investigated further with the aid of the Monte Carlo analysis which is discussed next.

\subsection{Low SNR Performance Analysis: Boxchart and Confusion Matrices}

To further analyze the performance of the classifiers at very low SNR, we performed 10 Monte Carlo experiments. In each Monte Carlo experiment, a set of 50 noisy RCS test data from each UAV is generated at 0 dB and -5 dB respectively. These noisy test data are fed into the 15 classifiers for UAV identification. The summary of the Monte Carlo experiment for SNR of 0 dB and -5 dB are depicted by the box charts (boxplots) shown in Fig.~\ref{Box_charts}. In Fig.~\ref{Box_charts}, each box, with its extending lines, is used to graphically display the maximum, minimum, median, lower and upper quartile, and an outlier in the Monte Carlo experiment for the given classifier. For the 0 dB SNR, Fig.~\ref{Box_charts} clearly shows that the classification tree algorithm is relatively better than the other models. The next best models are Peter Swerling's statistical models and the discriminant analysis algorithm. A similar trend is observed for the -5 dB SNR case, exception for HH 25 GHz RCS boxplot analysis shown in Fig.~\ref{HH_25GHz_BOXPLOT}. In general, the box charts show that the ML and SL algorithms outperformed the DL algorithms at the low SNR scenarios considered. This may be due to the loss in information feature when the RCS data is transformed to RCS CWT images used as inputs into the DL classifiers. Also, DL algorithms are believed to perform better with very large training data. However, such a requirement will demand longer training times as compared to the SL and ML algorithms.

Although the box charts show the average performance of each algorithm, it does not provide information about the weakness of each classifier and what accounts for the misclassifications in the Monte Carlo test experiments. To evaluate how our model performed and where it went wrong at very low SNR, say 0 dB SNR, we generate the confusion matrices shown in Fig.~\ref{Confusion_matrices}. In Fig.~\ref{Confusion_matrices}, we present the confusion matrices of the best and the worst classifiers after running 10 Monte Carlo test experiments with the VV 15 GHz UAV RCS test at 0 dB SNR. In each confusion matrix, the diagonal elements represent the instances of correct/accurate prediction of the identity of an unknown UAV by the given classifier while the off-diagonal elements are instances of misclassification.

\begin{table*}[t!]
\setlength{\tabcolsep}{5pt} 
\centering
\vspace{-2mm}
\caption{The average computational time for the RCS-based UAV classification algorithms after 10 Monte Carlo experiments. Results are provided for different frequency (\lowercase{Freq.}) and polarization (\lowercase{Polar.}). The classification algorithms are 1: GMM, 2: Gamma, 3: GPD, 4: Swerling 1 and 2, 5: Swerling 3 and 4,  6: kNN, 7: SVM, 8: Ensemble, 9: Naive Bayes, 10: Classification Tree, 11: Discriminant Analysis, 12: Squeezenet, 13: Googlenet, 14: Resnet, 15: Nasnet}
\label{computational_time}
\begin{tabular}{|c|c|c|c|c|c|c|c|c|c|c|c|c|c|c|c|c|}
\hline
\textbf{Polar.} & \textbf{Freq.} &   
\multicolumn{15}{c|}{\textbf{Average Computational Time (ms)}}  \\ 
& \textbf{(GHz)}& \multicolumn{15}{c|}{}  \\
\cline{3-17} 

& & \multicolumn{4}{c|}{SL} & & \multicolumn{5}{c}{ML} & & \multicolumn{4}{|c|}{DL}   \\
\cline{3-17}


 &  & 1 & 2 & 3 & 4 & 5 & 6 & 7 & 8 & 9 & 10 & 11 & 12 & 13 & 14 & 15 \\

\cline{1-17}
\multirow{2}{*}{\text{VV}}
& 15 & 626.02 & 660.09 & 709.58 & 612.18 & 588.79 & 0.92 & 2.95 & 38.39 &  2.82 & \cellcolor{blue!25}0.46 & 0.94 & 27.20 & 48.43 & 163.05 & 145.73\\
\cline{2-17}
&  25 & 667.09 & 706.75 & 659.94 & 685.89 & 686.64 & 0.95 &  3.00 & 10.80 & 2.89 & \cellcolor{blue!25}0.48 & 0.95 & 22.81 & 57.83 & 171.33 & 144.56\\

\hline
\multirow{2}{*}{\text{HH}}
& 15 & 580.26 & 589.77  & 569.80 & 583.02 &  570.14 &  0.89 & 3.78 & 200.97 & 2.80 & \cellcolor{blue!25}0.46 & 0.95 & 23.97 & 48.18 & 162.74 & 156.62\\
\cline{2-17}
& 25 & 675.33 & 712.28 & 668.68 &  689.67
 & 691.71 & 0.97 & 3.16 & 5.81 & 2.84 & \cellcolor{blue!25}0.50 & 0.99 & 23.08 & 45.10 & 153.33 & 144.00\\ 
 \hline
\end{tabular}
\end{table*}

Fig.~\ref{Confusion_matrix_CTree} and Fig.~\ref{Confusion_matrix_Swerling12} are the confusion matrix of the classification tree and the Swerling 1 and 2 models respectively. In these two classifiers, we see that the misclassification error at 0 dB SNR arises mainly from three instances: misclassifying DJI Phantom 4 Pro as DJI Matrice 100, misclassifying DJI Matrice 100 as DJI Inspire 1 Pro, and misclassifying DJI Inspire 1 Pro as Trimble zx5. However, the classification tree and the Swerling 1 and 2 models still achieved an average accuracy of 85.53$\%$ and 84$\%$ respectively at 0 dB SNR. The high accuracy shows that the classification tree and the Swerling 1 and 2 models are robust classifiers even at low SNR scenarios. On the other hand, Fig.~\ref{Confusion_matrix_SVM} and Fig.~\ref{Confusion_matrix_GPD} show the confusion matrix of the SVM and GPD classifiers respectively. From Fig.~\ref{Confusion_matrix_SVM} and Fig.~\ref{Confusion_matrix_GPD}, we see that at 0 dB SNR, the SVM and GPD classifiers perform badly in recognizing most of the UAV RCS test samples. Therefore, SVM and the GPD classifiers cannot be trusted at such a low SNR (or even at high noise instances) for the given dataset.

The low SNR performances of the SVM and GPD classifiers could be improved if we train these underperforming models with noisy UAV RCS samples. However, training models with noisy data may force the models to overfit the noise samples in the dataset. That is, the models could end up learning the noise distribution instead of the actual UAV RCS data or signal. These issues will be investigated in future work.

\subsection{Average Classification Time}
To analyze the time-complexity of the different algorithms, we perform 10 Monte Carlo experiments and average the classification time. In each Monte Carlo experiment, a set of 50 noisy RCS test data from each UAV is generated at 0 dB SNR and fed into the 15 classifiers. The average computation times are provided in Table~\ref{computational_time} for both the VV and HH-polarized RCS returns from the UAVs at 15 GHz and 25 GHz. The table shows that the classification tree algorithm is the fastest with an average classification time of about 0.46~ms. This is closely followed by the kNN and discriminant analysis with an average classification time in the range 0.89 - 0.97~ms and 0.94 - 0.99~ms, respectively.

Among the DL classifiers, the Resnet 101 algorithm took a relatively long time to perform the classification. This is because, compared to the other DL classifiers, Resnet 101 has more depth (highly dense layers), more parameters, and occupies more space in the PC memory as shown in Table~\ref{Transfer_Learning_DL}. On the other hand, the Squeezenet DL classifier, which is relatively less dense with fewer parameters, obtained relatively faster classification times as compared to the other DL classifiers. The next fastest DL classifier is the Googlenet followed by the Nasnet. On the other hand, the SL classifiers obtain relatively longer classification times as compared to the ML and DL classifiers. This is probably because the SL algorithms have to fit all the test data to class conditional distributions or densities. Fitting RCS data to SL models requires estimating model parameters using MLE or EM method for each test data using the different class conditional densities. This is an exhaustive process and accounts for the relatively longer computational times recorded by the SL classifiers. In general, Table~\ref{computational_time} shows that for the test datasets, the ML classifiers are much faster than the DL and the SL algorithms.

\subsection{Classification Performance with Limited RCS Data Extracted from a Segment of the Target UAV}

In Section~\ref{complete_azimuth}, we presented the plot of the average classification accuracy versus SNR using all the RCS measured from all around the UAV (($\phi\in[0^{\circ}, 360^{\circ}$) with a 2$^{\circ}$ increment). However, in practice, a radar may only be able to capture RCS reflection data from a limited area of the target UAV. For this reason, we decide to evaluate the performance of the classification algorithms in the case where the radar only captures RCS data from a 120$^{\circ}$ segment of the target. We assume the target is at the center of the segment (centered around 0$^{\circ}$ or ($\phi\in[-60^{\circ}, 60^{\circ}$]). 

\begin{figure}[t]
 \center
 \includegraphics[width=0.45\textwidth]{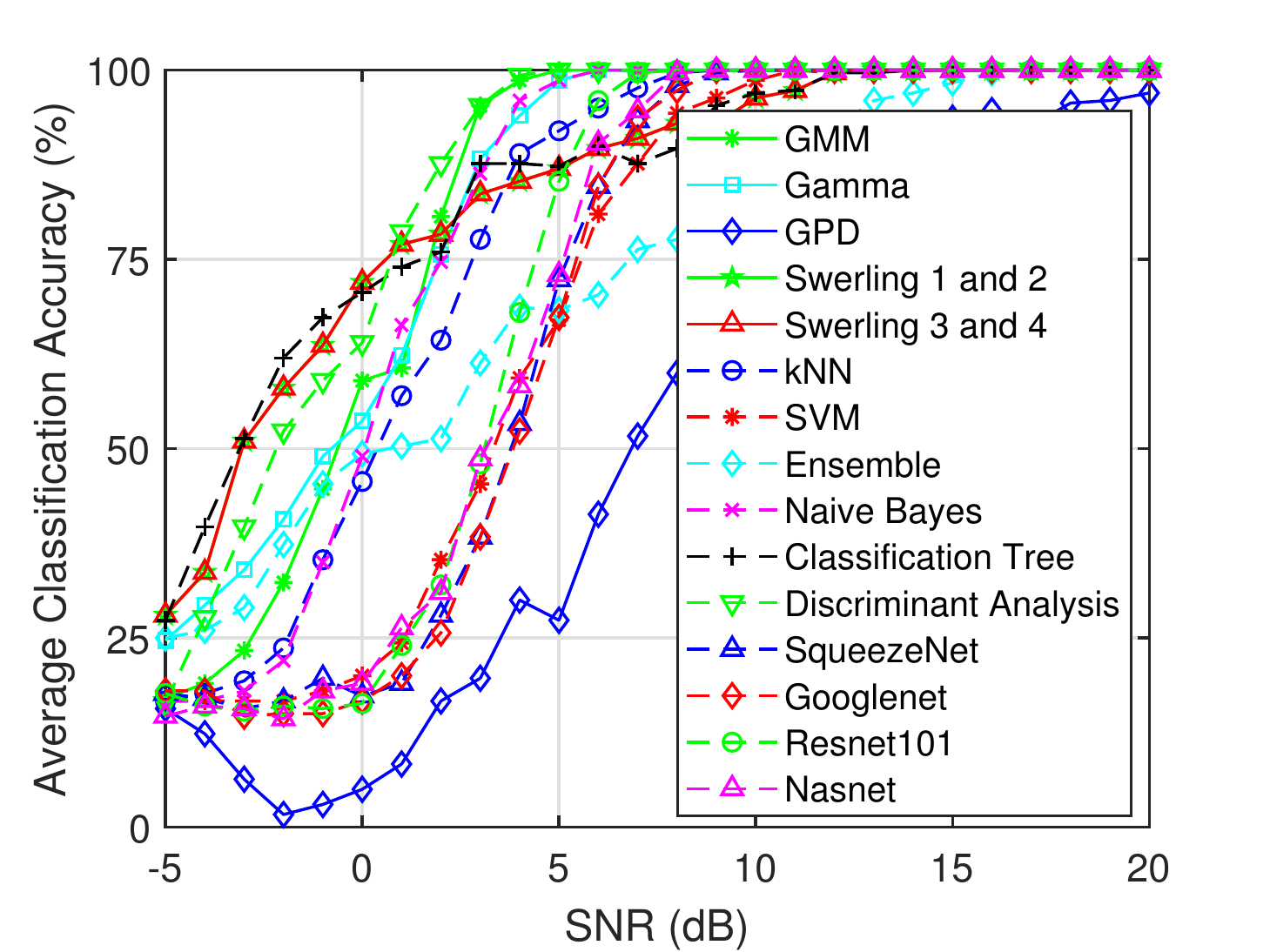}
\caption{Average classification accuracy versus SNR using limited azimuth HH-polarized RCS data.}
\label{classification_incomplete_azimuth}
\end{figure}

Fig.~\ref{classification_incomplete_azimuth} shows the average classification versus SNR for the 15~GHz HH-polarized RCS using the limited segment RCS returns from radar. Although with a little disparity in the accuracy values, Fig.~\ref{classification_incomplete_azimuth} is similar to Fig.~\ref{HH_15GHZ} which is obtained by using the RCS from all angles of the target UAVs. For example, at 0~dB SNR, Fig.~\ref{HH_15GHZ} shows that the average classification accuracy of the classification tree, discriminant analysis, Swerling 3 and 4, and the GPD algorithms are 79\%, 61\%, 80.3\%, and 19\% respectively. On the other hand,  Fig.~\ref{classification_incomplete_azimuth} shows that the average classification accuracy of the classification tree, discriminant analysis, Swerling 3 and 4, and the GPD algorithms are 70.67\%, 64\%, 72\%, and 5\% respectively. Also, at 5~dB SNR, Fig.~\ref{HH_15GHZ} shows that the average classification accuracy of the classification tree, discriminant analysis, Swerling 3 and 4, and the GPD algorithms are 100\%, 100\%, 100\%, and 58.33\% respectively. On the other hand, at 5~dB SNR, Fig.~\ref{classification_incomplete_azimuth} shows that the average classification accuracy of the classification tree, discriminant analysis, Swerling 3 and 4, and the GPD algorithms are 87.33\%, 100\%, 87\%, and 27.33\% respectively. Therefore, we can conclude that the classification algorithms perform better with RCS data from every part of the UAV as compared to the RCS data from a limited segment of the UAV.

\section{Conclusion}
The article presents a comparative analysis of RCS-based techniques for UAV classification. We evaluate 15 different classifiers using the RCS measured from six UAVs at 15 GHz and 25 GHz. A brief description of the UAV RCS measurement setup and the classification methods is provided. Each of the 15 classifiers considered falls under one of three categories: SL, ML, and DL techniques. Our study shows that on average, the classification tree and Swerling statistical models achieved the best classification accuracy at low SNR scenarios. On the other hand, the SVM and GPD performed relatively worse compared to all other approaches. 

In addition, we study  the computational or time complexity of all the 15 classifiers by averaging the results of 10 Monte Carlo experiments. The  results show that the classification tree ML classifier is the fastest with an average classification time of 0.46~ms, followed by the kNN and discriminant analysis classifiers with an average classification time of around 0.9~ms. Also, the time complexity analysis shows that the SL classifiers were the slowest due to the cost of model fitting and parameter estimation. In comparison, the average classification time of the SL classifiers is above 500~ms. 

Lastly, we investigate the case where the surveillance radar only obtains the RCS or scattered data from a limited segment of the target UAV. In this case, the average classification accuracy was a little lower as compared to the situation where the radar obtains RCS reflection from all sides on the target UAV. Even with a segment RCS data, the 15 classifiers could still recognize the UAV with good accuracy, especially at SNR above 5~dB. However, the average classification accuracy is better when the radar has obtains RCS from all segment of the target UAV. In the future, we hope to investigate how to discriminate UAVs and birds using RCS-based techniques. Also, we will consider the possibility of fusing data from radars and other sensor modalities (such as radio frequency, video camera, and microphone/acoustics) for improved UAV detection and identification.



\bibliography{IEEEabrv,references}
\bibliographystyle{IEEEtran}
\end{document}